\documentclass{tlp}
\usepackage{aopmath}

\newtheorem{definition}{Definition} 
\newtheorem{example}{Example}       
\newtheorem{remark}{Remark}         

\usepackage{amsfonts}
\usepackage{amssymb}

\providecommand{\LyX}{L\kern-.1667em\lower.25em\hbox{Y}\kern-.125emX\@}

\newlength{\LyXMinipageIndent}
\newcommand{\U}{\raisebox{1.pt}{\bf \b{ }}}
\newcommand{\eop}{\hfill $\Box$}
\renewcommand{\mathit}{\displaystyle}  

\newcommand{\ar}{\longrightarrow}
\newcommand{\arr}{\ {\stackrel{\geqslant}{\longrightarrow}}\ }
\newcommand{\larr}{\ {\stackrel{\geqslant}{\longleftrightarrow}}\ }
\newcommand{\llarr}{\ {\stackrel{_{_{\textstyle =}}}{\longleftrightarrow}}\ }

\newcommand{\feq}{$\begin{array}{rlll}}
\newcommand{\nex}{\end{array}$\par $\begin{array}{rlll}}
\newcommand{\eeq}{\end{array}$}

\newcommand{\num}[1]{\hspace*{-.8cm}\makebox[.8cm][r]{#1}&}
\newcommand{\Feq}[1]{\feq\num{#1}}
\newcommand{\Nex}[1]{\nex\num{#1}}

\newcommand{\nc}{&\hspace*{-.2cm}} 

\newcommand{\bdy}{\\ \nc\nc\nc}

\newcommand{\oeq}[2]{\par\smallskip
$\begin{array}{rl}\hspace*{-.8cm}
\makebox[.8cm][r]{#1}&{#2}\end{array}$
\par\smallskip}

\begin{document}
\bibliographystyle{acmtrans}

\long\def\comment#1{}

\title[Transformations of Logic Programs
with Goals as Arguments]
{Transformations of Logic Programs\\
with Goals as Arguments}

\author[A. Pettorossi and M. Proietti]
{ALBERTO PETTOROSSI\\
Dipartimento di Informatica, Sistemi e Produzione,\\
Universit\`a di Roma Tor Vergata,
Via del Politecnico 1, I-00133 Roma, Italy\\
\email{alberto.pettorossi@uniroma2.it}
\and
MAURIZIO PROIETTI\\
IASI-CNR, Viale Manzoni 30, I-00185 Roma, Italy\\
\email{proietti@iasi.rm.cnr.it} }

\pagerange{\pageref{firstpage}--\pageref{lastpage}}
\volume{\textbf{nn} (n):}
\jdate{month 2003}
\setcounter{page}{1}
\pubyear{2003}

\maketitle

\label{firstpage}

\begin{abstract}
We consider a simple extension of logic programming where
variables may range over goals and goals may be arguments of predicates.
In this language we can write logic programs which use goals as
data. We give practical evidence that, by exploiting this capability
when transforming programs, we can improve program efficiency. 

We propose a set of program transformation rules which extend the
familiar unfolding and folding rules and allow us to manipulate clauses
with goals which occur as arguments of predicates. 
In order to prove the correctness
of these transformation rules, we formally define the operational
semantics of our extended logic programming language. This semantics
is a simple variant of LD-resolution. When suitable conditions are
satisfied this semantics agrees with LD-resolution and, thus, the
programs written in our extended language can be run by ordinary Prolog
systems. 

Our transformation rules are shown to preserve the operational semantics
and termination.
\end{abstract}


\section{Introduction}

\noindent Program transformation is a very powerful and widely recognized
methodology for deriving programs from specifications. The {\it rules} $\!+\!$
{\it strategies} approach to program transformation was 
advocated in the 1970s by
Burstall and Darlington~\citeyear{BuD77} for developing first order functional
programs. Since then Burstall and Darlington's approach has been followed
in a variety of language paradigms, including logical languages~\cite{TaS84}
and higher order functional languages~\cite{San96}. The distinctive
feature of the rules $\!+\!$ strategies approach is that it allows
us to separate the concern of proving the correctness of programs
with respect to specifications from the concern of achieving computational
efficiency. Indeed, the correctness of the derived programs is ensured
by the use of semantics preserving transformation rules, 
whereas the computational
efficiency is achieved through the use of suitable strategies which
guide the application of the rules. The preservation of the semantics
is proved once and for all, for some given sets of transformation
rules, and if we restrict ourselves to suitable classes of programs,
we can also guarantee the effectiveness of the strategies for improving
efficiency.

In this paper we will argue through some examples, that a simple extension
of logic programming may give extra power to the program transformation
methodology based on rules and strategies. This extension consists
in allowing the use of variables which range over goals, called {\it goal
variables}, and the use of goals which are arguments of predicates,
called {\it goal arguments}.

In the practice of logic programming the idea of having goal variables
and goal arguments is not novel. The reader may look, for instance,
at~\cite{StS86,War82}.
Goal variables and goal arguments
can be used for expressing the meaning of logical connectives
and for writing programs in a {\it continuation passing
style}~\cite{TaB90,Wan80} as the following example shows.

\begin{example}
The following program \( P1 \):

\smallskip
\Feq{} F\vee G\leftarrow F
\Nex{} F\vee G\leftarrow G
\eeq
\smallskip

\noindent expresses the meaning of the {\it or} connective. The following
program \( P2 \): 

\smallskip
\Feq{}p([\, ],\, \mathit{Cont})\leftarrow \mathit{Cont}
\Nex{}p([X|\mathit{Xs}],\, \it{Cont})\leftarrow
   p(\mathit{Xs},q(X,\, \mathit{Cont}))
\Nex{}q(0,\, \mathit{Cont})\leftarrow \mathit{Cont}
\eeq

\smallskip
\noindent uses the goal variable {\it Cont} which denotes
a continuation. The goal $p(l,\, \mathit{true})$  succeeds in \( P2 \)
iff the list \( l \) consists of 0's only.\eop
\end{example}

\noindent
Programs with goal variables and goal arguments, such as
\( P1 \) and \( P2 \) in the above example, are not allowed by the
usual first order syntax of Horn clauses, where variables cannot occur
as atoms and predicate symbols are distinct from function symbols.
Nevertheless, these programs can be run by ordinary Prolog systems
whose operational semantics is based on LD-resolution, that
is, SLD-resolution with the leftmost selection rule. For the concepts
of {\it LD-resolution}, {\it LD-derivation}, and {\it LD-tree}
the reader may refer to~\cite{Apt97}

The extension of logic programming we consider in this paper, allows
us to write programs which use goals as data. This extension turns
out to be useful for performing program manipulations which are required
during program transformation and are otherwise impossible. For instance,
we will see that by using goal variables and goal arguments, we are
able to perform  goal rearrangements (also called {\it goal reorderings}
in~\cite{Bo&95}) which are often required for folding, without affecting
program termination and without increasing nondeterminism. 

Goal rearrangement is a long standing issue in logic program transformation.
Indeed, although the unfold/fold transformation rules by Tamaki and
Sato~\citeyear{TaS84} preserve the least Herbrand model, they may require
goal rearrangements and thus, they may not preserve the operational
semantics based on LD-resolution. Moreover, goal rearrangements may
increase nondeterminism by requiring that predicate calls have to
be evaluated before their arguments are sufficiently instantiated,
and in many Prolog systems, insufficiently instantiated calls of built-in
predicates may cause errors at run-time. In~\cite{BoC94} it
has been proved that by ruling out goal rearrangements, if some suitable
conditions hold, then the unfolding, folding, and goal replacement
transformation rules preserve the operational semantics of logic programs
based on LD-resolution and, in particular, these rules preserve {\it universal
termination}, that is, the finiteness of all LD-derivations~\cite{Apt97,VaP86}.
But, unfortunately, if we forbid goal rearrangements, many useful
program transformations are no longer possible.

In this paper we will show through some examples that in our simple
extension of logic programming we can restrict goal rearrangements
to leftward moves of goal equalities. We will also show that these
moves preserve universal termination and do not increase nondeterminism,
and thus, the deterioration of
performance of the derived program is avoided.

The following simple example illustrates the essential idea of our
technique which is based on the use of goal equalities. More complex
examples will be presented in Sections~\ref{sec:motivating_ex} 
and~\ref{sec:examples}. 

\begin{example}
\label{example:goal_variable}Suppose that during program transformation
we are required to fold a clause of the form:

\oeq{1.}{p(X)\leftarrow a(X),\ b(X),\ c(X)}

\noindent by using a clause of the form:

\oeq{2.}{q(X)\leftarrow a(X),\ c(X)}

\noindent We can avoid a leftward move of the atom $c(X)$ by
introducing, instead, an {\it equality between a goal variable and
a goal}, thereby transforming clause 1 into the following clause:

\oeq{3.}{ p(X)\leftarrow a(X),\ G\! =\! c(X),\ b(X),\ G}

\noindent Now we introduce the following predicate \( q' \)
which takes the goal variable \( G \) as an argument:

\oeq{4.}{\mathit{q}'(X,G)\leftarrow a(X),\ G\! =\! c(X)}

\noindent Then we fold clause 3 using clause 4, thereby getting the
clause:

\oeq{5.}{p(X)\leftarrow \mathit{q}'(X,G),\ b(X),\ G}

\noindent At this point we may continue the program transformation
process by transforming clause 4, which defines the predicate \( q' \),
instead of clause 2, which defines the predicate \( q \). For instance,
we may want to unfold clause 4 w.r.t.~the goal \( c(X) \) occurring
as an argument of the equality predicate.\eop 
\end{example}
\noindent As this example indicates, during program transformation
we need to have at our disposal some transformation rules which can
be used when goals occur as arguments. Indeed, in this paper:

\noindent
(i) we will introduce transformation rules for our logic language
which allows goals as arguments, 

\noindent
(ii) we will show through some examples that the use of these rules
makes it possible to improve efficiency without performing goal rearrangements
which increase nondeterminism, and

\noindent
(iii) we will prove that, under suitable conditions,
our transformation rules are correct in the
sense that they preserve the operational semantics of our logic language
and, in particular, they preserve universal termination. 

\medskip 

In order to show our correctness result, we will first define the
operational semantics of our logic language with goal arguments and
goal variables. This semantics will be given in terms of ordinary
LD-resolution, except for the following two important cases which
we now examine. 

The first case occurs when, during the construction of an LD-derivation,
we generate a goal which has an occurrence of an unbound goal variable
in the leftmost position. In this case we say that the LD-derivation
gets stuck. This treatment of unbound goal variables is in accordance
with that of most Prolog systems which halt with error when trying
to evaluate a call consisting of an unbound variable.

The second case occurs when we evaluate a goal equality of the form:
\( g_{1}\! =\! g_{2} \). In this case we stipulate that \( g_{1}\! =\! g_{2} \)
succeeds iff \( g_{1} \) {\it is a goal variable which does not
occur in} \( g_{2} \) and it gets stuck otherwise. (In particular,
for any goal \( g \) the evaluation of the equality \( g\! =\! g \)
gets stuck.) This somewhat restricted rule for the evaluation of goal
equalities is required for the correctness of our transformation rules,
as the following example shows. 

\begin{example}
\label{example:equality}Let us consider the program \( Q1 \):

\smallskip
\Feq{1.}{h\leftarrow p(q)}
\Nex{2.}{p(G)\leftarrow G\! =\! q}
\Nex{3.}{q\leftarrow s}
\eeq

\smallskip

\noindent where \( h,p,q, \) and \( s \) are predicate symbols and
\( G \) is a goal variable. If we unfold the goal argument \( q \)
in clause 1 using clause 3, we get the clause:

\oeq{4.}{h\leftarrow p(s)}

\noindent and we have the new program \( Q2 \) made out of clauses
2, 3, and 4. By using ordinary LD-resolution and unification, the
goal \( h \) succeeds in the original program \( Q1 \), while it
fails in the derived program \( Q2 \), because \( s \) does not
unify with \( q \).\eop 
\end{example}
\noindent This example shows that the set of successes is {\it not}
preserved by unfolding w.r.t.~a goal argument. Similar incorrectness
problems also arise with other transformation rules, such as folding
and goal replacement. These problems come from the fact that operationally
equivalent goals (such as \( q \) and \( s \) in the above example)
are not syntactically equal. 

In contrast, if we consider our restricted rule for the evaluation
of goal equalities, the LD-derivation which starts from the goal \( h \)
and uses the program \( Q1 \), gets stuck when the goal \( q\! =\! q \)
is selected. Also the LD-derivation which starts from the goal \( h \)
and uses the derived program \( Q2 \), gets stuck when the goal \( s=q \)
is selected. Thus, the unfolding w.r.t.~the argument \( q \) has
preserved the operational semantics based on LD-resolution with our
restricted rule for evaluating goal equalities.

In this paper we will consider two forms of correctness for our program
transformations: weak correctness and strong correctness. Suppose
that we have transformed a program \( P_{1} \) into a program \( P_{2} \)
by applying our transformation rules. We say that this transformation
is {\it weakly} {\it correct} iff, for any ordinary goal, that is, a goal
without occurrences of goal variables and goal arguments,
the following two properties hold: (i)~if \( P_{1} \) universally terminates,
then \( P_{2} \) universally terminates, and (ii)~if both \( P_{1} \)
and \( P_{2} \) universally terminate, then they compute the same
set of {\it most general} answer substitutions. The transformation
from \( P_{1} \) to \( P_{2} \) is {\it strongly} {\it correct}
iff (i)~it is weakly correct, and (ii) for any ordinary goal, if \( P_{2} \)
universally terminates, then \( P_{1} \) universally terminates.

Thus, when a transformation is weakly correct, the transformed program
may be more defined than the original program in the sense that there
may be some goals which have no semantic value in the original program
(that is, either their evaluation does not terminate or it gets stuck),
whereas they have a semantic value in the transformed program (that
is, their evaluation terminates and it does not get stuck).

\medskip{}
This paper is organized as follows. In Section \ref{sec:motivating_ex}
we present an introductory example to motivate the language extension
we will propose in this paper, and the transformation rules for this
extended language. In Section \ref{sec:syntax} we give the definition
of the syntax of our extended logic language with goal variables and
goal arguments. In Section \ref{sec:semantics} we introduce the operational
semantics of our extended language. 

In Sections \ref{sec:rules} and
\ref{sec:correctness} we present the transformation rules and the
conditions under which these rules are either weakly correct or strongly
correct.
For this purpose it is crucial that we assume that: (i)~the evaluation
of any goal variable gets stuck if that variable is unbound, and (ii)~the
evaluation of goal equalities is done according to the restricted
rule we mentioned above. We will also show that, if a goal does not
get stuck in a program, and we transform this program by using our
rules, then the given goal does not get stuck in the transformed program.
In this case, as it happens in the examples given in this paper, our
operational semantics agrees with LD-resolution, and we can execute
our transformed program by using ordinary Prolog systems. 

In Section
\ref{sec:examples} we give some more examples of program transformation
using our extended logic language and our transformation rules. We
also give practical evidence that these transformations improve program
efficiency. In Section \ref{sec:final} we make some final remarks
and we compare our results with related work.

\section{A Motivating Example\label{sec:motivating_ex}}

In order to present an example which motivates the introduction of
goal variables and goal arguments, we begin by recalling a well-known
program transformation strategy, called {\it tupling}
{\it strategy}~\cite{PeP94}.
Given a program where some predicate calls require common subcomputations
(detected by a suitable program analysis), the tupling strategy is
realized by the following three steps.

\smallskip
\noindent \hrulefill \nopagebreak

\noindent
\textit{The Tupling Strategy} \nopagebreak

\smallskip

\noindent ({\it Step} A) We introduce a new predicate defined by
a clause, say \( T \), whose body is the conjunction of the predicate
calls with common subcomputations. 

\smallskip{}
\noindent ({\it Step} B) We derive a program for the newly defined
predicate which avoids redundant common subcomputations. This step
can be divided into the following three substeps: (B.1) first, we unfold
clause \( T \), (B.2) then, we apply the goal replacement rule to
avoid redundant goals, and (B.3) finally, we fold using clause \( T \).

\smallskip{}
\noindent ({\it Step} C) By suitable folding steps using clause \( T \),
we express the predicates which are inefficiently computed by the
initial program, in terms of the predicate introduced at Step~(A).

\noindent \hrulefill

\smallskip{}
\noindent A difficulty encountered when applying the tupling strategy
is that, in order to apply the folding rule as indicated at Steps
(B) and (C), it is often necessary to rearrange the atoms in the body
of the clauses and, as already discussed in the Introduction, these
rearrangements may affect program termination or increase nondeterminism. 

The following example shows that this difficulty in the application
of the tupling strategy can be overcome by introducing goal variables
and goal arguments. 

\begin{example}
\label{example:deepest_1}Let us consider the following program {\it Deepest}:

\smallskip

\Feq{1.}\mathit{deepest}(l(N),N)\leftarrow
\Nex{2.}\mathit{deepest}(t(L,R),X) \nc \leftarrow \nc
        \mathit{depth}(L,\mathit{DL}),\ \mathit{depth}(R,\mathit{DR}),\
        \mathit{DL}\geq \mathit{DR},
        \bdy \mathit{deepest}(L,X)
\Nex{3.}\mathit{deepest}(t(L,R),X)\nc \leftarrow \nc
        \mathit{depth}(L,\mathit{DL}),\
        \mathit{depth}(R,\mathit{DR}),\ \mathit{DL}\leq \mathit{DR},\
        \bdy \mathit{deepest}(\mathit{R},X)
\Nex{4.}\mathit{depth}(l(N),1)\leftarrow
\Nex{5.}\mathit{depth}(t(L,R),D)\nc \leftarrow
        \nc\mathit{depth}(L,\mathit{DL}),\
        \mathit{depth}(R,\mathit{DR}),\
        \mathit{max}(\mathit{DL},\mathit{DR},M),\
        \bdy \mathit{plus}(M,1,D)
\eeq

\smallskip
\sloppy

\noindent
where \( \mathit{deepest}(T,X) \) holds iff \( T \) is
a binary tree and \( X \) is the label of one of the deepest leaves
of \( T \). The two calls \( \mathit{depth}(L,\mathit{DL}) \) and
\( \mathit{deepest}(L,X) \) in clause~2 may generate common redundant
calls of the \( \mathit{depth} \) predicate.
Indeed, both $\mathit{depth}(t(L1,R1),N)$
and $\mathit{deepest}(t(L1,R1),X)$ generate two calls of the
form $\mathit{depth}(L1,\mathit{DL})$ and
$\mathit{depth}(R1,\mathit{DR})$. In accordance with the tupling strategy, we
transform the given program as follows.

\fussy

\medskip

\noindent ({\it Step} A) We introduce the following new predicate:

\oeq{6.}{\mathit{dd}(T,D,X)\leftarrow \mathit{depth}(T,D),\,
\mathit{deepest}(T,X)}

\medskip

\noindent ({\it Step} B.1) We apply a few times the unfolding rule,
and we derive:

\smallskip
\Feq{7.}\mathit{dd}(l(N),1,N)\leftarrow
\Nex{8.} \mathit{dd}(t(L,R),D,X)\!\nc \leftarrow \nc
\mathit{depth}(L,\mathit{DL}),\ \mathit{depth}(R,\mathit{DR}),\
\bdy \mathit{max}(\mathit{DL},\mathit{DR},M),\ \mathit{plus}(M,1,D),\
\bdy \mathit{depth}(L,\mathit{DL}1),\ \mathit{depth}(R,\mathit{DR}1),\
\bdy \mathit{DL}1\geq \mathit{DR}1,\ \mathit{deepest}(L,X)
\Nex{9.}\mathit{dd}(t(L,R),D,X)\!\nc \leftarrow \nc
\mathit{depth}(L,\mathit{DL}),\ \mathit{depth}(R,\mathit{DR}),\
\bdy \mathit{max}(\mathit{DL},\mathit{DR},M),\ \mathit{plus}(M,1,D),\
\bdy \mathit{depth}(L,\mathit{DL}1),\ \mathit{depth}(R,\mathit{DR}1),\
\bdy \mathit{DL}1\leq \mathit{DR}1,\ \mathit{deepest}(R,X)
\eeq

\medskip
\noindent ({\it Step} B.2) Since \( \mathit{depth} \) is {\it functional}
with respect to its first argument, by applying the goal replacement
rule we delete the atoms \( \mathit{depth}(L,\mathit{D}L1) \) and
\( \mathit{depth}(R,\mathit{DR}1) \), in clauses 8 and 9, and we
replace the occurrences of \( \mathit{D}L1 \) and \( \mathit{DR}1 \)
by \( \mathit{D}L \) and \( \mathit{DR} \), respectively, thereby
getting the following clauses 10 and 11:

\smallskip
\Feq{10.}\mathit{dd}(t(L,R),D,X)\nc  \leftarrow \nc
         \mathit{depth}(L,\mathit{DL}),\
         \mathit{depth}(R,\mathit{DR}),\
         \mathit{max}(\mathit{DL},\mathit{DR},M),
\bdy     \mathit{plus}(M,1,D),\ \mathit{DL}\geq \mathit{DR},\
         \mathit{deepest}(L,X)
\Nex{11.}\mathit{dd}(t(L,R),D,X)\nc \leftarrow \nc
         \mathit{depth}(L,\mathit{DL}),\
         \mathit{depth}(R,\mathit{DR}),\
         \mathit{max}(\mathit{DL},\mathit{DR},M),\
\bdy  \mathit{plus}(M,1,D),\ \mathit{DL}\leq \mathit{DR},\
      \mathit{deepest}(R,X)
\eeq

\medskip
\noindent ({\it Step} B.3) In order to fold clause 10 using clause~6,
we move \( \mathit{deepest}(L,X) \) immediately to the right of
\( \mathit{depth}(L,\mathit{DL}) \). Similarly, in the body of clause~11 we
move \( \mathit{deepest}(R,X) \) immediately to the right of \(
\mathit{depth}(R,\mathit{DR}) \). Then, by folding we derive:

\smallskip
\Feq{12.}\mathit{dd}(t(L,R),D,X)\nc  \leftarrow \nc
         \mathit{dd}(L,\mathit{DL},X),\
         \mathit{depth}(R,\mathit{DR}),\
         \mathit{max}(\mathit{DL},\mathit{DR},M),
\bdy     \mathit{plus}(M,1,D),\ \mathit{DL}\geq \mathit{DR}
\Nex{13.}\mathit{dd}(t(L,R),D,X)\nc \leftarrow \nc
         \mathit{depth}(L,\mathit{DL}),\
         \mathit{dd}(R,\mathit{DR},X),\
         \mathit{max}(\mathit{DL},\mathit{DR},M),\
\bdy  \mathit{plus}(M,1,D),\ \mathit{DL}\leq \mathit{DR}
\eeq

\medskip
\noindent ({\it Step} C) Finally, we fold clauses 2 and 3 using clause
6, so that to evaluate the predicates \( \mathit{depth} \) and \( \mathit{deepest} \)
we use the predicate \( dd \), instead. Also for these folding steps
we have to suitably rearrange the order of the atoms. By folding,
we derive the following program {\it Deepest}1:

\smallskip

\Feq{1.}\mathit{deepest}(l(N),N)\leftarrow
\Nex{14.}\mathit{deepest}(t(L,R),D,X)\nc  \leftarrow \nc
         \mathit{dd}(L,\mathit{DL},X),\
         \mathit{depth}(R,\mathit{DR}),\
         \mathit{DL}\geq \mathit{DR}
\Nex{15.}\mathit{deepest}(t(L,R),D,X)\nc \leftarrow \nc
         \mathit{depth}(L,\mathit{DL}),\
         \mathit{dd}(R,\mathit{DR},X),\
         \mathit{DL}\leq \mathit{DR}
\Nex{7.}\mathit{dd}(l(N),1,N)\leftarrow
\Nex{12.}\mathit{dd}(t(L,R),D,X)\nc  \leftarrow \nc
         \mathit{dd}(L,\mathit{DL},X),\
         \mathit{depth}(R,\mathit{DR}),\
         \mathit{max}(\mathit{DL},\mathit{DR},M),
\bdy     \mathit{plus}(M,1,D),\ \mathit{DL}\geq \mathit{DR}
\Nex{13.}\mathit{dd}(t(L,R),D,X)\nc \leftarrow \nc
         \mathit{depth}(L,\mathit{DL}),\
         \mathit{dd}(R,\mathit{DR},X),\
         \mathit{max}(\mathit{DL},\mathit{DR},M),\
\bdy  \mathit{plus}(M,1,D),\ \mathit{DL}\leq \mathit{DR}
\eeq

\smallskip
\noindent In order to evaluate a goal of the form \( \mathit{deepest}(t,X) \),
where \( t \) is a ground tree and \( X \) is a variable, we may
construct an LD-derivation using the program {\it Deepest}1 which
does not generate redundant calls of \( \mathit{depth} \). This LD-derivation
performs only one traversal of the tree \( t \) and has linear length with
respect to~the size of~\( t \). However, this LD-derivation is
constructed in a nondeterministic way, and if the corresponding LD-tree
is traversed in a depth-first manner, like most Prolog systems do,
the program exhibits an inefficient generate-and-test behaviour. Thus,
in practice, the tupling strategy may diminish program efficiency. 

The main reason  of this decrease of efficiency is that, in order to fold
clause 10, we had to move the atom \( \mathit{deepest}(L,X) \) to a position
to the left
of \( \mathit{DL}\geq \mathit{DR} \), and this move forces the evaluation
of calls of \( \mathit{deepest}(L,X) \) even  when \( \mathit{DL}\geq
\mathit{DR} \) fails. (Notice that the move of \( \mathit{deepest}(R,X) \) to
the left of \( \mathit{DL}\leq \mathit{DR} \) is harmless because \(
\mathit{DL}\leq \mathit{DR} \) is evaluated after the failure of \(
\mathit{DL}\geq \mathit{DR} \) and, thus, \( \mathit{DL}\leq \mathit{DR} \)
never fails.)\eop \end{example}
\smallskip{}
In the following example we will present an alternative program derivation
which starts from the same initial program {\it Deepest}. In this alternative
derivation we will use our extended logic language which will be formally defined
in the following Section~\ref{sec:syntax}. As already mentioned in the
Introduction, when writing programs in our extended language, we may use:
(i)~the goal equality predicate \( = \), (ii)~goal variables occurring at top
level in the body of a clause, and (iii)~the disjunction predicate
$\vee$. This alternative program derivation avoids harmful goal rearrangements
and produces an efficient program without redundant subcomputations.

\begin{example}
\label{example:deepest_2}Let us consider the program {\it Deepest}
listed at the beginning of Example~\ref{example:deepest_1} consisting of
clauses 1--5. By using disjunction in the body of a clause, clauses 2 and 3
can be rewritten as follows:

\smallskip{}
\Feq{16.}\mathit{deepest}(t(L,R),X)\nc \leftarrow
 \mathit{depth}(L,\mathit{DL}),\ \mathit{depth}(R,\mathit{DR}),
\Nex{}\hspace*{1.5cm}
((\mathit{DL}\!\geq\!\mathit{DR},\,\mathit{deepest}(L,\!X))\,\vee
\,(\mathit{DL}\!\leq\!\mathit{DR},\,\mathit{deepest}(\mathit{R},\!X)))
\eeq

\smallskip{}
\noindent After this initial transformation step the derived program, call it
{\it DeepestOr}, consists of clauses 1, 4, 5, and 16.

\medskip

Now we consider an extension of the tupling strategy which
makes use of the transformation rules for logic programs with goal
arguments and goal variables. These rules will be formally presented
in Section~\ref{sec:rules}. We proceed as follows. 

\medskip

\noindent ({\it Step} A) We introduce the following new predicate
\( g \) which takes a goal variable \( G \) as an argument:

\oeq{17.}{g(T,D,X,G)\leftarrow \mathit{depth}(T,D),\, G\! =\! 
\mathit{deepest}(T,X)}

\noindent
Notice also that in clause 17
the goal \( \mathit{deepest}(T,X) \) occurs as an argument
of the equality predicate.

\medskip
\noindent ({\it Step} B) We derive a set of clauses for the newly
defined predicate \( g \) as follows. 

\medskip
\noindent ({\it Step} B.1)
We unfold clause 17 w.r.t.~\( \mathit{depth}(T,D) \)
and we derive: 

\smallskip
\Feq{18.}g(l(N),1,X,G)\leftarrow G\! =\! \mathit{deepest}(l(N),X)
\Nex{19.}g(t(L,R),D,X,G)\nc \leftarrow \nc
\mathit{depth}(L,\mathit{DL}),\ \mathit{depth}(R,\mathit{DR}),\
\mathit{max}(\mathit{DL},\mathit{DR},M),
\bdy \mathit{plus}(M,1,D),\ G\!=\!\mathit{deepest}(t(L,R),X)
\eeq

\smallskip
\noindent Now, by unfolding clauses 18 and 19 w.r.t.~the atoms
with the \( \mathit{deepest} \) predicate, we derive:

\smallskip
\Feq{20.}g(l(N),1,N,\mathit{true})\leftarrow
\Nex{21.}g(t(L,R),D,X,G)\nc \leftarrow \nc
\mathit{depth}(L,\mathit{DL}),\ \mathit{depth}(R,\mathit{DR}),
\Nex{}\hspace*{1.5cm}\mathit{max}(\mathit{DL},\mathit{DR},M),\
\mathit{plus}(M,1,D),
\Nex{}\hspace*{1.5cm}G\! =\!
(\mathit{depth}(L,\mathit{DL}1),\ \mathit{depth}(R,\mathit{DR}1),
\Nex{}\hspace*{2.3cm}
((\mathit{DL}1\!\geq\!\mathit{DR}1,\,\mathit{deepest}(L,X) )  \vee
 (\mathit{DL}1\!\leq\!\mathit{DR}1,\,\mathit{deepest}(R,X))))
\eeq

\medskip
\noindent ({\it Step} B.2) We perform two goal replacement steps
based on the functionality of \( \mathit{depth} \), and from clause 21 we
derive:

\smallskip
\Feq{22.}g(t(L,R),D,X,G)\nc \leftarrow \nc
\mathit{depth}(L,\mathit{DL}),\ \mathit{depth}(R,\mathit{DR}),
\Nex{}\hspace*{1.5cm}\mathit{max}(\mathit{DL},\mathit{DR},M),\
\mathit{plus}(M,1,D),
\Nex{}\hspace*{1.5cm}G\!=\!
((\mathit{DL}\!\geq\!\mathit{DR},\,\mathit{deepest}(L,X)) \vee
 (\mathit{DL}\!\leq\!\mathit{DR},\,\mathit{deepest}(R,X)))
\eeq

\medskip
\noindent ({\it Step} B.3) In order to fold clause 22 using clause
17, we first introduce goal equalities and we
then perform suitable leftward moves of those goal equalities. We derive
the following clause:

\smallskip
\Feq{23.}g(t(L,R),D,X,G)\nc \leftarrow \nc
\mathit{depth}(L,\mathit{DL}),\ \mathit{GL}\! =\!
  \mathit{deepest}(L,X),
\Nex{}\hspace*{1.5cm}\mathit{depth}(R,\mathit{DR}),\
\mathit{GR}\! =\! \mathit{deepest}(\mathit{R},X),
\Nex{}\hspace*{1.5cm}\mathit{max}(\mathit{DL},\mathit{DR},M),\
\mathit{plus}(M,1,D),
\Nex{}\hspace*{1.5cm}G\!=\!
((\mathit{DL}\!\geq\!\mathit{DR}, \mathit{GL}) \vee
 (\mathit{DL}\!\leq\!\mathit{DR}, \mathit{GR}))
\eeq

\smallskip
\noindent Notice that we can move the goal equality
\( \mathit{GL}\! =\! \mathit{deepest}(L,X) \)
to the left of the test \( \mathit{DL}\! \geq \! \mathit{DR} \) without
altering the operational semantics of our program. Indeed, this goal
equality succeeds and binds the goal variable \( \mathit{GL} \) to
the goal \( \mathit{deepest}(L,X) \) without evaluating
it. The goal \( \mathit{deepest}(L,X) \) will be evaluated
only when \( \mathit{GL} \) is called. A similar remark holds for
the goal equality \( \mathit{GR}\! =\! \mathit{deepest}(L,X) \).
Now, by folding twice clause 23 using clause 17, we get:

\smallskip
\Feq{24.}g(t(L,R),D,X,G) \leftarrow g(L,\mathit{DL},X,\mathit{GL}),\
g(R,\mathit{DR},X,\mathit{GR}),
\Nex{}\hspace*{1.5cm}\mathit{max}(\mathit{DL},\mathit{DR},M),\
\mathit{plus}(M,1,D),
\Nex{}\hspace*{1.5cm}G\!=\!
((\mathit{DL}\!\geq\!\mathit{DR},\mathit{GL}) \vee
 (\mathit{DL}\!\leq\!\mathit{DR},\mathit{GR}))
\eeq

\medskip
\noindent ({\it Step} C) Now we express the predicate \( \mathit{deepest} \)
in terms of the new predicate \( g \) by transforming clause 16 as
follows: (i)~we first replace the two \( \mathit{deepest} \) atoms by the
goal variables \( GL \) and \( GR, \) (ii)~we then introduce suitable
goal equalities, (iii)~we then suitably move
to the left the goal equalities, and (iv)~we finally fold using clause
17. We derive the following clause: 

\smallskip
\Feq{25.}\mathit{deepest}(t(L,R),X)\leftarrow
    g(L,\mathit{DL},X,\mathit{GL}),\
    g(R,\mathit{DR},X,\mathit{GR}),
\Nex{}\hspace*{1.5cm}
((\mathit{DL}\!\geq\!\mathit{DR},\mathit{GL}) \vee
 (\mathit{DL}\!\leq\!\mathit{DR},\mathit{GR}))
\eeq

\smallskip
\noindent Our final program {\it Deepest}2 is as follows:

\smallskip
\Feq{1.}\mathit{deepest}(l(N),N)\leftarrow
\Nex{25.}\mathit{deepest}(t(L,R),X)\leftarrow
    g(L,\mathit{DL},X,\mathit{GL}),\
    g(R,\mathit{DR},X,\mathit{GR}),
\Nex{}\hspace*{1.5cm}
((\mathit{DL}\!\geq\!\mathit{DR},\mathit{GL}) \vee
 (\mathit{DL}\!\leq\!\mathit{DR},\mathit{GR}))
\Nex{20.}g(l(N),1,N,\mathit{true})\leftarrow
\Nex{24.}g(t(L,R),D,X,G) \leftarrow
\Nex{}\hspace*{1.5cm}g(L,\mathit{DL},X,\mathit{GL}),\
g(R,\mathit{DR},X,\mathit{GR}),
\Nex{}\hspace*{1.5cm}\mathit{max}(\mathit{DL},\mathit{DR},M),\
\mathit{plus}(M,1,D),
\Nex{}\hspace*{1.5cm}G\!=\!
((\mathit{DL}\!\geq\!\mathit{DR},\mathit{GL}) \vee
 (\mathit{DL}\!\leq\!\mathit{DR},\mathit{GR}))
\eeq

\smallskip
\noindent 
Now, when we evaluate a goal of the form \( \mathit{deepest}(t,X) \),
where \( t \) is a ground tree and \( X \) is a variable, {\it Deepest}2
does not generate redundant
calls and it performs only one traversal of the tree \( t \).
{\it Deepest}2 is more efficient than {\it Deepest} because
in the worst case {\it Deepest}2
performs \( O(n) \) LD-resolution steps to compute
an answer to \( \mathit{deepest}(t,X) \), where $n$ is the number of nodes 
of $t$, while the initial program {\it Deepest}
takes \( O(n^{2}) \)  LD-resolution steps. 
The program {\it Deepest}2
can be run by an ordinary Prolog system and computer experiments confirm
substantial efficiency improvements with respect to~the initial program
{\it Deepest} (see Section~\ref{sec:speedups}). 

Efficiency improvements, although smaller, 
are obtained also when comparing the
final program {\it Deepest}2 with respect to the intermediate program
{\it DeepestOr} which has been obtained from the initial program
{\it Deepest} by replacing clauses~2
and 3 by clause~16, thereby avoiding the repetition of the common goals
in clauses~2 and 3. Indeed, although more
efficient than {\it Deepest}\/ in the worst case, the program {\it DeepestOr}
still takes a quadratic number of LD-resolution steps
to compute an answer to \( \mathit{deepest}(t,X) \).
\eop
\end{example}
\noindent In Section~\ref{sec:examples}
we will present more examples of program derivation and
we will also provide some experimental results.

\section{The Extended Logic Language with Goals as Arguments
\label{sec:syntax}}

Let us now formally define our extended logic language. 
Suppose that the following pairwise disjoint sets are given: 
(i){\it ~individual
variables}: \( X,X_{1},X_{2},\ldots  ,\) (ii){\it ~goal variables}:
\( G,G_{1},G_{2},\ldots  ,\) (iii)~{\it function symbols}
(with arity): \( f,f_{1},f_{2},\ldots , \) (iv)~{\it primitive
predicate symbols}: \( \mathit{true} \), \( \mathit{false} \), 
\( =_{\mathit{t}} \)
(denoting equality between terms), \( =_{\mathit{g}} \) (denoting
equality between goals), and (v) {\it predicate symbols} (with arity):
\( p,p_{1},p_{2},\ldots  \) Individual and goal variables are collectively
called {\it variables}, and they are ranged over by
\( V,V_{1},V_{2},\ldots  \)
Occasionally, we will feel free to depart from these naming conventions,
if no confusion arises.

\noindent \textit{Terms}: \( t,t_{1},t_{2},\ldots  \), \textit{goals}:
\textit{\( g,g_{1},g_{2},\ldots  \)}, and \textit{arguments}: 
\textit{\( u,u_{1},u_{2},\ldots  \),}
have the following syntax: 

\smallskip{}
\( t::=X\, \, |\, \, f(t_{1},\ldots ,t_{n}) \)

\( g::=G\, \, |\, \, \mathit{true}\, \, |\, \, \mathit{false}\, \, |\, \,
t_{1}\! =_{t}\! t_{2}\, \, |\, \, g_{1}\! =_{g}\! g_{2}\, \,
|\, \, p(u_{1},\ldots ,u_{m})\, \, |\, \, g_{1}\wedge g_{2}\, \, |\, \,
g_{1}\vee g_{2} \)

\( u::=t\, \, |\, \, g \)

\smallskip{}
\noindent The binary operators \( \wedge  \) (conjunction) and \( \vee  \)
(disjunction) are assumed to be associative with neutral elements
{\it true} and {\it false}, respectively. Thus, a goal \( g \)
is the same as \( \mathit{true}\wedge g \) and \( g\wedge \mathit{true} \).
Similarly, \( g \) is the same as \( \mathit{false}\vee g \) and
\( g\vee \mathit{false} \). Goals of the form \( p(u_{1},\ldots ,u_{m}) \)
are also called {\it atoms}. In the sequel, for reasons of simplicity,
we will write \( = \), instead of \( =_{t} \) or \( =_{g} \),
and we leave it to the reader to distinguish between the two equalities
according to the context of use.
Notice that, according to our operational
semantics (see Section~\ref{sec:semantics}), 
$\vee$ is commutative, $\wedge$ is not
commutative, \( =_{t} \) is symmetric, and \( =_{g} \) is not symmetric.

\noindent {\it Clauses} \( c,c_{1},c_{2},\ldots  \) have the following
syntax:

\smallskip{}
\( c::=p(V_{1},\ldots ,V_{m})\leftarrow g \)
\smallskip{}

\noindent where \( p \) is a non-primitive predicate symbol and
\( V_{1},\ldots ,V_{m} \) are distinct variables. 
The atom \( p(V_{1},\ldots ,V_{m}) \)
is called the {\it head} of the clause and the goal \( g \) is called
the {\it body} of the clause. A clause of the form:
\( p(V_{1},\ldots ,V_{m})\leftarrow \mathit{true} \)
will also be written as \( p(V_{1},\ldots ,V_{m})\leftarrow  \).
\smallskip{}

\noindent \textit{Programs} \( P,P_{1},P_{2},\ldots  \) are sets
of clauses of the form: 

\smallskip{}
\( p_{1}(V_{1},\ldots ,V_{m1})\leftarrow g_{1} \) 

~~~~~~~~~~~~~~~~~~~~~~\( \vdots  \)

\( p_{k}(V_{1},\ldots ,V_{mk})\leftarrow g_{k} \)
\smallskip{}

\sloppy

\noindent where \( p_{1},\ldots ,p_{k} \) are distinct non-primitive
predicate symbols, and every non-primitive predicate symbol occurring
in \( \{g_{1},\ldots ,g_{k}\} \) is an element of 
\( \{p_{1},\ldots ,p_{k}\} \).
Each clause head has distinct variables as arguments. Given a program
\( P \) and a non-primitive predicate \( p \) occurring in \( P \),
the unique clause in \( P \) of the form 
\( p(V_{1},\ldots ,V_{m})\leftarrow g \),
is called the {\it definition} of \( p \) in \( P \). We say that
a predicate \( p \) is {\it defined} in a program \( P \) iff \( p \)
has a definition in \( P \).

\fussy

An {\it ordinary goal} is a goal without goal variables
or goal arguments. Formally, an ordinary goal has the following
syntax:

\smallskip
\( g::= \mathit{true}\, \, |\, \, \mathit{false}\, \, |\, \,
t_{1}\! =_{t}\! t_{2}\, \,
|\, \, p(t_{1},\ldots ,t_{m})\, \, |\, \, g_{1}\wedge g_{2}\, \, |\, \,
g_{1}\vee g_{2} \)

\smallskip
\noindent
where $t_1, t_2,\ldots,t_m$ are terms. {\it
Ordinary programs} are programs whose goals are ordinary goals.

\medskip{}
\noindent {\it Notes on syntax}.

\noindent (1)~When no confusion arises, we also use comma, instead
of \( \wedge  \), for denoting conjunction.

\noindent (2)~The assumption that in our programs clause heads have
only variables as arguments is not restrictive, because we may always
replace a non-variable argument, say \( u \), by a variable argument,
say \( V \), in the head of a clause, at the expense of adding the
extra equality \( V\! =\! u \) in the body.

\smallskip{}
\noindent (3)~The assumption that in every program there exists at
most one clause for each predicate symbol is not restrictive, because
one may use disjunctions in the body of clauses. In particular, every
definite logic program written by using the familiar syntax~\cite{Llo87},
can be rewritten into an equivalent program of our language by suitable
introductions of equalities and \( \vee  \) operators in the bodies
of clauses. 

\sloppy

\smallskip{}
\noindent (4)~Our logic language is a typed language in the sense
that: (i)~every individual variable has type \( \mathit{term} \),
(ii)~every function symbol of arity \( n \) has type 
\mbox{\( \mathit{term}^{n}\rightarrow \mathit{term} \)},
(iii)~{\it true}, {\it false}, and every goal variable have type
\( bool \), (iv.1)~\( =_{\mathit{t}} \) has type 
\mbox{\( \mathit{term}\times \mathit{term}\, \rightarrow \mathit{bool} \)},
(iv.2)~\( =_{\mathit{g}} \) has type 
\( \mathit{bool}\times \mathit{bool}\, \rightarrow \mathit{bool} \),
and (v)~every predicate symbol of arity \( n \) has a unique type
of the form: 
\( (\mathit{term}\mid \mathit{bool})^{n}\rightarrow \mathit{bool} \).
We assume that all our programs can be uniquely typed according to
the above rules. 

\fussy

\section{The Operational Semantics\label{sec:semantics}}

In this section we define the operational semantics of our extended
logic language. We choose a syntax-directed style of presentation
which makes use of {\it deduction rules}. For an elementary presentation
of this technique, sometimes called {\it structural operational semantics}
or {\it natural semantics}, the reader may refer to~\cite{Win93}.

Before defining the semantics of our logic language, we recall the
following notions.
By \( \{V_{1}/u_{1},\ldots ,V_{m}/u_{m}\} \) we denote
the substitution of \( u_{1},\ldots ,u_{m} \) for the variables 
\( V_{1},\ldots ,V_{m} \).
As usual, we assume that the \( V_{i} \)'s are all distinct and
for \( i\! =\! 1,\ldots ,m \), \( u_{i} \) is distinct from \( V_{i} \).
By \( \varepsilon  \) we denote the identity substitution. 
By \( \vartheta \! \upharpoonright \! S \)
we denote the {\it restriction} of the substitution \( \vartheta  \)
to set \( S \) of variables, that is,
\( \vartheta \! \upharpoonright \! S=\) \(\{V\!/\!u\, \, |\, \,
V\!/\!u \in \vartheta\, {\rm and}\, V\!\!\in\! S\} \).
Given the substitutions \( \vartheta ,\eta _{1},\ldots ,\eta _{k} \),
by \( \vartheta \circ \{\eta _{1},\ldots ,\eta _{k}\} \) we denote
the set of substitutions 
\( \{\vartheta \eta _{1},\ldots ,\vartheta \eta _{k}\} \)
(where, as usual, juxtaposition of substitutions denotes 
composition~\cite{Llo87}).
By \( g\vartheta  \) we denote the application of the substitution
\( \vartheta  \) to the goal \( g \). By \( \mathit{mgu}(t_{1},t_{2}) \)
we denote a relevant, idempotent, most general unifier of the terms
\( t_{1} \) and \( t_{2} \).

\sloppy

The set of all substitutions is denoted by {\it Subst} and the set
of all {\it finite} subsets of {\it Subst} is denoted by
\( \mathcal{P}(\mathit{Subst}) \).
Given \( A,B\in \mathcal{P}(\mathit{Subst}) \), we say that \( A \)
and \( B \) are {\it equally general} with respect to~a goal \( g \)
iff (i)~for every \( \alpha \in A \) there exists \( \beta \in B \)
such that \( g\alpha  \) is an instance of \( g\beta  \), and symmetrically,
(ii)~for every \( \beta \in B \) there exists $\alpha$~$\in$~$A$
such that \( g\beta  \) is an instance of \( g\alpha  \). For example,
\( A=\{\{X/t\},\, \{X/Y\},\, \{X/Z\}\} \) and \( B=\{\{X/W\}\} \)
are equally general with respect to the goal \( p(X) \).

Given a set of substitutions \( A\in \mathcal{P}(\mathit{Subst}) \)
and a goal \( g \), let \( \mathit{mostgen}(A,g) \) denote a largest
subset of \( \{g\vartheta \, |\, \vartheta \in A\} \) such that for
any two goals \( g_{1} \) and \( g_{2} \) in \( \mathit{mostgen}(A,g) \),
\( g_{1} \) is not an instance of \( g_{2} \). For example,
 $\mathit{mostgen}(\{\{X/t\}, \{X/Y\}, \{X/Z\}\},  p(X))$ = $\{p(Y)\}$.
Notice that the set denoted by {\it mostgen} is not uniquely determined.
However, it can be shown that, whatever choice we make for the set
denoted by {\it mostgen}, any two sets of substitutions \( A \)
and \( B \) are equally general with respect to~a goal \( g \)
iff there exists a bijection \( \rho  \) from \( \mathit{mostgen}(A,g) \)
to \( \mathit{mostgen}(B,g) \) such that for any goal
\( h\in \mathit{mostgen}(A,g) \),
\( \rho (h) \) is a variant of \( h \). In this case we write
\( \mathit{mostgen}(A,g) \)
\( \approx  \) \( \mathit{mostgen}(B,g) \).

\fussy

We use \( g[u] \) to denote a goal \( g \) in which we have
selected an occurrence of its subconstruct \( u \), where \( u \)
may be either a term or a goal. By \( g[\_] \) we denote the goal
\( g[u] \) without the selected occurrence of its subconstruct \( u \).
We say that \( g[\_] \) is a {\it goal context}.
For any syntactic construct \( r \), we use \( \mathit{vars}(r) \)
to denote the set of variables occurring in \( r \) and, 
for any set \( \{r_{1},\ldots ,r_{m}\} \) of syntactic constructs,
we use \( \mathit{vars}(r_{1},\ldots ,r_{m}) \) to denote the set
of variables \( \mathit{vars}(r_{1})\cup\ldots\cup\mathit{vars}(r_{m}) \). 
In particular, given a substitution
\( \vartheta  \), a variable belongs to \( \mathit{vars}(\vartheta ) \)
iff it occurs either in the domain of \( \vartheta  \) or in the
range of \( \vartheta  \). Given two goals \( g \) and \( g_{1} \)
and a clause \( c \) of the form 
\( p(V_{1},\ldots ,V_{m})\leftarrow g[g_{1}] \),
the {\it local variables} of \( g_{1} \) in \( c \) are those in
the set \( \mathit{vars}(g_{1})-(\{V_{1},\ldots ,V_{m}\}\cup 
\mathit{vars}(g[\_])) \). 

Given a program \( P \), we define the semantics of \( P \) as a
ternary relation \( P\vdash g\mapsto A \), where \( g \) is a goal
and \( A \) is a finite set of substitutions, meaning that for \( P \)
and \( g \) all derivations are finite and \( A \) is the finite
set of {\it answer substitutions} which are computed by these derivations.
The relation \( P\vdash g\mapsto A \) is defined by the
deduction rules given in Figure~\ref{fig:deduction_rules}.

\begin{figure}
\normalsize
\begin{centering}
\begin{minipage}{12.6cm}

\vspace*{5mm}
\noindent 
$(\mathit{tt})\, \, \, \, \, \, \, \, \, \, \, \frac{}{\mathit{P}\, \vdash
\mathit{true}\, \mapsto \, \{\varepsilon \}}$

\bigskip

\noindent $(\mathit{ff})\, \, \, \, \, \, \, \, \, \, \, \frac{}{\mathit{P}\, \vdash
\mathit{false}\wedge g\, \mapsto \, \emptyset }$

\bigskip

\noindent $(\mathit{teq}1)\, \, \, \, \,  \frac{}{\mathit{P}\, \vdash \, (t_{1}\! =\!
t_{2})\wedge g\, \mapsto \, \emptyset }$
\hfill $\textrm{if
}t_{1}\textrm{ and }t_{2}\textrm{ are non}\mbox {-}\textrm{unifiable terms}$

\bigskip

\noindent 
$(\mathit{teq}2)\, \, \, \, \, \frac{\mathit{P}\, \vdash g\, 
\mathit{mgu}(t_{1},t_{2})\,
\mapsto \, A}{\mathit{P}\, \vdash \, (t_{1}\! =\! t_{2})\wedge g\, \mapsto \,
(\mathit{mgu}(t_{1},t_{2})\! \circ \! A)}$ \hfill$\textrm{if
}t_{1}\textrm{ and }t_{2}\textrm{ are unifiable terms}$

\noindent 

\bigskip

$(\mathit{geq})\, \, \, \, \, \, \, \displaystyle{\frac{\mathit{P}\,
\vdash \, g_{2}\{G/g_{1}\}\,
\mapsto \, A}{\mathit{P}\, \vdash \, (G\! =\! g_{1})\wedge g_{2}\, \mapsto \,
(\{G/g_{1}\}\! \circ \! A)}} $

\smallskip

\hfill $\textrm{if the goal variable }G$
\textrm{is not in} \it{vars}$(g_{1})$

\bigskip

\noindent  
$(\mathit{at})\, \, \, \, \, \, \, \, \, \, \frac{\mathit{P}\, \vdash \,
g_{1}\{V_{1}/u_{1},\ldots ,V_{m}/u_{m}\}\wedge g\, \mapsto 
\, A}{\mathit{P}\, \vdash \,
p(u_{1},\ldots ,u_{m})\wedge g\, \mapsto \, A\! \upharpoonright \! S}$ 

\smallskip

\hfill {\rm where $p(V_{1},\ldots ,V_{m})\leftarrow g_{1}$
is a renamed apart clause of \( P \)}

\hfill {\rm and
\( S \) is $\mathit{vars}(p(u_{1},\ldots ,u_{m})\wedge g)$}

\bigskip

\noindent 
$(\mathit{or})\, \, \, \, \, \, \, \, \, \, \frac{\mathit{P}\, \vdash \,
g_{1}\wedge g\, \mapsto \, A_{1}\, \, \, \, \, \, \, \, \, \mathit{P}\, \vdash
\, g_{2}\wedge g\, \mapsto \, A_{2}}{\mathit{P}\, \vdash \, (g_{1}\vee
g_{2})\wedge g\, \mapsto \, (A_{1}\cup A_{2})}$

\end{minipage}
\end{centering}
\caption{Operational Semantics\label{fig:deduction_rules}}
\end{figure}

A {\it deduction tree} \( \tau  \) for \( P\vdash g\mapsto A \)
is a tree such that: (i) the root of \( \tau  \) is \( P\vdash g\mapsto A \),
and (ii) for every node \( n \) of \( \tau  \) with sons 
\( n_{1},\ldots ,n_{k} \)
(with \( k\geq 0 \)), there exists an instance of a deduction rule,
say {\it r}, whose conclusion is \( n \) and whose premises are
\( n_{1},\ldots ,n_{k} \). We say that \( n \) {\it is derived
by applying rule} \( r \) to \( n_{1},\ldots ,n_{k} \). A {\it proof}
of \( P\vdash g\mapsto A \) is a finite deduction tree for
\( P\vdash g\mapsto A \) where every leaf is a deduction rule which
has no premises.

We say that \( P\vdash g\mapsto A \) {\it holds} iff there exists
a proof of \( P\vdash g\mapsto A \). If \( P\vdash g\mapsto A \) holds
and \( A\not =\emptyset  \),
then we say that \( g \) {\it succeeds} in \( P \), written
\( P\vdash g\! \downarrow \! \mathit{true} \). Otherwise,  if 
\( P\vdash g\mapsto\emptyset\) holds, then
we say that \( g \) {\it fails} in \( P \), written \( P\vdash g\!
\downarrow \! \mathit{false} \). If \( g \) either succeeds or fails in \( P
\) we say that \( g \) {\it terminates} in \( P \). We say that a goal \( g
\) is {\it stuck} iff it is either of the form \( G\wedge g_{1} \), where
\( G \) is a goal variable, or of the form 
\( (g_{0}\! =\! g_{1})\wedge g_{2} \),
where either \( g_{0} \) is a non-variable goal 
or \( g_{0} \) is a goal variable
occurring in \( g_{1} \). We say that \( g \) {\it gets stuck}
in \( P \) iff there exist a set \( A \) of substitutions and a
(finite or infinite) deduction tree \( \tau  \) for \( P\vdash g\mapsto A \)
such that a leaf of \( \tau  \) is of the form 
\( P\, \vdash \, g_{1}\mapsto B \)
and \( g_{1} \) is stuck. For instance, the goal \( (G\! =\! p)\wedge 
(G\! =\! q) \)
gets stuck in any program \( P \). We say that \( g \) is {\it safe}
in \( P \) iff \( g \) does not get stuck in \( P \). 

For every program \( P \) and goal \( g \), the three cases: (i)~\( g \)
succeeds in \( P \), (ii)~\( g \) fails in \( P \), and (iii)~\( g \)
gets stuck in \( P \), are pairwise mutually exclusive, but not exhaustive.
Indeed, there is a fourth case in which 
the unique maximal deduction tree with root \( P\vdash g\mapsto A \)
is infinite and each of its leaves, if any, is the conclusion of a
deduction rule which has no premises.
In this case no \( A \) exists such that \( P\vdash g\mapsto A \)
holds and \( g \) does not get stuck in \( P \).

\medskip{}

\noindent {\it Notes on semantics.}

\nopagebreak
\noindent (1)~In our presentation of the deduction rules we have
exploited the assumption that $\wedge$ and $\vee$ are associative
operators with neutral elements {\it true} and {\it false}, respectively.
For instance, we have not introduced the rule 
\( \frac{}{
{\mathit P\, \vdash \, \mathit{false}\, \mapsto \, \emptyset }}\) 
because it is an instance of rule $({\it ff})$ for $g\!=\!\mathit{true}$.

\smallskip{}

\noindent (2)~Given a program \( P \) and a goal \( g \), if there
exists a proof for \( P\, \vdash \, g\mapsto A \) for some \( A \),
then the proof is unique up to isomorphism. More precisely, given
two proofs, say \( \pi _{1} \) for \( P\, \vdash \, g\mapsto A_{1} \)
and \( \pi _{2} \) for \( P\, \vdash \, g\mapsto A_{2} \), there
exists a bijection \( \rho  \) from the nodes of \( \pi _{1} \)
to the nodes of \( \pi _{2} \) which preserves the application of
the deduction rules and if 
\( \rho (P\, \vdash \, g_{1}\mapsto B_{1})=P\, 
\vdash \, g_{2}\mapsto B_{2} \) then 

\smallskip{}
\hspace*{-2cm}
\begin{tabular}{rl}
(i)&
\( g_{1} \) is a variant of \( g_{2} \),
and~~~~~~~~~~~~~~~~~~~~~~~~~~~~~~~~~~~~~\\
(ii)&
\( \forall \beta _{1}\! \in \! B_{1}\, \, \exists \beta _{2}\! \in \! 
B_{2} \) such that \( g_{1}\beta _{1} \)
is a variant of \( g_{2}\beta _{2} \),~and\\
(iii)&
\( \forall \beta _{2}\! \in \! B_{2}\, \, \exists \beta _{1}\! \in \! 
B_{1} \) such that \( g_{2}\beta _{2} \)
is a variant of \( g_{1}\beta _{1} \). \\
\end{tabular}

\smallskip{}
\noindent This property is a consequence of the fact that: (i)~for
any program \( P \) and goal \( g \), there exists at most one rule
instance whose conclusion is of the form \( P\vdash g\mapsto A \)
for some \( A \), and (ii)~our rules for the operational semantics
are deterministic, in the sense that no choice has to be made when
one applies them  during the construction of a proof, 
apart from the choice of how to compute the
most general unifiers and how to rename apart the clauses.

In particular, any two sets \( A_{1} \) and \( A_{2} \) of answer
substitutions for a program \( P \) and a goal \( g \), are related
as follows: if \( P\vdash g\mapsto A_{1} \) and \( P\vdash g\mapsto A_{2} \)
then  \( \forall \alpha _{1}\! \in \! A_{1}\, \exists \alpha _{2}\! \in \!
A_{2} \)~\( g\alpha _{1} \) is a variant of \( g\alpha _{2} \) and \( \forall
\alpha _{2}\! \in \! A_{2}\, \exists \alpha _{1}\! \! \in A_{1} \)~\( g\alpha
_{2} \) is a variant of \( g\alpha _{1} \). Thus, \( A_{1} \) and \( A_{2} \)
are equally general with respect to~\( g \). The same property holds also for
any two sets of computed answer substitutions which are constructed by
LD-resolution (recall that by LD-resolution we can construct different sets of
computed answer substitutions by choosing different most general unifiers and
different variable renamings).

Notice that, if \( P\vdash g\mapsto A_{1} \) and \( P\vdash g\mapsto A_{2} \)
hold, then \( A_{1} \) and \( A_{2} \) may have different cardinality.
Indeed, let us consider the program \( P \) consisting of the following
clause only:

\smallskip{}
\( p(X,Y,Z)\leftarrow (X\! =\! Y\wedge Z\! =\! Y)\, \vee \, 
(X\! =\! Z\wedge Y\! =\! Z) \) 

\smallskip{}
\noindent In this case, since both \( Z/Y \) and \( Y/Z \) are most
general unifiers of \( Y\! =\! Z \), we have that both
\( P\vdash p(X,Y,Z)\mapsto \{\{X/Y,Z/Y\},\, \{X/Z,Y/Z\}\} \)
and \( P\vdash p(X,Y,Z)\mapsto \{\{X/Y,Z/Y\}\} \) hold. Notice also
that the substitution \( \{X/Y,Z/Y\} \) is more general than the
substitution \( \{X/Z,Y/Z\} \) and vice versa. 

\smallskip{}
\noindent (3)~If \( P\vdash g\mapsto A \) and \( \vartheta \in A \),
then the domain of \( \vartheta  \) is a subset of \( \mathit{vars}(g) \).

\smallskip{}
\noindent (4)~In the presentation of the deduction rules for the
ternary relation \mbox{$P\vdash g\mapsto A$}, the program \( P \) never
changes and thus, it could have been omitted. However, the explicit
reference to \( P \) is useful for presenting our Correctness Theorem
(see Theorem~\ref{th:correctness} in Section~\ref{sec:correctness}). 

\smallskip{}
\noindent (5)~We assume that in any relation \( P\vdash g\mapsto A \),
the program \( P \) and the goal \( g \) have consistent types,
that is, the type of every function and predicate symbol should be
the same in \( P \) and in \( g \). For instance, if 
\( P=\{p(G)\leftarrow \} \)
where \( G \) is a goal variable, then \( P\vdash p(0)\mapsto 
\{\varepsilon \} \)
does {\it not} hold, because in the program \( P \) 
the predicate \( p \) has type
\( \mathit{bool}\rightarrow \mathit{bool} \),
while in the goal \( p(0) \) the predicate $p$ 
has type \( \mathit{term}\rightarrow \mathit{bool} \).
Moreover, for any relation \( P\vdash g_{1}\mapsto A_{1} \) occurring
in the proof of \( P\vdash g\mapsto A \), we have that program \( P \)
and goal \( g_{1} \) have consistent types. 

\medskip{}
Now we discuss the relationship between LD-resolution and the operational
semantics defined in this section. Apart from the style of presentation
(usually LD-resolution is presented by means of the notions of 
{\it LD-derivation}
and {\it LD-tree}~\cite{Apt97,Llo87}), LD-resolution differs from
our operational semantics only in the treatment of goal equality.
Indeed, by using LD-resolution, the goal equality \( g_{1}\! =\! g_{2} \)
is evaluated by applying the ordinary unification algorithm also in
the case where \( g_{1} \) is not a goal variable or \( g_{1} \)
is a goal variable occurring in \( \mathit{vars}(g_{2}) \). 
In contrast, according to our
operational semantics, a goal of the form \( g_{1}\! =\! g_{2} \)
is evaluated by unifying \( g_{1} \) and \( g_{2} \), only if \( g_{1} \)
is a variable which does not occur in \( \mathit{vars}(g_{2}) \)
(see rule ({\it geq}) above). 

Thus, if a goal \( g \) is safe in \( P \), then the evaluation
of \( g \) according to our operational semantics agrees with the
one which uses LD-resolution in the following sense: if \( g \) is
safe in \( P \), then there exists a set \( A \) of answer substitutions
such that \( P\vdash g\mapsto A \) holds iff: (i)~all LD-derivations
starting from \( g \) and using \( P \) are finite (that is, \( g \)
{\it universally terminates} in \( P \)~\cite{Apt97,VaP86}), and
(ii)~\( A \) is the set of the computed answer substitutions obtained
by LD-resolution. Point~(i) follows from the fact that in our operational
semantics, the evaluation of a disjunction of goals (see the ({\it or})
rule) requires the evaluation of each disjunct. Thus, in order to
compute the relation \( P\vdash g\mapsto A \) in the case where \( g \)
is safe in \( P \), we can use any ordinary Prolog system which implements
LD-resolution.

Notice that, given a program \( P \) and a goal \( g \), if the
LD-tree has an infinite LD-derivation, then no set \( A \) of answer
substitutions exists such that \( P\vdash g\mapsto A \). In particular,
for the program \( P=\{p(0)\leftarrow ,\, \, p(X)\leftarrow p(X)\} \)
no \( A \) exists such that \( P\vdash p(X)\mapsto A \), while the
set of computed answer substitutions constructed by LD-resolution
for the program \( P \) and the goal \( p(X) \) is the singleton
consisting of the substitution \( \{X/0\} \) only.

It may also be the case that a goal \( g \) is not safe in a program
\( P \) (thus, there exists no set \( A \) of answer substitutions
such that \( P\vdash g\mapsto A \) holds) while, by using LD-resolution,
\( g \) succeeds or fails in \( P \). For instance, for any program
and for any two distinct nullary predicates \( p \) and \( q \),
(i)~the goal \( p\! =\! p \) is not safe, while it succeeds by using
LD-resolution and (ii)~the goal \( p\! =\! q \) is not safe, while
it fails by using LD-resolution.

We recall that our interpretation of goal equality is motivated by
the fact that we want the operational semantics to be preserved by
program transformations and, in particular, by unfolding. As already
shown in the Introduction, unfortunately, unfolding does not preserve
the operational semantics based on ordinary LD-resolution.

\medskip

The following Proposition~\ref{proposition:cas} establishes an important
property of our operational semantics. This property is useful for
the proof the correctness results in Section~\ref{sec:correctness}
(see Theorem~\ref{th:correctness}). The proof of this proposition
is similar to the one in the case of LD-resolution for definite programs~(see,
for instance,~\cite{Llo87}) and will be omitted.

\begin{proposition}
\textup{\label{proposition:cas} Let \( P \) be a program, \( g \)
be an ordinary goal, and \( A \) be a set of substitutions such that
\( P\vdash g\mapsto A \). Then, for all \( \vartheta \in \mathit{Subst} \),
the following hold:}

\medskip

\noindent
\begin{tabular}{rl}
(i)&\hspace*{-.3cm}\( g\vartheta  \) terminates, that is,
either \( P\vdash g\vartheta
\downarrow \mathit{true} \) or~\( P\vdash g\vartheta \downarrow
\mathit{false} \), and\\
(ii.1)&\hspace*{-.3cm}\( P\vdash g\vartheta \downarrow \mathit{true} \)
iff there exists \( \alpha \in A \) such that \( g\vartheta  \)
is an instance of \( g\alpha  \), and\\
(ii.2)&\hspace*{-.3cm}\( P\vdash g\vartheta \downarrow \mathit{false} \)
iff it does {\it not} exist \( \alpha \in A \) such that \( g\vartheta  \)
is an instance of \( g\alpha  \).\\
\end{tabular}
\end{proposition}

\noindent Let us conclude this section by introducing the notions
of {\it refinement} and {\it equivalence} between programs which
we will use in Section~\ref{sec:correctness} to state the weak and
strong correctness of the program transformations that can be
realized by applying our transformation rules. These rules
are presented in the next section.

\begin{definition}[Refinement and Equivalence]\label{def:ref_equiv}
Given
two programs \( P_{1} \) and \( P_{2} \), we say that \( P_{2} \)
is a {\it refinement} of \( P_{1} \), written
\( P_{1}\sqsubseteq P_{2} \), iff for every ordinary goal \( g \)
and for every \( A\in \mathcal{P}(\mathit{Subst}) \), if
\( P_{1}\vdash g\mapsto A \) then there exists  \( B\in
\mathcal{P}(\mathit{Subst}) \) such that:

\smallskip{}

(1)~\( P_{2}\, \vdash \, g\mapsto B \) and 

(2)~\( A \)
and \( B \) are equally general with respect to~\( g \).

\smallskip{}

\noindent \textup{We say that \( P_{1} \) is} \textup{{\it equivalent}}
\textup{to \( P_{2} \), written \( P_{1}\equiv P_{2} \), iff 
\( P_{1}\sqsubseteq P_{2} \)
and \( P_{2}\sqsubseteq P_{1} \).}
\end{definition}

\begin{remark}
\label{re:most_gen_cas}
Recall that Condition~(2) can be written
as \( \mathit{mostgen}(A,g)\approx \mathit{mostgen}(B,g) \). In this
sense we will say that if \( P_{1}\sqsubseteq P_{2} \) and the ordinary
goal \( g \) terminates in \( P_{1} \), then the most general answer
substitutions for \( g \) are the same in \( P_{1} \) and \( P_{2} \),
modulo variable renaming.\eop
\end{remark}

\begin{remark}
\noindent \( P_{1}\sqsubseteq P_{2} \) implies that, for every ordinary
goal \( g \),

- if \( g \) succeeds in \( P_{1} \) then \( g \) succeeds in \( P_{2} \),
and

- if \( g \) fails in \( P_{1} \) then \( g \)
fails in \( P_{2} \).\eop
\end{remark}
\noindent Theorem~\ref{th:correctness} stated in
Section~\ref{sec:correctness} shows that, if from program \( P_{1} \) we
derive program \( P_{2} \) by using our transformation rules and suitable
conditions hold, then \( P_{1}\sqsubseteq P_{2} \). In this case we say that
the transformation is {\it weakly correct}. If additional conditions hold,
then we may have that \( P_{1}\equiv P_{2} \) and we say that the
transformation is {\it strongly correct}.

In Section~\ref{sec:correctness} we will also show that our transformation
rules preserve safety, that is, if from program \( P_{1} \) we derive
program \( P_{2} \) by using the transformation rules and goal \( g \)
is safe in \( P_{1} \), then goal \( g \) is safe also in \( P_{2} \).

\section{The Transformation Rules \label{sec:rules}}

In this section we present the transformation rules for our extended
logic language. We assume that starting from an initial program \( P_{0} \)
we have constructed the {\it transformation sequence}  \( P_{0},\ldots ,P_{i}
\)~\cite{PeP94,TaS84}. 
By an application of a transformation rule, from program \(
P_{i} \) we derive a new program \( P_{i+1} \).

\medskip{}
\noindent \textit{Rule R1} (\textit{Definition Introduction})

\noindent We derive the
new program \( P_{i+1} \) by adding to program \( P_{i} \) a new
clause, called a {\it definition}, of the form:
\smallskip{}

\( \mathit{newp}(V_{1},\ldots ,V_{m})\leftarrow g \)

\smallskip{}
\noindent where: (i) \textit{newp} is a new non-primitive predicate
symbol not occurring in any program of the sequence \( P_{0},\ldots ,P_{i} \),
(ii) the non-primitive predicate symbols occurring in \( g \) are
defined in \( P_{0} \), and (iii) \( V_{1},\ldots ,V_{m} \) are
some of (possibly all) the distinct variables occurring in \( g \).
\\
The set of all definitions introduced during the transformation sequence
\( P_{0},\ldots ,P_{i} \), is denoted by \( \mathit{Def}\! _{i} \).
Thus,~\( \mathit{Def}\! _{0}=\emptyset  \).
\medskip{}

\noindent
\textit{Rule R2} (\textit{Unfolding})

\noindent
Let \( c_{1} \): \( h\leftarrow \mathit{body}[p(u_{1},\ldots ,u_{m})] \)
be a renamed apart clause in program \( P_{i} \) where \( p \) is
a non-primitive predicate symbol. Let  \( d \): 
\( p(V_{1},\ldots ,V_{m})\leftarrow g \)
be a clause in \( P_{0}\cup \mathit{Def}\! _{i} \).
By \textit{unfolding} \( c_{1} \) {\it w.r.t.}~\( p(u_{1},\ldots ,u_{m}) \)
{\it using} \( d \) we derive the new clause \( c_{2} \):~\( h\leftarrow 
\mathit{body}[g\{V_{1}/u_{1},\ldots ,V_{m}/u_{m}\}] \).
We derive the new program \( P_{i+1} \) by replacing in program \( P_{i} \)
clause \( c_{1} \) by clause~$c_{2}$.
\medskip{}

\sloppy
\noindent
\textit{Rule R3} (\textit{Folding})

\noindent
Let \( c_{1} \): \( h\leftarrow \mathit{body}[g\vartheta ] \)
be a renamed apart clause in program \( P_{i} \) and let \( d \): 
\( \mathit{p}(V_{1},\ldots ,V_{m})\leftarrow g \)
be a clause in \( \mathit{Def}\! _{i} \). Suppose that, for every
local variable \( V \) of \( g \) in \( d \), we have that:
\vspace{-.17cm}
\begin{itemize}
\item[(1)]\( V\vartheta  \) is a local variable of \( g\vartheta  \) in \( c_{1} \),
and 
\item[(2)] the variable \( V\vartheta  \) does not occur in \( W\vartheta  \),
for any variable \( W \) occurring in \( g \) and different from
\( V \). 
\end{itemize}
\vspace{-.17cm} 
\noindent Then, by \textit{folding} \( c_{1} \)  \textit{using} \( d \)
we derive the new clause \( c_{2} \):  \( h\leftarrow
\mathit{body}[\mathit{p}(V_{1},\ldots ,V_{m})\vartheta ] \). We derive the new
program \( P_{i+1} \) by replacing in program \( P_{i} \) clause \( c_{1} \)
by clause~$c_{2}$. 

\fussy
\medskip{}

In order to present the goal replacement rule (see rule R4 below)
we introduce the notion of {\it replacement law}. Basically, a replacement
law denotes two goals which can be replaced one for the other in
the body of a clause. We have two kinds of replacement laws: the {\it weak}
and the {\it strong} replacement laws, which ensure weak and strong
correctness, respectively (see the end of this section for an informal
discussion and Section~\ref{sec:correctness} for a formal proof
of this fact). 

First we need the following
definition.

\begin{definition}[Depth of a Deduction Tree]
Let \( \tau  \)
be a finite deduction tree and let \( m \) be the maximal number
of applications of the ({\it at})~rule in a root-to-leaf path of
\( \tau  \). Then we say that \( \tau  \) has {\it depth} \( m \).\\
Let \( \pi  \) be a proof for \( P\, \vdash \, g\mapsto A \),
for some program \( P \), goal \( g \), and set \( A \) of substitutions,
and let \( m \) be the depth of \( \pi  \). If \( A\! =\! \emptyset  \)
we write \( P\, \vdash \, g\downarrow _{m}\mathit{false} \); otherwise,
if \( A\! \neq \! \emptyset  \) we write \( P\, \vdash \,  g\downarrow
_{m}\mathit{true} \).
\end{definition}
\noindent Recall that, given a program \( P \) and a goal \( g \),
if for some set \( A \) of substitutions there exists a proof for
\( P\, \vdash \, g\mapsto A \), then the proof is unique up to isomorphism.
In particular, given a proof for \( P\, \vdash \, g\mapsto A_{1} \)
and a proof for \( P\, \vdash \, g\mapsto A_{2} \), they have the
same depth.

\begin{definition}[Replacement Laws]
\noindent\label{def:repl_laws}Let \( P \) be a program, let \( g_{1} \) and
\( g_{2} \) be two goals, and let \( V \) be a set of
variables.

\smallskip{}
\noindent (i) The relation \( P\vdash\, \forall V \, (g_{1}\ar g_{2}) \)
holds iff for every goal context \( g[\_] \) such that 
\( \mathit{vars}(g[\_])\cap  \mathit{vars}(g_1,g_2) \subseteq V \),
and for every \( b\in \{\mathit{true},\mathit{false}\} \), we have
that:

\smallskip{}
if \( P\, \vdash \, g[g_{1}]\downarrow b \) ~then~ \( P\, \vdash \, 
g[g_{2}]\downarrow b \).
\hfill{}\( (\dagger ) \)~~~

\smallskip{}
\noindent (ii) The relation \( P\vdash \, \forall V \, (g_{1}\arr g_{2}) \), 
called a {\it weak replacement law}, holds
iff for every goal context \( g[\_] \) such that 
\( \mathit{vars}(g[\_])\cap  \mathit{vars}(g_1,g_2) \subseteq V \), 
and for every \( b\in
\{\mathit{true},\mathit{false}\} \), we have that:

\smallskip{}
if \( P\, \vdash \, g[g_{1}]\downarrow _{m}b \) ~then~ \( P\, \vdash \, 
g[g_{2}]\downarrow _{n}b \)~~with
\( m\! \geq \! n \). \hfill{}\( (\dagger \! \dagger ) \)~~~

\smallskip{}
\noindent (iii) The relation \( P\vdash \, \forall V \, (g_{1}\larr g_{2}) \),
called a {\it strong replacement law}, holds iff 
\( P\vdash \, \forall V \, (g_{1}\arr g_{2}) \) and \( P\vdash \, \forall V \,
(g_{2}\ar g_{1}) \).

\smallskip{}
\noindent (iv)~We write \( P\vdash \, \forall V \, (g_{1}\llarr g_{2}) \)
to mean that the strong replacement laws
 \( P\vdash \, \forall V \, (g_{1}\arr g_{2}) \)
and \( P\vdash \, \forall V \, (g_{2}\arr g_{1}) \) hold.
\end{definition}

\noindent 
If \( V\! =\! \emptyset  \) then 
\( P\vdash \, \forall V \, (g_{1}\arr g_{2}) \)
is also written as \( P\vdash \, g_{1}\arr g_{2} \). 
If \( V\! =\! \{V_1,\ldots,V_n\}  \) then 
\( P\vdash \, \forall V \, (g_{1}\arr g_{2}) \)
is also written as \( P\vdash \,\forall V_1,\ldots,V_n\,(g_{1}\arr g_{2}) \).
If \( V\! =\! {\it vars}(g_1,g_2)  \) then 
\( P\vdash \, \forall V \, (g_{1}\arr g_{2}) \)
is also written as \( P\vdash \, \forall \, (g_{1}\arr g_{2}) \). 

\medskip

A few comments on the above Definition~\ref{def:repl_laws} are now
in order. 

\noindent 
(1)~In the relation \( P\vdash\, \forall V \, (g_{1}\ar g_{2}) \)
we have used the set $V$ of universally quantified
variables  as a notational device for indicating that when we replace 
\( g_{1} \) by \( g_{2} \) in a clause 
\( h\leftarrow \mathit{body}[g_{1}] \), the variables
in common between \( h\leftarrow \mathit{body}[\_] \) 
and \( (g_{1}, g_{2})\) are those in $V$
(see the goal replacement rule R4 below). Thus, 
\( {\it vars}(g_{1})-V \) is the set of the local variables of $g_1$ in
\( h\leftarrow \mathit{body}[g_{1}] \) 
and \( {\it vars}(g_{2})-V \) is the set of  
the local variables of $g_2$ in
\( h\leftarrow \mathit{body}[g_{2}] \).

\noindent (2)~Implication \( (\dagger \! \dagger ) \) implies Implication
\( (\dagger ) \).

\noindent (3)~Every strong replacement law is also a weak replacement
law.

\noindent (4)~If \( P\vdash \, \forall V \, (g_{1}\llarr 
g_{2}) \) then there exists \( A_{1}\in \mathcal{P}(\mathit{Subst}) \) such
that \mbox{\(P\vdash g_{1}\mapsto A_{1}\)} has a proof of depth
\( m \) iff there exists \( A_{2}\in \mathcal{P}(\mathit{Subst}) \)
such that \( P\, \vdash \, g_{2}\mapsto A_{2} \) has a proof of depth
\( m \). Moreover, if both proofs exist, \( A_{1}\! =\! \emptyset  \)
iff \( A_{2}\! =\! \emptyset  \).

\medskip

The properties listed in the next proposition follow
directly from Definition~\ref{def:repl_laws}.

\begin{proposition}
\label{proposition:repl}
Let $P$ be a program, let \(g_{1}\) and \(g_{2}\) be goals, and
let $V$ be a set of variables. 

\smallskip

\noindent (i)~\( P\vdash \, \forall V\, (g_{1}\ar  g_{2}) 
\) holds iff for every goal context \( g[\_] \) such that \(
\mathit{vars}(g[\_])\cap {\it vars}(g_1,g_2) \subseteq V  \), 
 \( P\vdash \, \forall W \, (g[g_{1}]\ar g[g_{2}]) \) holds,
where $W=V\cup \mathit{vars}(g[\_])$.

\smallskip

\noindent
(ii)~\( P\vdash \, \forall V\, (g_{1}\ar  g_{2}) \) holds iff
\( P\vdash \, \forall W\, (g_{1}\ar  g_{2}) \) holds, 
where $W=V\cap \mathit{vars}(g_1,g_2)$.

\smallskip

\noindent
(iii)~\( P\vdash \, \forall V\, (g_{1}\ar  g_{2}) \) holds iff
for every $W\subseteq V$,
\( P\vdash \, \forall W\, (g_{1}\ar  g_{2}) \) holds.

\smallskip

\noindent
(iv)~\( P\vdash \, \forall V\, (g_{1}\ar  g_{2}) \) holds iff
for every substitution $\vartheta$ such that 
${\it vars}(\vartheta)\cap {\it vars}(g_1,g_2)\subseteq V$,
\( P\vdash \, \forall W\, (g_{1}\vartheta\ar  g_{2}\vartheta) \) holds, 
where $W={\it vars}(V\vartheta)$.

\smallskip

\noindent
(v)~\( P\vdash \, \forall V\, (g_{1}\ar  g_{2}) \) holds iff
for every renaming substitution $\rho$ such that 
${\it vars}(\rho)\cap V=\emptyset$,
\( P\vdash \, \forall V\, (g_{1}\rho\ar  g_{2}\rho) \) holds.

The properties obtained from (i) -- (v) 
by replacing $\ar$ by $\arr$ are also true. 
We will refer to them 
as Properties (i$'$) -- (v$'$), respectively.

\end{proposition}

\sloppy

\begin{definition}
We  say that a weak replacement law \( P\vdash \, \forall V \,
(g_{1}\arr g_{2}) \) (or a strong replacement law \(
P\vdash \, \forall V \, (g1\larr g_{2}) \))
{\it preserves safety}
iff for every goal context \( g[\_] \) such that 
\( \mathit{vars}(g[\_])\cap  \mathit{vars}(g_1,g_2) \subseteq V \), 
we have that:

\smallskip{}
if \( g[g_{1}] \) is safe in \( P \) then \( g[g_{2}] \)
is safe in \( P \).
\end{definition}

\fussy

\noindent
\textit{Rule R4} (\textit{Goal Replacement})

\noindent
Let \( c_{1} \):
\( h\leftarrow \mathit{body}[g_{1}] \) be a clause in program \( P_{i} \)
and let \( g_{2} \) be a goal such that: (i)~all non-primitive predicate
symbols occurring in \( g_{1} \) or \( g_{2} \) are defined in \( P_{0} \),
and either (ii.1)~\( P_{0}\vdash \, \forall V \, (g_{1}\arr g_{2}) \),
or (ii.2)~\( P_{0}\vdash \, \forall V \, (g_{1}\larr g_{2} )\),
where \( V = \mathit{vars}(h,\mathit{body}[\_]) \cap \mathit{vars}(g_1,g_2)\). 

\noindent By {\it goal replacement} we derive the new clause \( c_{2} \):
\( h\leftarrow \mathit{body}[g_{2}] \), and we derive the new program
\( P_{i+1} \) by replacing in program \( P_{i} \) clause \( c_{1} \)
by clause \( c_{2} \).

\noindent In case (ii.1)~we say that the goal replacement is {\it based
on a weak replacement law}. In case (ii.2)~we say that the goal replacement
is {\it based on a strong replacement law}. We say that the goal replacement
{\it preserves safety} iff it is based on a (weak or strong) replacement
law which preserves safety.

\smallskip{}
Implication~\( (\dagger \! \dagger ) \)
of Definition~\ref{def:repl_laws}
makes \( \arr  \) and \( \larr  \) to be {\it improvement} relations
in the sense of~\cite{San96}. As stated in Theorem~\ref{th:correctness}
of Section~\ref{sec:correctness}, 
Implication~\( (\dagger \! \dagger ) \)
is required for ensuring the weak correctness of a goal replacement
step, while Implication~\( (\dagger ) \) of
Definition~\ref{def:repl_laws}
does not suffice. This fact is illustrated by the following example.

\begin{example}
Let us consider the program \( P_{1} \): 

\Feq{1.} p\leftarrow q
\Nex{2.} q\leftarrow
\eeq

\smallskip{}
\noindent We have that \( P_{1}\vdash q\ar p \) and thus, 
Implication~\( (\dagger ) \)
holds by taking \( g_{1} \) to be \( q \), \( g_{2} \) to be \( p \),
and $g[\_]$ to be the empty goal context. The
replacement of \( q \) by \( p \) in clause 1 produces the following
program \( P_{2} \): 

\hspace{.7mm}
\Feq{1*.} p\leftarrow p
\Nex{2.} ~q\leftarrow
\eeq

\noindent This replacement is not an application of rule R4, because
Implication~\( (\dagger \! \dagger ) \) does not hold. (Indeed,
we have that the depth of the proof for \( P_{1}\vdash q\mapsto 
\{\varepsilon \} \)
is smaller than the depth of the proof for \( P_{1}\vdash p\mapsto 
\{\varepsilon \} \)).
The transformation from program \( P_{1} \) to program \( P_{2} \)
is not weakly correct (nor strongly correct), 
because \( p \) succeeds in \( P_{1} \),
while \( p \) does not terminate in \( P_{2} \), and thus, it is
not the case that \( P_{1}\sqsubseteq P_{2} \).\eop
\end{example}
The reader may check that, for any program \( P \), and goals \( g \),
\( g_{1} \), \( g_{2} \), and \( g_{3} \), we have the following
replacement laws. It can be shown that these replacement laws 
preserve safety.

\smallskip{}

\noindent 1.{\it ~~Boolean Laws}:
\nopagebreak

\smallskip{}
\noindent
\begin{tabular}{lll}
\( P\vdash \, \forall\, (g\wedge \mathit{true}\, \llarr \, g) \)&

\( P\vdash \, \forall\, (g\wedge g\, \arr \, g) \)\\

\( P\vdash \, \forall\, (\mathit{true}\wedge g\, \llarr \, g) \) &

\( P\vdash \, \forall\, (g\vee g\, \llarr \, g) \)\\

\( P\vdash \, \forall\, (\mathit{true}\vee g\, \arr \, \mathit{true}) \)&

\( P\vdash \, \forall\, (g_{1}\vee g_{2}\, \llarr \, g_{2}\vee g_{1}) \)\\

\( P\vdash \, \forall\, (g\wedge \mathit{false}\, \arr \, \mathit{false}) \)&

\( P\vdash \, \forall\, ((g_{1}\wedge g_{2})\vee (g_{1}\wedge g_{3})\, \llarr \,
g_{1}\wedge (g_{2}\vee g_{3}))\)\\

\( P\vdash \, \forall\, (\mathit{false}\wedge g\, \llarr \, \mathit{false}) \)&

\( P\vdash \, \forall\, ((g_{1}\wedge g_{2})\vee (g_{3}\wedge g_{2})\, \llarr \,
(g_{1}\vee g_{3})\wedge g_{2})\)\\

\( P\vdash \, \forall\, (\mathit{false}\vee g\, \llarr \, g)\)&

\( P\vdash \, \forall\, ((g_{1}\vee g_{2})\wedge (g_{1}\vee g_{3})\, \arr \, g_{1}\vee
(g_{2}\wedge g_{3})) \)\\

\end{tabular}
\smallskip{}

\noindent In the following replacement laws 2.1 and 2.2, according
to our conventions, \( V \) stands for either an individual variable
or a goal variable, and \( u \) stands for either a term or a goal,
respectively.
\smallskip{}

\noindent 2.1~~{\it Introduction and elimination of equalities}:

\smallskip{}
\noindent {\it ~~~~~~}\( P\vdash \, \forall U \, (g[u] \llarr 
((V\! =\! u)\wedge g[V]) )\)~~~\hfill where $U = \mathit{vars}(g[u])$ and
\( V\not \in U \).
\smallskip{}

\noindent 2.2~~{\it Rearrangement of equalities}:

\smallskip{}
\noindent {\it ~~~~~~}\( P\vdash  \forall U\,  (g[(V\! =\! u)\wedge
g_{1}] \llarr ((V\! =\! u)\wedge g[g_{1}])) \)

\hfill where
\( U=\mathit{vars}(g[g_1],u) \)
and \( V\not \in U \).
\smallskip{}

\noindent When referring to goal variables, laws 2.1 and 2.2 will
also be called `Introduction and elimination of goal equalities' and
`Rearrangement of goal equalities', respectively.

\smallskip{}
\noindent 3.~~{\it Rearrangement of term equalities}:

\smallskip{}
\( P\vdash \, \forall\, (g\wedge (t_{1}\! =\! t_{2})\, \arr \, (t_{1}\! =\! t_{2})
\wedge g )\)
\smallskip{}

\smallskip{}
\noindent 4.~~{\it Clark Equality Theory} (also called CET, see~\cite{Llo87}):

\smallskip{}
\( P\vdash \,  \forall X\,(eq_{1} \llarr eq_{2}) \)
~~~~~~~if \( \textrm{CET}\vdash \, \forall X\, \, (\exists Y\, 
\mathit{eq}_{1}\leftrightarrow \exists Z\, \mathit{eq}_{2}) \)
\smallskip{}

\noindent where: (i)~\( eq_{1} \) and \( eq_{2} \) are goals constructed
by using {\it true}, {\it false}, term equalities, conjunctions,
and disjunctions, and (ii)~\( Y\!=\!(\mathit{vars}(eq_{1})-\!X)\) and
\(Z\!=\!(\mathit{vars}(eq_{2})-\!X)\). 

\smallskip{}

Notice that, for some program \( P \) and for some goals
\( g,g_{1},g_{2} \), and \( g_{3} \), the following do {\it not}\/
hold:

\smallskip

\( P\vdash \, \forall \, (\mathit{true} \ar  \mathit{true}\vee g) \)

\( P\vdash \, \forall \, (\mathit{false}\ar g\wedge \mathit{false}) \)

\( P\vdash \, \forall \, ((t_{1}\! =\! t_{2})\wedge g \ar g\wedge (t_{1}\! =\!
t_{2})) \)

\( P\vdash \, \forall \, (g_{1}\vee (g_{2}\wedge g_{3}) \ar   
(g_{1}\vee g_{2})\wedge (g_{1}\vee g_{3})) \)

\( P\vdash \,  \forall V\, (g_{2}[g_{1}] \ar  g_{2}[G]\wedge (G\! =\!
g_{1})) \)
\hfill
where $V\!=\!\mathit{vars}(g_{2}[g_{1}])$ and $G\not \in V$

\( P\vdash \,  \forall V\, (g[(G\! =\! g_{1})\wedge g_{2}] \ar
(G\! =\! g_{1})\wedge g[g_{2}])\)

\hfill
where $V\!=\!(\mathit{vars}(g[g_{2}],g_1)-\{G\})$ and 
\(G\in \mathit{vars}(g[\_],g_{1})\)

\( P\vdash \, \forall\, ( g[(G\! =\! g_{1})\wedge g_{2}] \ar  (G\! =\!
g_{1})\wedge g[g_{2}]) \)
\hfill
where \(G\not \in \mathit{vars}(g[\_],g_{1})\)

\medskip{}
\noindent Let us now make some remarks on the goal replacement rule. 

In the Weak Correctness part of Theorem~\ref{th:correctness} (see
Section~\ref{sec:correctness}) we will prove that if program \( P_{2} \)
is derived from program \( P_{1} \) by an application of the goal
replacement rule based on a weak replacement law, then \( P_{2} \)
is a refinement of \( P_{1} \), that is, \( P_{1}\sqsubseteq P_{2} \).
Thus, there may be some ordinary goal
\( g \) which either succeeds or fails in  \( P_{2} \), while \( g \)
does not terminate in \( P_{1} \), as shown by the following example.

\begin{example}
\smallskip{}
\label{example:refinement} Let us consider the following two programs
\( P_{1} \) and \( P_{2} \), where \( P_{2} \) is derived from
\( P_{1} \) by applying the goal replacement rule based on the weak
(and not strong) replacement law
\( P_{1}\vdash \, \forall\,(\mathit{true}\vee g\, \arr \, \mathit{true}) \):

\smallskip{}\hspace*{-1cm}
\begin{tabular}{lllll}
\( P_{1} \):~~~&
\( p\leftarrow \mathit{true}\vee q \)&
~~~~~~~~~~~~~~~~~~~~&
\( P_{2} \):~~~&
\( p\leftarrow \mathit{true} \)\\
&
\( q\leftarrow q \)&
&
&
\( q\leftarrow q \)\\
\end{tabular}
\smallskip{}

\noindent We have that \( p \) does not terminate in \( P_{1} \)
and $p$ succeeds in \( P_{2} \).

\smallskip{}
\noindent Next, let us consider the following programs:

\smallskip{}\hspace*{-1cm}
\begin{tabular}{lllll}
\( P_{3} \):~~~&
\( p\leftarrow q\wedge \mathit{false} \)&
~~~~~~~~~~~~~~~~~~~~&
\( P_{4} \):~~~&
\( p\leftarrow \mathit{false} \)\\
&
\( q\leftarrow q \)&
&
&
\( q\leftarrow q \)\\
\end{tabular}

\smallskip{}
\noindent where \( P_{4} \) is derived from \( P_{3} \) by a goal
replacement rule based on a weak (and not strong) replacement law
\( P\vdash \,\forall\,( g\wedge \mathit{false}\arr \mathit{false} )\).
We have that \( p \) does not terminate in \( P_{3} \), while $p$ 
fails in \( P_{4} \).\eop 

\end{example}
In the Strong Correctness part of Theorem~\ref{th:correctness} we
will prove that if program \( P_{2} \) is derived from program \( P_{1} \)
by an application of the goal replacement rule based on a strong replacement
law, then \( P_{1} \) and \( P_{2} \) are equivalent, that is \( P_{1}\equiv 
P_{2} \).
Thus, in particular, for any goal \( g \), \( g \) terminates in
\( P_{1} \) iff \( g \) terminates in \( P_{2} \).

Moreover, in Theorem~\ref{th:safety} of Section~\ref{sec:correctness}
we will prove that if program \( P_{2} \) is derived from program
\( P_{1} \) by goal replacements which preserve safety, then every
goal which is safe in \( P_{1} \), is safe also in \( P_{2} \).

\section{Correctness of Program Transformations\label{sec:correctness}}

The unrestricted use of our rules for transforming programs 
may allow the construction of 
incorrect transformation sequences, as the following example shows.

\begin{example}
Let us consider the following initial program:

\smallskip

\makebox[1.5cm]{$P_0$:}$p \leftarrow q$

\makebox[1.5cm]{}$q \leftarrow $

\smallskip

\noindent
By two definition introduction steps, 
we get:

\smallskip

\makebox[1.5cm]{$P_1$:}$p \leftarrow q$

\makebox[1.5cm]{}$q \leftarrow $

\makebox[1.5cm]{}${\it newp}1 \leftarrow q$

\makebox[1.5cm]{}${\it newp}2 \leftarrow q$

\smallskip

\noindent
By three folding
steps, from program $P_1$  we get the final program:

\smallskip

\makebox[1.5cm]{$P_2$:}$p \leftarrow {\it newp}1$

\makebox[1.5cm]{}$q \leftarrow $

\makebox[1.5cm]{}${\it newp}1 \leftarrow {\it newp}2$

\makebox[1.5cm]{}${\it newp}2 \leftarrow {\it newp}1$

\smallskip
\noindent
We have that $p$ succeeds in $P_0$,
while $p$ does not terminate in $P_2$. \eop

\end{example}

In this section we will present some conditions which ensure
that every transformation sequence $P_0,\ldots,P_k$ constructed by using
our rules, is:

\noindent
(i)~weakly correct, in the sense that $P_0\cup {\it Def_k}\sqsubseteq P_k$
(see Point~(1) of Theorem~\ref{th:correctness}),

\noindent
(ii)~strongly correct, in the sense that $P_0\cup {\it Def_k}\equiv P_k$
 (see Point~(2) of Theorem~\ref{th:correctness}),

\noindent
(iii)~preserves safety, in the sense that, for every
goal \( g \),  if \( g \)
is safe in \( P_{0}\cup \mathit{Def}\! _{k} \) then \( g \) is safe
also in \( P_{k} \) (see Theorem~\ref{th:safety}).

Similarly to other correctness results presented in the
literature~\cite{BoC94,PeP94,San96,TaS84},
some of the conditions which ensure (weak or strong) correctness, 
require that the transformation sequences are constructed
by performing suitable unfolding steps before performing folding steps.

In particular, Theorem~\ref{th:correctness} below ensures the (weak or
strong) correctness of a given transformation sequence in the case where
this sequence is {\it admissible}, that is, it is constructed
by performing {\it parallel leftmost} 
unfoldings (see Definition~\ref{def:plunfold}) 
on all definitions which are used for performing 
subsequent foldings.

In order to present our correctness results 
it is convenient to consider
admissible transformation sequences which are {\it ordered}, 
that is, transformation sequences
constructed by:

\noindent (i)~first, applying the definition
introduction rule,

\noindent (ii)~then, performing
parallel leftmost unfoldings of the definitions that are
used for subsequent foldings, 
and

\noindent (iii)~finally, performing unfoldings, foldings,
and goal replacements in any order.

Thus, an ordered, admissible transformation sequence has all its definition
introductions performed  at the beginning, and it can be written in
the form \( P_{0},\ldots ,P_{0}\cup\mathit{Def}\! _{k}, \ldots , P_{k} \),
where \( \mathit{Def}\! _{k} \) is the set of all definitions
introduced during the entire transformation
sequence \( P_{0},\ldots ,P_{0}\cup\mathit{Def}\! _{k}, \ldots , P_{k} \).
By Proposition~\ref{prop:ordered} below we may assume,
without loss of generality, that all 
admissible transformation sequences are ordered.

In order to prove that an admissible transformation sequence
is weakly correct (see Point~(1) of Theorem~\ref{th:correctness}),
we proceed as follows. 

\smallskip

\noindent 
(i) In Lemma~\ref{lemma:improvement} we consider
a generic transformation by which
we derive a program \( \mathit{NewP} \) from a program \( P \) by replacing
the bodies of the clauses of \( P \) by new bodies. 
We show that, if these body replacements
can be viewed as goal replacements based on weak replacement laws, 
then the transformation from $P$ to {\it NewP}
preserves successes and failures, that is,

\noindent
- if a goal $g$ succeeds in \( P \) then $g$
succeeds in $ \mathit{NewP} $, and

\noindent
- if a goal $g$ fails in \( P \) then $g$
fails in $ \mathit{NewP} $.

\smallskip
\noindent
(ii) Then, in Lemma~\ref{lemma:improving_rules} we prove that in an ordered,
admissible transformation sequence 
\( P_{0},\ldots ,P_{0}\cup\mathit{Def}\! _{k}, \ldots , P_{k} \),
any application of the unfolding, folding, and goal replacement rule
is an instance of the generic transformation considered in
Lemma~\ref{lemma:improvement}, that is, it
consists in the replacement of the body of a clause by a new body,
and this replacement can be viewed as a goal replacement 
based on a weak replacement law. 

\smallskip

\noindent
(iii) Thus, by using Lemmata~\ref{lemma:improvement} 
and~\ref{lemma:improving_rules} we get Point~(1) of
Theorem~\ref{th:preservation}. In particular, we have that
in any admissible transformation sequence 
\( P_{0},\ldots ,P_{0}\cup\mathit{Def}\! _{k}, \ldots , P_{k} \),
successes and failures are preserved, that is:

\noindent
- if a goal $g$ succeeds in \( P_{0}\cup\mathit{Def}\! _{k} \) then $g$ 
succeeds in $P_{k}$, and

\noindent
- if a goal $g$ fails in \( P_{0}\cup\mathit{Def}\! _{k} \) then $g$ 
fails in $P_{k}$.

\smallskip

\noindent
(iv) Finally, Proposition~\ref{proposition:cas} allows us to 
infer the preservation of most general answer substitutions
from the preservation of successes and failures.
Indeed, by Proposition~\ref{proposition:cas} and 
Point (1) of Theorem~\ref{th:preservation} we prove that
if an ordinary goal $g$ succeeds in \( P_{0}\cup\mathit{Def}\! _{k} \)
then the set of answer substitutions for
$g$ in \( P_{0}\cup\mathit{Def}\! _{k} \) and
the set of answer substitutions for $g$ in
\( P_{k} \) are equally general.

\smallskip

\noindent According to Definition~\ref{def:ref_equiv}, 
Points~(iii) and (iv)
mean that \( P_{0}\cup\mathit{Def}\! _{k} \sqsubseteq P_{k}\), that is,
the ordered, admissible transformation sequence 
\( P_{0},\ldots ,P_{0}\cup\mathit{Def}\! _{k}, \ldots , P_{k} \)
is weakly correct
(see Point~(1) of Theorem~\ref{th:correctness}).

\smallskip

In order to prove that an admissible transformation sequence
is strongly correct (see Point~(2) of Theorem~\ref{th:correctness}),
we make the additional hypothesis that all
goal replacements performed during the construction of the transformation
sequence are based on strong replacement laws. Analogously to the proof of
weak correctness which is based on Lemmata~\ref{lemma:improvement} and
\ref{lemma:improving_rules}, the proof of strong correctness is based on
Lemmata~\ref{lemma:soundness} and~\ref{lemma:sound_rules}  which we give
below. By using these lemmata, we prove Point~(2) of
Theorem~\ref{th:preservation}, that is:

\noindent
- if a goal $g$ succeeds in $P_{k}$ then $g$
succeeds in \( P_{0}\cup\mathit{Def}\! _{k} \), and

\noindent
- if a goal $g$ fails in \( P_k \) then $g$
fails in \( P_{0}\cup\mathit{Def}\! _{k} \).

\noindent
Finally, by Proposition~\ref{proposition:cas}
and Theorem~\ref{th:preservation}, we prove that any admissible
transformation sequence in which all goal replacements are based on strong
replacement laws, is strongly correct
(see Point~(2) of Theorem~\ref{th:correctness}), that is,
\( P_{0}\cup\mathit{Def}\! _{k} \equiv P_{k}\).

\medskip

Now let us formally define the notions of {\it parallel
leftmost unfolding} of a clause, 
{\it admissible} transformation sequence, and {\it ordered} 
admissible transformation sequence
as follows. 

\begin{definition}\label{def:plunfold}
Let \( c \) be a clause in a program \( P \). If \( c \) is of
the form:

\smallskip{}
~~~~\( p(V_{1},\ldots ,V_{m})\leftarrow (a_{1}\wedge g_{1})\vee \ldots 
\vee (a_{s}\wedge g_{s}) \) 

\smallskip{}
\noindent where \( a_{1},\ldots ,a_{s} \) are atoms with non-primitive
predicates, \( g_{1},\ldots ,g_{s} \) are goals, and \( s>0 \),
then the {\it parallel leftmost unfolding} of clause \( c \) in
program \( P \) is the program \( Q \) obtained from \( P \) by
applying \( s \) times the unfolding rule w.r.t.~\( a_{1},\ldots ,a_{s} \),
respectively. 
\end{definition}

If clause \( c \) is not of the form indicated in
Definition~\ref{def:plunfold} above, then the parallel leftmost
unfolding of \( c \) is not defined. 

\begin{definition}
A transformation sequence \( P_{0},\ldots ,P_{k} \) is
said to be {\it admissible} iff for every \( h \), with \( 0\! \leq \!
h\! <\! k \), if \( P_{h+1} \) has been obtained from \( P_{h} \) by folding
clause \( c \) using clause \( d \), then there exist \( i,j \), with
\( 0\! \leq \! i\! <\! j\! \leq \! k \), such that \(d\in P_i\)
and \( P_{j} \) is
obtained from \( P_{i} \) by parallel leftmost unfolding of \( d \).
\end{definition}

\begin{definition}
\label{def:ordered_transf_seq}An admissible transformation
sequence \( P_{0},\ldots ,P_{k} \) is said to be {\it ordered}\/
iff it is of the form \( P_{0},\ldots ,P_{i},\ldots ,P_{j},\ldots, P_{k}
\), where: (i)~the sequence \( P_{0},\ldots ,P_{i} \) is constructed
by applying the definition introduction rule,
(ii)~the sequence \( P_{i},\ldots ,P_{j} \) is constructed by parallel
leftmost unfolding of all definitions which have been introduced during
the sequence \( P_{0},\ldots ,P_{i} \) and are used for folding
during the sequence \( P_{j},\ldots ,P_{k} \), and (iii)~the definition
introduction rule is never applied in the sequence
\( P_{j},\ldots ,P_{k} \).
\end{definition}

Given an ordered, admissible transformation sequence
$P_{0},$ $\ldots ,$ $P_{i},$ $\ldots ,P_{j},$ $\ldots ,P_{k} \),
the set of definitions introduced during \( P_{0},\ldots ,P_{i} \)\/
is the same as the set of definitions introduced during the entire
sequence \( P_{0},\ldots ,P_{k} \), and thus, in the
above Definition~\ref{def:ordered_transf_seq} we have that
\( P_{i} \) is  \( P_{0}\cup \mathit{Def}\!_{k} \).

An admissible transformation sequence \( P_{0},\ldots ,P_{k} \)
which is ordered, is also denoted by \( P_{0},\ldots ,  P_{i},
\ldots, P_{j},\ldots ,P_{k} \), where we explicitly indicate the
program \( P_{i} \) after the introduction of the definitions, and the program
\( P_{j} \) after the parallel leftmost unfolding steps.

\begin{proposition}
\label{prop:ordered}For any admissible transformation sequence
\( P_{0},\ldots ,P_{n} \) there exists an ordered, admissible transformation
sequence \( P_{0},\ldots ,P_{i},\ldots , \) \( P_{j},\ldots ,P_{k} \)
such that \(P_n= P_k\) and \(\mathit{Def}\! _{n}=\mathit{Def}\! _{k}\).
\end{proposition}
Now, in order to prove the correctness of transformation sequences,
we state the following Lemmata~\ref{lemma:improvement}, 
\ref{lemma:improving_rules},
\ref{lemma:soundness}, and \ref{lemma:sound_rules}, whose proofs
are given in the Appendix. As already mentioned,
these Lemmata~\ref{lemma:improvement},
\ref{lemma:improving_rules}, \ref{lemma:soundness}, and 
\ref{lemma:sound_rules}
will allow us to show that, under suitable conditions, for every admissible
transformation sequence \( P_{0},\ldots ,P_{k} \), (i)~successes
and failures are preserved (see Theorem~\ref{th:preservation} below),
and (ii)~weak correctness holds (that is, \( P_{0}\cup
\mathit{Def}\! _{k}\sqsubseteq P_{k} \))
or strong correctness holds (that is,
 \( P_{0}\cup \mathit{Def}\! _{k}\equiv P_{k} \))
(see Theorem~\ref{th:correctness} below).

\smallskip{}

\begin{lemma}
\noindent\label{lemma:improvement}Let \( P \) and {\it NewP} be
programs of the form:

\smallskip{}\hspace*{-1cm}
\begin{tabular}{ccccc}
$P:$ &
\( hd_{1}\leftarrow bd_{1} \) &
~~~~~~~~~~~~~~~&
$\mathit{NewP}:$ &
\( hd_{1}\leftarrow \mathit{newbd}_{1} \)\\
&
~\( \vdots  \)&
&
&
\( \vdots  \)~~~~~\\
&
\( hd_{s}\leftarrow bd_{s} \)&
&
&
\( hd_{s}\leftarrow \mathit{newbd}_{s} \)\\
\end{tabular}
\smallskip{}

\noindent For \( r=1,\ldots ,s \), let \( V_{r} \) be \(
\mathit{vars}(hd_{r}) \) and suppose that 
\( P\, \vdash \, \forall V_{r}\, (bd_{r}\arr \mathit{newbd}_{r}) \).

\noindent Then, for every goal \( g \) and for every \( b\in
\{\mathit{true},\mathit{false}\} \), we have that:
\smallskip{}

if~ \( P\, \vdash \, g\downarrow _{m}b \) ~then~ \( \mathit{NewP}\, 
\vdash \, g\downarrow _{n}b \)
with \( m\geq n \).
\end{lemma}

\begin{lemma}
\noindent\label{lemma:improving_rules}Let us consider an
ordered, admissible transformation sequence \( P_{0},\ldots , \)
\( P_{i},\ldots , \) \( P_{j},\ldots ,P_{k} \), where \( P_{i} \)
is \( P_{0}\cup \mathit{Def}\! _{k} \).

\noindent
(i)~For $h=i,\ldots,j\!-\!1$ and for any pair of clauses
$c_1$: $\mathit{hd} \leftarrow \mathit{bd}$ in program $P_{h}$ and
$c_2$: $\mathit{hd} \leftarrow \mathit{newbd}$ in program $P_{h+1}$,
such that $c_2$ is derived from $c_1$ by
applying the unfolding rule, we have that:

\( P_{i}\, \vdash \, \forall V\, (\mathit{bd}\arr \mathit{newbd}) \)

\noindent
where $V = \mathit{vars}(\mathit{hd})$.
(Notice that the unfolding rule does not change the heads of the clauses.)

\noindent
(ii)~For $h=j,\ldots,k\!-\!1$ and for any pair of clauses
$c_1$: $\mathit{hd} \leftarrow \mathit{bd}$ in program $P_{h}$ and
$c_2$: $\mathit{hd} \leftarrow \mathit{newbd}$ in program $P_{h+1}$,
such that $c_2$ is derived from $c_1$ by
applying the unfolding, or folding, or goal replacement rule,  we have that:

\( P_{j}\, \vdash \, \forall V\, (\mathit{bd}\arr \mathit{newbd}) \)

\noindent
where $V = \mathit{vars}(\mathit{hd})$.
(Notice that the unfolding, folding, and goal replacement rules do
not change the heads of the clauses.)
\end{lemma}

\begin{lemma}
\noindent\label{lemma:soundness}Let \( P \)
and {\it NewP} be programs of the form:

\smallskip{}\hspace*{-1cm}
\begin{tabular}{ccccc}
\( P: \) &
\( hd_{1}\leftarrow bd_{1} \) &
~~~~~~~~~~~~~~&
\( \mathit{NewP}: \) &
\( hd_{1}\leftarrow newbd_{1} \)\\
&
~\( \vdots  \)&
&
&
\( \vdots  \)~~~~~\\
&
\( hd_{s}\leftarrow bd_{s} \)&
&
&
\( hd_{s}\leftarrow newbd_{s} \)\\
\end{tabular}
\smallskip{}

\noindent For \( r=1,\ldots ,s \),  let \( V_{r} \) be 
\(\mathit{vars}(hd_{r}) \) and suppose that \( P\,
\vdash \, \forall V_{r}\, (newbd_{r}\ar bd_{r}) \).

\noindent Then, for
every goal \( g \) and for every \( b\in \{\mathit{true},\mathit{false}\} \),
we have that if \( \mathit{NewP}\, \vdash \, g\downarrow b \) ~then~ \( P\,
\vdash \, g\downarrow b \).
\end{lemma}

Notice that Lemma~\ref{lemma:soundness} is a partial converse
of Lemma~\ref{lemma:improvement}. These two
lemmata imply that if we derive a program \( \mathit{NewP} \) from a
program \( P \) by replacing the bodies of the clauses of \( P \) by new
bodies, and these body replacements are goal replacements based on
strong replacement laws, then every goal terminates
in \( \mathit{NewP} \) iff it terminates in \( P \). 
 
\begin{lemma}
\noindent \label{lemma:sound_rules}Let us
consider a transformation sequence \( P_{0},\ldots ,P_{k} \)
and let  \(\mathit{Def}\! _{k} \) be the set of definitions
introduced during that sequence. 
For $h=0,\ldots,k\!-\!1$ and for any pair of clauses
$c_1$: $\mathit{hd} \leftarrow \mathit{bd}$ in program $P_{h}$ and
$c_2$: $\mathit{hd} \leftarrow \mathit{newbd}$ in program $P_{h+1}$,
such that $c_2$ is derived from $c_1$ by
applying the unfolding rule, or the folding rule, or the goal replacement rule
based on strong replacement laws, we have that:

\( P_{0}\cup \mathit{Def}\! _{k} \, \vdash \, \forall V\, (\mathit{newbd}\ar
\mathit{bd} )\)

\noindent
where $V = \mathit{vars}(\mathit{hd})$.
\end{lemma}

In particular, as a consequence of Lemma~\ref{lemma:improving_rules} and
Lemma~\ref{lemma:sound_rules}, we have  that in any ordered, admissible
transformation sequence the unfolding and folding rules can be viewed as goal
replacements based on strong replacement laws.

The following theorem states that for every admissible transformation
sequence successes and failures are preserved.

\begin{theorem}[Preservation of Successes and Failures]
\noindent\label{th:preservation}Let \( P_{0},\ldots ,P_{k} \) be an admissible
transformation sequence and let \( \mathit{Def}\! _{k} \) be the
set of definitions introduced during that sequence. Then for every
goal \( g \) and for every \( \mathit{b}\in 
\{\mathit{true},\mathit{false}\} \),
we have that:

\noindent (1) if \( P_{0}\cup \mathit{Def}\! _{k}\vdash g\downarrow _{m}b \)
~then~ \( P_{k}\vdash g\downarrow _{n}b \) with \( m\geq n \),~~and

\noindent (2) if all applications of the goal replacement rule
are based on strong replacement laws and \( P_{k}\vdash g\downarrow b \),
then \( P_{0}\cup \mathit{Def}\! _{k}\vdash g\downarrow b \).
\end{theorem}
\begin{proof}[Proof of Theorem~\ref{th:preservation}]
See Appendix. The proof
of (1) is based on Proposition~\ref{prop:ordered} and
Lemmata~\ref{lemma:improvement} and~\ref{lemma:improving_rules}, and the proof
of (2) is based on Proposition~\ref{prop:ordered} and
Lemmata~\ref{lemma:soundness} and~\ref{lemma:sound_rules}.
\hfill\end{proof}

\smallskip{}
The following theorem establishes the weak correctness and,
under suitable conditions, the strong
correctness of admissible transformation sequences.

\begin{theorem}[Correctness Theorem]
\noindent\label{th:correctness}Let
\( P_{0},\ldots ,P_{k} \) be an admissible transformation sequence.
Let \( \mathit{Def}\! _{k} \) be the set of definitions introduced
during that sequence. We have that:

\noindent (1)~(Weak Correctness) \( P_{0}\cup \mathit{Def}\! _{k}\sqsubseteq 
P_{k} \),
that is, \( P_{k} \) is a refinement of \( P_{0}\cup \mathit{Def}\! _{k} \),
and

\noindent (2)~(Strong Correctness) if all applications
of the goal replacement rule are based on strong replacement laws
then \( P_{0}\cup \mathit{Def}\! _{k}\equiv P_{k} \), that is, \( P_{k} \)
is equivalent to \( P_{0}\cup \mathit{Def}\! _{k} \).
\end{theorem}
\begin{proof}[Proof of Theorem~\ref{th:correctness}]
 See Appendix. The proof of (1) is based
on Proposition~\ref{proposition:cas} and Theorem~\ref{th:preservation}~(Point~1),
and the proof of (2) is based on Proposition~\ref{proposition:cas}
and Theorem~\ref{th:preservation}~(Points~1 and 2).\hfill\end{proof}

The following two examples show that in the statement
of Theorem~\ref{th:correctness} we cannot drop the admissibility condition.
Indeed, in these examples we construct transformation sequences which are
not admissible and not weakly correct.

\begin{example}
Let us construct a transformation sequence as follows. The initial program is:

\smallskip

\makebox[1.5cm]{$P_0$:}$p \leftarrow p \wedge q$

\makebox[1.5cm]{}$q \leftarrow {\it false}$

\smallskip

\noindent
By definition introduction we get:

\smallskip

\makebox[1.5cm]{$P_1$:}$p \leftarrow p\wedge q$

\makebox[1.5cm]{}$q \leftarrow {\it false}$

\makebox[1.5cm]{}${\it newp} \leftarrow {\it false} \wedge p$

\smallskip

\noindent
Then we perform the unfolding of 
${\it newp} \leftarrow {\it false} \wedge p$ w.r.t. $p$. (Notice that this
is not a parallel leftmost unfolding.) We get:

\smallskip

\makebox[1.5cm]{$P_2$:}$p \leftarrow p\wedge q$

\makebox[1.5cm]{}$q \leftarrow {\it false}$

\makebox[1.5cm]{}${\it newp} \leftarrow {\it false} \wedge p \wedge q$

\smallskip

\noindent
By folding we get the final program:

\smallskip

\makebox[1.5cm]{$P_3$:}$p \leftarrow p\wedge q$

\makebox[1.5cm]{}$q \leftarrow {\it false}$

\makebox[1.5cm]{}${\it newp} \leftarrow {\it newp} \wedge q$

\smallskip
\noindent
We have that {\it newp} fails in $P_0\cup {\it Def_3}$ (that is, $P_1$),
while {\it newp} does not terminate in $P_3$. \eop
\end{example}

\begin{example}
Let us construct a transformation sequence as follows. The initial program is:
\smallskip

\makebox[1.5cm]{$P_0$:}$p \leftarrow {\it false}$

\makebox[1.5cm]{}$q \leftarrow {\it true} \vee q$

\smallskip
\noindent
By definition introduction we get:

\smallskip

\makebox[1.5cm]{$P_1$:}$p \leftarrow {\it false}$

\makebox[1.5cm]{}$q \leftarrow {\it true} \vee q$

\makebox[1.5cm]{}${\it newp} \leftarrow p \vee (p \wedge q)$

\smallskip
\noindent
Then we perform the unfolding of ${\it newp} \leftarrow p \vee (p \wedge q)$ w.r.t.
$q$. (Notice that this is not a parallel leftmost unfolding.)   We get:

\smallskip

\makebox[1.5cm]{$P_2$:}$p \leftarrow {\it false}$

\makebox[1.5cm]{}$q \leftarrow {\it true} \vee q$

\makebox[1.5cm]{}${\it newp} \leftarrow {\it false} \vee
			(p \wedge ({\it true} \vee q))$

\smallskip
\noindent
By goal replacement based on boolean laws we get:

\smallskip

\makebox[1.5cm]{$P_3$:}$p \leftarrow {\it false}$

\makebox[1.5cm]{}$q \leftarrow {\it true} \vee q$

\makebox[1.5cm]{}${\it newp} \leftarrow {\it p} \vee (p \wedge q)$

\smallskip
\noindent
By folding we get the final program:

\smallskip

\makebox[1.5cm]{$P_4$:}$p \leftarrow {\it false}$

\makebox[1.5cm]{}$q \leftarrow {\it true} \vee q$

\makebox[1.5cm]{}${\it newp} \leftarrow {\it newp}$

\smallskip
\noindent
We have that 
{\it newp} fails in $P_0\cup {\it Def_4}$ (that is, $P_1$), while {\it newp}
does not terminate in $P_4$. \eop
\end{example}

Finally, the following theorem states that a (possibly not admissible)
transformation sequence preserves safety, if all goal replacements 
performed during that sequence preserve safety.

\begin{theorem}[Preservation of Safety]
\label{th:safety}Let
\( P_{0},\ldots ,P_{k} \) be a transformation sequence and let \( 
\mathit{Def}\! _{k} \)
be the set of definitions introduced during that sequence. Let us
also assume that all applications of the goal replacement rule R4
preserve safety. Then, for every goal \( g \), if \( g \)
is safe in \( P_{0}\cup \mathit{Def}\! _{k} \) then \( g \) is safe
in \( P_{k} \).
\end{theorem}
\begin{proof}[Proof of Theorem~\ref{th:safety}] See Appendix. The proof
is based on Lemmata~\ref{lemma:safety} and~\ref{lemma:safe_rules}
given in the Appendix.\hfill\end{proof}

We end this section by making some comments about our correctness
results. Let us consider an admissible transformation sequence
\( P_{0} , \ldots, P_{k} \), during which we introduce the set
${\it Def_k}$ of definitions. 
Then, by Point~(1) of Theorem~\ref{th:preservation}
program \( P_{k} \) may be {\it more defined}\/ than program
\( P_{0}\cup {\it Def_k} \)
in the sense that there may be a goal which terminates (i.e., succeeds
or fails) in \( P_{k} \), while it does not terminate in 
\( P_{0}\cup {\it Def_k} \).
This `increase of termination' is often desirable when transforming programs
 and it may be achieved by goal replacements which are
not based on strong replacement laws (see, for instance, 
Example~\ref{example:refinement}
in Section~\ref{sec:rules}). 

Now suppose that during the construction of the
admissible transformation sequence
\( P_{0} , \ldots, P_{k} \) all applications of the goal replacement rule
are based on strong replacement laws. Then, by 
Theorem~\ref{th:preservation} we have that 
for all goals \( g \), \( g \) terminates in \( P_{0}\cup {\it Def_k} \)
iff \( g \) terminates in \( P_{k} \). However, safety may be not
preserved, in the sense that there may be a goal \( g \) which is
safe in \( P_{0}\cup {\it Def_k} \) (but \( g \) neither 
succeeds nor fails in \( P_{0}\cup {\it Def_k} \))
and $g$ is not safe in \( P_{k} \) (or vice versa), as shown by the
following example.

\begin{example}
Let us consider the following two programs \( P_{1} \) and \( P_{2} \):

\smallskip{}

\hspace*{-1cm}
\begin{tabular}{lllll}
\( P_{1} \):~~&
\( p\leftarrow p \)&
~~~~~~~~~~~~~~~~~&
\( P_{2} \):~~~&
\( p\leftarrow G \)\\
\end{tabular}
\smallskip{}

\noindent Program \( P_{2} \) is derived from \( P_{1} \) 
by applying the goal
replacement rule based on the strong replacement law \( P_{1}\vdash \, p \llarr G \),
which does not preserve safety.
We have that \( p \) is safe, $p$ does not terminate
in \( P_{1} \), and \( p \) is not safe in \( P_{2} \). Notice that
the replacement law \( P_{1}\vdash \, p \llarr  G \)
trivially holds because, for any \( b\in \{\mathit{true},\mathit{false}\} \),
\( P_{1}\vdash \, p\downarrow b \) does not hold and
\( P_{1}\vdash \, G\downarrow b \) does not hold.\eop
\end{example}

In order to ensure that if \( g \) is safe in \( P_{1} \) then
\( g \) is safe in \( P_{2} \), it is enough to use replacement laws
which preserve safety (see Theorem~\ref{th:safety}). 
Indeed, unfolding and folding always preserve
safety (see Lemma~\ref{lemma:safe_rules} in the Appendix).

We have not presented any result which guarantees that if a goal is
safe in the final program \( P_{k} \) then it is safe in the
program \( P_{0}\cup {\it Def_k} \). This result could have been achieved by
imposing further restrictions on the goal replacement rule. However, we
believe that this `inverse preservation of safety' is not important
in practice, because usually we start from an initial program where
all goals of interest are safe and we want to derive a final program
where those goals of interest are still safe. 
In particular, if in the transformation
sequence \( P_{0},\ldots ,P_{k} \) the 
initial program \( P_{0} \) is an ordinary
program, then every ordinary goal \( g \) is safe in \( P_{0} \)
and, by Theorem~\ref{th:safety}, we have that \( g \) is safe also
in \( P_{k} \). Thus, as discussed in Section~\ref{sec:semantics},
we can use ordinary implementations of LD-resolution to compute the
relation \( P_{k}\models g\mapsto A \). 

Notice also that, if \( P_{0}\cup \mathit{Def}\! _{k}\sqsubseteq P_{k} \)
and an ordinary goal \( g \) terminates in \( P_{0} \), then \( g \)
has the same most general answer substitutions in 
\( P_{0}\cup \mathit{Def}\! _{k} \)
and \( P_{k} \), modulo variable renaming (see Point (i) of 
Remark~\ref{re:most_gen_cas}
at the end of Section~\ref{sec:semantics}). However, the set of {\it all}
answer substitutions may not be preserved, and in particular, there
are programs \( P_{1} \) and \( P_{2} \) such that \( P_{1}\sqsubseteq P_{2} \)
and, for some goal \( g \), we have that \( P_{1}\vdash g\mapsto A_{1} \)
and \( P_{2}\vdash g\mapsto A_{2} \), where \( A_{1} \) and \( A_{2} \)
have different cardinality, as shown by the following example adapted
from~\cite{Bo&92a}. A similar property holds if we assume that \( P_{1}\equiv
P_{2} \), instead of \( P_{1}\sqsubseteq P_{2} \).

\begin{example} \sloppy
\label{example:g_and_g->g}Let us consider the following two programs
\( P_{1} \) and \( P_{2} \), where \( P_{2} \) is derived from
\( P_{1} \) by applying the goal replacement rule based on the weak
replacement law
\mbox{\( P\vdash \, \forall\,(g\wedge g\, \arr \, g )\),}
which holds
for every program \( P \) and and goal \( g \):

\fussy

\smallskip{}

\smallskip{}\hspace*{-1cm}
\begin{tabular}{lllll}
\( P_{1} \):~~&
\( p(X)\leftarrow q(X)\wedge q(X) \)&
~~~~~~~~~&
\( P_{2} \):~~~&
\( p(X)\leftarrow q(X) \)\\
&
\( q(X)\leftarrow X\! =\! f(a,Z) \)&
&
&
\( q(X)\leftarrow X\! =\! f(a,Z) \)\\
&
\( q(X)\leftarrow X\! =\! f(Y,a) \)&
&
&
\( q(X)\leftarrow X\! =\! f(Y,a) \)\\
\end{tabular}
\smallskip{}

\noindent We have that:
\smallskip{}

\( P_{1}\vdash p(X)\mapsto \{\{X/f(a,Z)\},\{X/f(a,a)\},\{X/f(Y,a)\}\} \),
and

\( P_{2}\vdash p(X)\mapsto \{\{X/f(a,Z)\},\{X/f(Y,a)\}\} \).\eop 
\end{example}
\noindent The above example shows that, if
during program transformation we want to preserve the
set of answer substitutions, then
we should not apply goal replacements based on the replacement law
\( P\vdash \, \forall\,(g\wedge g\, \arr \, g) \) which, 
however, may be useful
for avoiding the computation of redundant goals and improving program
efficiency. 

Another replacement law which is very useful in many examples of program
transformation, is the law which expresses the functionality of a
predicate. For instance, in the {\it Deepest} example of
Section~\ref{sec:motivating_ex},
the \( \mathit{depth} \) predicate is functional with respect to its first
argument in the sense that, for every goal context $g[\_]$, 
the following replacement law holds:

\smallskip{}
\noindent \hspace*{.8cm}\(  \mathit{Deepest}\vdash \forall\,
(\mathit{depth}(T,X)\wedge
g[\mathit{depth}(T,Y)]\larr \mathit{depth}(T,X)\wedge g[X\! =\! Y]) \).

\smallskip{}
\noindent The following example, similar to Example~\ref{example:g_and_g->g},
shows that in general the functionality law does not preserve the
set of answer substitutions.

\begin{example}
\label{example:functionality}Let us consider the following two programs
\( P_{1} \) and \( P_{2} \), where \( P_{2} \) is derived from
\( P_{1} \) by applying the goal replacement rule based on the (strong)
replacement law \( P_{1}\vdash \, \forall\, (q(X,Y)\wedge q(X,Z)\, \larr \, 
q(X,Y)\wedge Y\! =\! Z) \): 
\smallskip{}

\smallskip{}
\begin{tabular}{lllll}
\( P_{1} \):~&
\( p(X)\leftarrow q(X,Y)\wedge q(X,Z) \)&
~~~~~~~&
\( P_{2} \):~&
\( p(X)\leftarrow q(X,Y)\wedge Y\! =\! Z \)\\
&
\( q(f(a,Z),b)\leftarrow  \)&
&
&
\( q(f(a,Z),b)\leftarrow  \)\\
&
\( q(f(Y,a),b)\leftarrow  \)&
&
&
\( q(f(Y,a),b)\leftarrow  \)\\
\end{tabular}
\smallskip{}

\noindent As in Example~\ref{example:g_and_g->g}, we have that:
\smallskip{}

\( P_{1}\vdash p(X)\mapsto \{\{X/f(a,Z)\},\{X/f(a,a)\},\{X/f(Y,a)\}\} \)
~~and

\( P_{2}\vdash p(X)\mapsto \{\{X/f(a,Z)\},\{X/f(Y,a)\}\} \).\eop 
\end{example}
\noindent Finally, notice that Theorem~\ref{th:correctness}
 ensures the
preservation of most general answer substitutions for ordinary goals
only. Thus, the answer substitutions computed for goals with
occurrences of goal variables, may not be preserved, as shown by the
following example.

\begin{example}
\label{example:goal_answers}Let us consider the following two programs
\( P_{1} \) and \( P_{2} \), where \( P_{2} \) is derived from
\( P_{1} \) by unfolding clause 1 w.r.t.~\( p \) using clause 2:
\smallskip{}

\begin{tabular}{lllllll}
\( P_{1} \):~&
1.~&
\( a(G)\leftarrow (G\! =\! p)\wedge G \) &
~~~~~~~&
\( P_{2} \):~&
1{*}.~&
\( a(G)\leftarrow (G\! =\! q)\wedge G \) \\
&
2.&
\( p\leftarrow q \)&
&
&
2.&
\( p\leftarrow q \)\\
&
3.&
\( q\leftarrow  \)&
&
&
3.&
\( q\leftarrow  \)\\
\end{tabular}
\smallskip{}

\noindent We have that \( P_{1}\vdash a(G)\mapsto \{\{G/p\}\} \),
and \( P_{2}\vdash a(G)\mapsto \{\{G/q\}\} \).\eop
\end{example}

\section{Program Derivation in the Extended Language \label{sec:examples}}

In this section we present some examples which illustrate the use
of our transformation rules. In these examples, by using goal variables
and goal arguments, we introduce and manipulate {\it continuations}.
For this reason we have measured the improvements of program efficiency
by running our programs using the BinProlog continuation passing
compiler~\cite{Tar96}. These run-time improvements have been reported in
Section~\ref{sec:speedups}. Compilers based on different implementation
methodologies, such as SICStus Prolog, may not give the same improvements.
However, it should be noticed that the efficiency improvements we get, do not
come from the use of continuations, but from the program transformations
performed by applying our transformation rules (see
Section~\ref{sec:rules}). Indeed, in
BinProlog the continuation passing style transformation in itself gives no
speed-ups.

Let us introduce the following terminology which will be useful in the
sequel. We say that:~(i)~a clause is in continuation passing style iff its
body has no occurrences of the conjunction operator, and (ii)~a program is in
continuation passing style iff all its clauses are in continuation passing
style. Thus, every program in continuation passing style is
a {\it binary} program in the sense of Tarau and Boyer~\citeyear{TaB90}, that
is, a program with at most one atom in the body of its clauses.

When writing  programs in this section we use the following
primitive predicates: $=$, $\neq$, $\geq$, and $<$.
For the derivation of programs in continuation passing style,
we assume that, for each of these predicates  there exists a corresponding
primitive predicate with an extra argument denoting a continuation. Let us
call these predicates ${\it eq}\U{c}$, ${\it diff}\U{c}$, ${\it geq}\U{c}$,
and ${\it lt}\U{c}$, respectively.

We assume that, for every program $P$, the following strong replacement laws
hold:

\smallskip
\noindent
~~~~~~$P\vdash\, \forall\,( (X\!=\!Y) \wedge C \, \llarr \, 
{\it eq}\U{c}(X,Y,C))$

\smallskip
\noindent
~~~~~~$P\vdash\,\forall\, ((M\!\neq\!N) \wedge C \, \llarr \, 
{\it diff}\U{c}(M,N,C))$

\smallskip
\noindent
~~~~~~$P\vdash\,\forall\, ( (M\!\geq\!N) \wedge C \, \llarr \, 
{\it geq}\U{c}(M,N,C))$

\smallskip
\noindent
~~~~~~$P\vdash\, \forall\, ((M\!<\!N) \wedge C \, \llarr \, 
{\it lt}\U{c}(M,N,C))$

\smallskip
\noindent
In this section we use the following syntactical conventions: 

\noindent (1) the conjunction operator \( \wedge  \) is replaced
by comma,

\noindent (2) a clause of the form \( h\leftarrow g_{1}\! \vee g_{2} \)
is also written as two clauses, namely, \( h\leftarrow g_{1} \) and
\( h\leftarrow g_{2} \), and

\noindent (3) a clause of the form \( h\leftarrow (V\! =\! u),\, g \)~
where the variable \( V \) does not occur in the argument \( u \), is
also written as \( (h\leftarrow g)\{V/u\} \).

\subsection{Tree Flipping}\label{sec:flipcheck}

This example is borrowed from~\cite{Jo&96} where it is used for showing
that {\it conjunctive partial deduction} may affect program termination
when transforming programs for eliminating multiple traversals of
data structures. A similar problem arises when multiple traversals
of data structures are avoided by applying Tamaki and Sato's unfold/fold
transformation rules~\cite{TaS84} according to the tupling strategy
(see Section~\ref{sec:motivating_ex}).
In this example by using goal arguments and introducing continuations,
we are able to derive a program in continuation passing style which
eliminates multiple traversals of data structures and, at the same time,
preserves universal termination.

\medskip
\noindent Let us consider the initial program {\it FlipCheck}:

\smallskip
\Feq{1.}\mathit{flipcheck}(X,Y)\leftarrow \mathit{flip}(X,Y),\
\mathit{check}(Y) \Nex{2.}\mathit{flip}(l(N),l(N))\leftarrow
\Nex{3.}\mathit{flip}(t(L,N,R),t(\mathit{FR},N,\mathit{FL}))\leftarrow 
\mathit{flip}(L,\mathit{FL}),\
        \mathit{flip}(R,\mathit{FR})
\Nex{4.}\mathit{check}(l(N))\leftarrow \mathit{nat}(N)
\Nex{5.}\mathit{check}(t(L,N,R))\leftarrow \mathit{nat}(N),\
        \mathit{check}(L),\ \mathit{check}(R)
\Nex{6.}\mathit{nat}(0)\leftarrow
\Nex{7.}\mathit{nat}(s(N))\leftarrow \mathit{nat}(N)
\eeq

\smallskip
\noindent where: (i)~the term \( l(N) \) denotes a leaf with label
\( N \) and the term \( t(L,N,R) \) denotes a tree with label \( N \)
and the two subtrees \( L \) and \( R \), (ii)~\( \mathit{nat}(X) \)
holds iff \( X \) is a natural number, (iii)~\( \mathit{check}(X) \)
holds iff all labels in the tree \( X \) are natural numbers, and
(iv)~\( \mathit{flip}(X,Y) \) holds iff the tree \( Y \) can be
obtained by flipping all subtrees of the tree~\( X \). 

We would like to transform this program so to avoid the double traversal
of trees (see the double occurrence of \( Y \) in the body of clause
1). By applying the tupling strategy (or, equivalently, 
{\it conjunctive
partial deduction}), we derive the following program {\it FlipCheck}1:

\smallskip
\Feq{8.}\mathit{flipcheck}(l(N),l(N))\leftarrow \mathit{nat}(N)
\Nex{9.}\mathit{flipcheck}(t(L,N,R),t(\mathit{FR},N,\mathit{FL}))\nc
        \leftarrow  \nc \mathit{nat}(N),\
\Nex{}\hspace*{2.5cm}\mathit{flipcheck}(L,\mathit{FL}),\
      \mathit{flipcheck}(R,\mathit{FR})
\eeq

\smallskip
\noindent Program {\it FlipCheck}1 performs only one traversal of
any input tree which is the first argument of \( \mathit{flipcheck} \).
However, as already mentioned, {\it FlipCheck}1 does not preserve
termination. Indeed, the goal \( \mathit{flipcheck}(t(l(N),0,l(a)),Y) \)
fails in {\it FlipCheck}, while this goal does not terminate in the
derived program {\it FlipCheck}1.

Now we present a second derivation starting from the same program
{\it FlipCheck} and producing a final program {\it FlipCheck}2 which:
(i)~is in continuation passing style, (ii)~traverses the input tree only once,
and (iii)~preserves termination. During this second derivation we introduce
goal arguments and we make use of the transformation rules introduced in
Section~\ref{sec:rules}. The initial step of this derivation is the
introduction of the following new clause:

\oeq{10.}{\mathit{newp}(X,Y,G,C,D) \leftarrow \mathit{flip}(X,Y),\   G\! =\!
(\mathit{check}(Y),C),\  D }

\noindent As already mentioned, in this paper we do not illustrate
the strategies needed for guiding the application of our transformation
rules and, in particular, we do not indicate how to construct the
new definitions to be introduced, such as clause~10 above.
For clause~10 we notice that: (i)~by introducing a definition with the
goal equality \( G\! =\! (\mathit{check}(Y),\, C) \), instead of
the goal \( \mathit{check}(Y) \), we will be able to apply the folding
rule by first performing leftward moves of goal equalities, instead
of (possibly incorrect) leftward moves of goals, and (ii)~by introducing
the continuations \( C \) and \( D \), we will avoid the expensive
use of the conjunction operator for constructing goal arguments.

We continue our derivation by unfolding clause 10 w.r.t.~\(
\mathit{flip}(X,Y) \) and we get:

\smallskip
\Feq{11.}\mathit{newp}(l(N),l(N),G,C,D)\leftarrow (G\! =\!
(\mathit{check}(l(N)),C)),\ D
\Nex{12.}\mathit{newp}(t(L,N,R),t(\mathit{FR},N,\mathit{FL}),G,C,D)\nc
\leftarrow \nc \mathit{flip}(L,\mathit{FL}),\
\mathit{flip}(R,\mathit{FR})
\Nex{}\hspace*{2.5cm}
  (G\!=\!(\mathit{check}(t(\mathit{FR},N,\mathit{FL})),C)),\ D \eeq

\smallskip
\noindent We then unfold clauses 11 and 12 w.r.t.~the {\it check}
atoms, and after some applications of the goal replacement rule based
on boolean laws and CET, we get:

\smallskip
\Feq{13.}\mathit{newp}(l(N),l(N),G,C,D)\leftarrow G\! =\!
(\mathit{nat}(N),C),\, \, D
\Nex{14.}\mathit{newp}(t(L,N,R),t(\mathit{FR},N,\mathit{FL}),G,C,D)\nc
\leftarrow \nc \mathit{flip}(L,\mathit{FL}),\ \mathit{flip}(R,\mathit{FR}),
\Nex{}\hspace*{2.5cm}
(G\!=\!(\mathit{nat}(N),\mathit{check}(\mathit{FR}),
\mathit{check}(\mathit{FL}),C)),\ D
\eeq

\smallskip
\noindent By introducing and rearranging goal equalities (see laws
2.1 and 2.2, respectively, in Section~\ref{sec:rules}), we transform
clause 14 into:

\Feq{15.}\mathit{newp}(t(L,N,R),t(\mathit{FR},N,\mathit{FL}),G,C,D)\leftarrow
\mathit{flip}(L,\mathit{FL}),\,  U\! =\!
(\mathit{check}(\mathit{FL}),C),
\Nex{}\hspace*{2.5cm}\mathit{flip}(R,\mathit{FR}),\,  V\! =\!
(\mathit{check}(\mathit{FR}),U),\ \ (G\! =\! (\mathit{nat}(N),V)),\ D
\eeq

\noindent Now we fold twice clause 15 using clause 10 and we get:

\smallskip
\Feq{16.} \mathit{newp}(t(L,N,R),t(\mathit{FR},N,\mathit{FL}),G,C,D)\leftarrow
\Nex{}\hspace*{2.5cm}
\mathit{newp}(L,\mathit{FL},U,C,\mathit{newp}(R,\mathit{FR},V,U,
(G\! =\! (\mathit{nat}(N),V),D)\, ))
\eeq

\smallskip
\noindent In order to express \( \mathit{flipcheck} \) in terms of
\( \mathit{newp} \) we introduce a goal equality into clause 1 and
we derive:
\smallskip{}

\oeq{17.}{\mathit{flipcheck}(X,Y)\leftarrow  \mathit{flip}(X,Y),\, \, G\! =\!
(\mathit{check}(Y),\mathit{true}),\, \, G}

\noindent Then we fold clause 17 using clause 10 and we get:
\smallskip{}

\oeq{18.}{\mathit{flipcheck}(X,Y)\leftarrow
\mathit{newp}(X,Y,G,\mathit{true},G)}

\noindent The program we have derived so far consists of clauses 13,
16, and 18. Notice that clauses 13 and 16 are not in continuation passing
style because the conjunction operator occurs in their bodies.
In order to derive clauses in
continuation passing style we
introduce the following new definition:

\smallskip
\Feq{19.}\mathit{nat}\U\mathit{c}(N,C)\leftarrow \mathit{nat}(N),\ C
\eeq

\smallskip

\noindent
By unfolding, folding, and goal replacement steps based on the replacement law  
${\it FlipCheck} \vdash\,\forall\, ( (X\!=\!Y), C \, \llarr \, 
{\it eq}\U{c}(X,Y,C))$, we derive the following
final program {\it FlipCheck}2:

\smallskip
\Feq{18.}\mathit{flipcheck}(X,Y)\leftarrow
\mathit{newp}(X,Y,G,\mathit{true},G)

\Nex{20.}\mathit{newp}(l(N),l(N),G,C,D) \leftarrow
         \mathit{eq}\U\mathit{c}(G,\mathit{nat}\U\mathit{c}(N,C),D)

\Nex{21.}\mathit{newp}(t(L,N,R),t(\mathit{FR},N,\mathit{FL}),G,C,D) \leftarrow
\Nex{}\hspace*{2.5cm}\mathit{newp}(L,\mathit{FL},U,C,\
                     \mathit{newp}(R,\mathit{FR},V,U,\
\Nex{}\hspace*{2.5cm}
         \mathit{eq}\U\mathit{c}(G,\mathit{nat}\U\mathit{c}(N,V),D)\ ))
\Nex{22.}\mathit{nat}\U\mathit{c}(0,C)\leftarrow C
\Nex{23.}\mathit{nat}\U\mathit{c}(s(N),C)\leftarrow
         \mathit{nat}\U\mathit{c}(N,C)
\eeq

\smallskip
\noindent Program {\it FlipCheck}2 traverses the input tree only
once. Moreover, Theorem~\ref{th:preservation} ensures that, for every goal
\( g \) of the form \( \mathit{flipcheck}(t_{1},t_{2}) \), where
\( t_{1} \) and \( t_{2} \) are any two terms, \( g \)
terminates in 
{\it FlipCheck} iff \( g \) terminates in {\it FlipCheck}2
(see also Section~\ref{subsec:correctness_of_ex} for a more detailed 
discussion of the correctness properties
of our program derivations).

\subsection{Summing the Leaves of a Tree}\label{sec:treesum}

Let us consider the following program {\it TreeSum} that, given a
binary tree \( t \) whose leaves are labeled by natural numbers,
computes the sum of the labels of the leaves of \( t \).

\smallskip
\Feq{1.}\mathit{treesum}(l(N),N)\leftarrow
\Nex{2.}\mathit{treesum}(t(L,R),N)\leftarrow \mathit{treesum}(L,\mathit{NL}),\
      \mathit{treesum}(R,\mathit{NR}),\
      \mathit{plus}(\mathit{NL},\mathit{NR},N)
\Nex{3.}\mathit{plus}(0,X,X)\leftarrow
\Nex{4.}\mathit{plus}(s(X),Y,s(Z))\leftarrow \mathit{plus}(X,Y,Z)
\eeq

\smallskip
\noindent
By using Tamaki and Sato's transformation rules, from program
{\it TreeSum} we may derive a more efficient program with {\it accumulator}
arguments. In particular, during this program derivation we introduce
the following new predicate:

\oeq{5.}{\mathit{acc}\U\mathit{ts}(T,Y,Z)\leftarrow \mathit{treesum}(T,X),\
\mathit{plus}(X,Y,Z)}

\noindent We also use the associativity of the predicate {\it plus},
that is, we use the following equivalence which holds in the least
Herbrand model \( M(\mathit{TreeSum}) \) of the given
program {\it TreeSum}:

\smallskip{}
\noindent \hspace*{.8cm}\( \mathit{M}(\mathit{TreeSum})\models 
\forall \, X1,X2,X3,S\, (\exists I\,
  (\mathit{plus}(X1,X2,I),\, \mathit{plus}(I,X3,S))\leftrightarrow  \)

\noindent \hspace*{5.833cm} \(
\exists J\, (\mathit{plus}(X1,J,S),\, \mathit{plus}(X2,X3,J))) \) \smallskip{}

\noindent During the derivation, we also make suitable goal rearrangements
needed for performing foldings that use clause 5. 
We derive the following program {\it TreeSum}1.

\smallskip
\Feq{6.}\mathit{treesum}(l(N),N)\leftarrow
\Nex{7.}\mathit{treesum}(t(L,R),N)\leftarrow
      \mathit{acc}\U\mathit{ts}(L,\mathit{NR},N),\
      \mathit{treesum}(R,\mathit{NR})
\Nex{8.}\mathit{acc}\U\mathit{ts}(l(N),\mathit{Acc},Z)\leftarrow
      \mathit{plus}(N,\mathit{Acc},Z)
\Nex{9.}\mathit{acc}\U\mathit{ts}(t(L,R),\mathit{Acc},N)\leftarrow
      \mathit{acc}\U\mathit{ts}(L,\mathit{Acc},\mathit{NewAcc}),\
      \mathit{acc}\U\mathit{ts}(R,\mathit{NewAcc},N)
\eeq

\smallskip
\noindent The least Herbrand models of programs {\it TreeSum} and
{\it TreeSum}1 define the same relation for the predicate {\it treesum}.
However, the two programs do not have the same termination behaviour.
For instance, the goal \( \mathit{treesum}(t(l(N),0),Z) \) fails
in {\it TreeSum} while it does not terminate in {\it TreeSum}1.

By  introducing
goal arguments and using the transformation rules
presented in Section~\ref{sec:rules}, we are able to derive a program
which: (i)~is in continuation passing style, (ii)~preserves termination, and
(iii)~is asymptotically more efficient than the original program {\it
TreeSum}. Our derivation begins by introducing the following new clause:

\oeq{10.}{\mathit{gen}\U\mathit{ts}(T,Y,Z,G,C,D)\leftarrow
\mathit{treesum}(T,X),\   (G\! =\!
(\mathit{plus}(X,Y,Z),C)),\  D}

\noindent We unfold clause~10 and we get:

\smallskip
\Feq{11.}\mathit{gen}\U\mathit{ts}(l(N),Y,Z,G,C,D)\leftarrow
    (G\! =\! (\mathit{plus}(N,Y,Z),C)),\  D
\Nex{12.}\mathit{gen}\U\mathit{ts}(t(L,R),Y,Z,G,C,D) \leftarrow
\mathit{treesum}(L,\mathit{LS}),\ \mathit{treesum}(R,\mathit{RS}),\
\Nex{}\hspace*{2.5cm}\mathit{plus}(\mathit{LS},\mathit{RS},S),\
(G\! =\! (\mathit{plus}(S,Y,Z),C)),\  D
\eeq
\smallskip

\noindent Now we may exploit the
following generalized associativity law for {\it plus}:

\smallskip
\noindent \hspace*{.8cm}\( \mathit{TreeSum}\vdash \forall V\,
((\mathit{plus}(X1,X2,I),\, g[\mathit{plus}(I,X3,S)])\larr  \)

\noindent \hspace*{3.3cm}\( (\mathit{plus}(X1,J,S),\,
\mathit{g}[\mathit{plus}(X2,X3,J)])) \)

\smallskip
\noindent  where $V= \{X1,X2,X3,S\}\cup {\it vars}(g[\_])$ and 
$\{I,J\}\cap {\it vars}(g[\_])=\emptyset$.
By this law, from clause~12 we get the following clause:

\smallskip
\Feq{13.}\mathit{gen}\U\mathit{ts}(t(L,R),Y,Z,G,C,D) \leftarrow
\mathit{treesum}(L,\mathit{LS}),\ \mathit{treesum}(R,\mathit{RS}),\
\Nex{}\hspace*{2.5cm}\mathit{plus}(\mathit{LS},\mathit{S1},Z),\
(G\! =\! (\mathit{plus}(\mathit{RS},Y,\mathit{S1}),C)),\  D
\eeq
\smallskip

\noindent By introducing and rearranging goal equalities (see laws~2.1 and
2.2 in Section~\ref{sec:rules}), we transform clause 13 into:

\smallskip
\Feq{14.}\mathit{gen}\U\mathit{ts}(t(L,R),Y,Z,G,C,D) \leftarrow
\Nex{}\hspace*{2.5cm}\mathit{treesum}(L,\mathit{LS}),\
(\mathit{GL}\!=\!(\mathit{plus}(\mathit{LS},S1,Z),\ G\!=\!\mathit{GR},\ D)),\
\Nex{}\hspace*{2.5cm}\mathit{treesum}(R,\mathit{RS}),\
(\mathit{GR}\!=\!(\mathit{plus}(\mathit{RS},Y,S1),C)),\ GL
\eeq
\smallskip

\noindent
In order to derive clauses in continuation passing style 
we introduce the
following new definitions:

\smallskip
\Feq{15.}\mathit{ts}\U\mathit{c}(T,N,C) \leftarrow \mathit{treesum}(T,N),\ C
\Nex{16.}\mathit{plus}\U\mathit{c}(X,Y,Z,C)\leftarrow \mathit{plus}(X,Y,Z),\ C
\eeq
\smallskip

\noindent By unfolding clauses 15 and 16 we get:

\smallskip
\Feq{17.}\mathit{ts}\U\mathit{c}(l(N),N,C) \leftarrow C
\Nex{18.}\mathit{ts}\U\mathit{c}(t(L,R),N,C)\nc \leftarrow \nc
         \mathit{treesum}(L,\mathit{LN}),\ \mathit{treesum}(R,\mathit{RN}),
\bdy\mathit{plus}(LN,RN,N),\ C
\Nex{19.}\mathit{plus}\U\mathit{c}(0,X,X,C)\leftarrow C
\Nex{20.}\mathit{plus}\U\mathit{c}(s(X),Y,s(Z),C)\leftarrow
         \mathit{plus}(X,Y,Z),\ C
\eeq
\smallskip

\noindent
By introducing and rearranging goal equalities, we transform
clause~18 into:

\smallskip
\Feq{21.}\mathit{ts}\U\mathit{c}(t(L,R),N,C) \nc \leftarrow \nc
\mathit{treesum}(L,\mathit{LN}),\ (G=(\mathit{plus}(LN,RN,N),C)),\
\bdy \mathit{treesum}(R,\mathit{RN}),\ G
\eeq
\smallskip

\noindent By folding steps and goal replacements (based on, among others, 
the replacement law
${\it TreeSum} \vdash\, \forall\, ((X\!=\!Y), C \, \llarr \, 
{\it eq}\U{c}(X,Y,C))$), we get the
following final program {\it TreeSum}2:

\smallskip
\Feq{22.}\mathit{treesum}(T,N)\leftarrow
         \mathit{ts}\U c(T,N,\mathit{true})
\Nex{18.}\mathit{ts}\U\mathit{c}(l(N),N,C) \leftarrow C
\Nex{23.}\mathit{ts}\U c(t(L,R),N,C)\leftarrow   \mathit{gen}\U
      \mathit{ts}(L,\mathit{RN},N,G,C,\mathit{ts}\U c(R,\mathit{RN},G))
\Nex{24.}\mathit{gen}\U\mathit{ts}(l(N),Y,Z,G,C,D)\leftarrow
    \mathit{eq}\U c(G, \mathit{plus}\U c(N,Y,Z,C), D)
\Nex{25.}\mathit{gen}\U \mathit{ts}(t(L,R),Y,Z,G,C,D)\nc \leftarrow \nc
      \mathit{gen}\U \mathit{ts}(L,\mathit{S1},Z,\mathit{GL},
              \mathit{eq}\U c(G,\mathit{GR},D),\
\bdy \mathit{gen}\U \mathit{ts}(R,Y,\mathit{S1},
              \mathit{GR},C,\mathit{GL}))
\Nex{19.}\mathit{plus}\U c(0,X,X,C)\leftarrow C
\Nex{20.}\mathit{plus}\U c(s(X),Y,s(Z),C)\leftarrow
         \mathit{plus}\U c(X,Y,Z,C)
\eeq

\smallskip
\noindent
This final program {\it TreeSum}2 is more efficient than
{\it TreeSum}. Indeed, in the worst case, {\it TreeSum}2 takes \( O(n) \)
steps for solving a goal of the form \( \mathit{treesum}(t,N) \),
where $t$ is a ground tree and \( s^{n}(0) \) is the sum of 
the labels of the leaves of \( t \), while
the initial program {\it TreeSum} takes \( O(n^{2}) \) steps. Moreover,
by our Theorem~\ref{th:preservation} of Section~\ref{sec:correctness}, 
for every goal \( g \) of the form \( \mathit{treesum}(t_1,t_2) \),
where $t_1$ and $t_2$ are any terms, 
$g$ terminates in {\it TreeSum} iff $g$ terminates in
{\it TreeSum}2 (see also Section~\ref{subsec:correctness_of_ex}).

\subsection{Matching a Regular Expression}\label{sec:matching}

Let us consider the following matching problem: given a string \( S \)
in \( \{0,1,2\}^{*} \), we want to find the position \( N \) of
an occurrence of a substring \( P \) of \( S \) such that \( P \)
is generated by the regular expression \( 0^{*}1 \). The following
program {\it RegExprMatch} computes such a position:

\smallskip
\Feq{1.}\mathit{match}(S,N)\leftarrow \mathit{pattern}(S),\ N\!=\!0
\Nex{2.}\mathit{match}([C|S],N)\leftarrow
      \mathit{char}(C),\ \mathit{match}(S,M),\ \mathit{plus}(s(0),M,N)
\Nex{3.}\mathit{pattern}([0|S])\leftarrow
      \mathit{pattern}(S)
\Nex{4.}\mathit{pattern}([1|S])\leftarrow
\Nex{5.}\mathit{char}(0)\leftarrow
\Nex{6.}\mathit{char}(1)\leftarrow
\Nex{7.}\mathit{char}(2)\leftarrow
\Nex{8.}\mathit{plus}(0,X,X)\leftarrow
\Nex{9.}\mathit{plus}(s(X),Y,s(Z))\leftarrow \mathit{plus}(X,Y,Z)
\eeq

\smallskip{}
\noindent If we assume the depth-first, left-to-right evaluation
strategy of Prolog, the running time of this program 
{\it RegExprMatch} is \( O(n^{2}) \)
in the worst case, where \( n \) is the length of the
input string. For a goal of the form \( \mathit{match}(s,N) \), where
\( s \) is a ground string made out of \( n \) 0's, the program {\it
RegExprMatch} performs one resolution step using clause 1 for the
call to {\it match}, and then \( n \)
resolution steps using clause 3 for the successive
calls to {\it pattern}. When the
computation backtracks, for the successive call of \( \mathit{match}(s1,N) \),
where \( s1 \) is the tail of \( s \), the program {\it RegExprMatch} performs
again \( n\! -\! 1 \) resolution steps using clause~3.

By using the transformation rules of Section~\ref{sec:rules}, we now
present the derivation of a new program {\it RegExprMatch}1 which: (i)~is in
continuation passing style, (ii)~preserves termination, and (iii)~is
asymptotically more efficient than the original program {\it RegExprMatch}.
Indeed, program {\it RegExprMatch}1 avoids the redundant resolution steps
performed by {\it RegExprMatch} using clause~3. For our derivation we
introduce the following new predicates with goal arguments which are
continuations:

\smallskip
\Feq{10.}\mathit{match}\U c(S,N,C)\leftarrow \mathit{match}(S,N),\ C
\Nex{11.}\mathit{newp}(S,N,C1,C2)\leftarrow
(\mathit{pattern}(S),\, C1)\  \vee \       (\mathit{match}(S,N),\, C2)
\Nex{12.}\mathit{plus}\U\mathit{c}(X,Y,Z,C)\leftarrow \mathit{plus}(X,Y,Z),\ C
\eeq

\smallskip{}
\noindent By unfolding clauses 10, 11, and 12 we get:

\smallskip
\Feq{13.}\mathit{match}\U c([0|S],N,C)\nc\leftarrow\nc(\mathit{pattern}(S),
         N\!=\!0, C)\ \vee  \
\bdy (\mathit{match}(S,M), \mathit{plus}(s(0),M,N), C)
\Nex{14.}\mathit{match}\U c([1|S],N,C)\nc\leftarrow\nc(N\!=\!0, C)\ \vee  \
\bdy (\mathit{match}(S,M), \mathit{plus}(s(0),M,N), C)
\Nex{15.}\mathit{match}\U c([2|S],N,C) \leftarrow \mathit{match}(S,M),
         \mathit{plus}(s(0),M,N), C
\Nex{16.}\mathit{newp}([0|S],N,C1,C2)\nc\leftarrow\nc(\mathit{pattern}(S),C1)\
            \vee  \
\bdy (\mathit{pattern}(S), N\!=\!0, C2)\ \vee  \
\bdy (\mathit{match}(S,M),\mathit{plus}(s(0),M,N), C2)
\Nex{17.}\mathit{newp}([1|S],N,C1,C2)\nc\leftarrow\nc C1\
            \vee  \
\bdy (N\!=\!0, C2)\ \vee  \
\bdy (\mathit{match}(S,M),\mathit{plus}(s(0),M,N), C2)
\Nex{18.}\mathit{newp}([2|S],N,C1,C2) \leftarrow
            \mathit{match}(S,M),\mathit{plus}(s(0),M,N), C2
\Nex{19.}\mathit{plus}\U\mathit{c}(0,X,X,C)\leftarrow C
\Nex{20.}\mathit{plus}\U\mathit{c}(s(X),Y,s(Z),C)\leftarrow
         \mathit{plus}(X,Y,Z),\ C
\eeq

\smallskip
\noindent By goal replacement using boolean laws, from clause 16 we get:

\smallskip
\Feq{21.}\mathit{newp}([0|S],N,C1,C2)\nc\leftarrow\nc(\mathit{pattern}(S),
(C1 \ \vee  \ (N\!=\!0, C2)))\ \vee  \
\bdy (\mathit{match}(S,M),\mathit{plus}(s(0),M,N), C2)
\eeq

\smallskip
\noindent By performing folding and goal replacement steps 
(based on the replacement law
${\it RegExprMatch} \vdash\,\forall\, ( (X\!=\!Y), C \, \llarr \, 
{\it eq}\U{c}(X,Y,C))$ and other laws), we derive the
following program {\it RegExprMatch}1:

\smallskip
\Feq{22.}\mathit{match}(S,N)\leftarrow \mathit{match}\U
\mathit{c}(S,N,\mathit{true})
\Nex{23.}\mathit{match}\U
\mathit{c}([0|S],N,C)\leftarrow \mathit{newp}(S,M,\, \mathit{eq}\U
\mathit{c}(N,0,C),\,  \mathit{plus}\U \mathit{c}(s(0),M,N,C))
\Nex{24.}\mathit{match}\U \mathit{c}([1|S],N,C)\leftarrow  \mathit{eq}\U
   c(N,0,C)
\Nex{25.}\mathit{match}\U \mathit{c}([1|S],N,C)\leftarrow
\mathit{match}\U \mathit{c}(S,M,\, \mathit{plus}\U \mathit{c}(s(0),M,N,C))
\Nex{26.}\mathit{match}\U \mathit{c}([2|S],N,C)\leftarrow  \mathit{match}\U
\mathit{c}(S,M,  \,       \mathit{plus}\U \mathit{c}(s(0),M,N,C))
\Nex{27.}\mathit{newp}([0|S],N,C1,C2)\leftarrow
\Nex{}\hspace*{2.5cm}\mathit{newp}(S,M,\,(C1\vee\mathit{eq}\U\mathit{c}(N,0,C2)),
      \mathit{plus}\U\mathit{c}(s(0),M,N,C2))
\Nex{28.}\mathit{newp}([1|S],N,C1,C2)\leftarrow C1
\Nex{29.}\mathit{newp}([1|S],N,C1,C2)\leftarrow  \mathit{eq}\U
\mathit{c}(N,0,C2)
\Nex{30.}\mathit{newp}([1|S],N,C1,C2)\leftarrow
\mathit{match}\U \mathit{c}(S,M,\,       \mathit{plus}\U
\mathit{c}(s(0),M,N,C2))
\Nex{31.}\mathit{newp}([2|S],N,C1,C2)\leftarrow
\mathit{match}\U \mathit{c}(S,M,\,       \mathit{plus}\U
\mathit{c}(s(0),M,N,C2))
\Nex{19.}\mathit{plus}\U\mathit{c}(0,X,X,C)\leftarrow C
\Nex{32.}\mathit{plus}\U\mathit{c}(s(X),Y,s(Z),C)\leftarrow
         \mathit{plus}\U\mathit{c}(X,Y,Z,C)
\eeq

\smallskip{}
\noindent This program {\it RegExprMatch}1 is in continuation
passing style, avoids redundant
calls in case of backtracking, and takes \( O(n) \) resolution steps
in the worst case, to find an occurrence of a substring of the form
\( 0^{*}1 \), where $n$ is the length of the input string.
Moreover, by our Theorem~\ref{th:preservation} of 
Section~\ref{sec:correctness},
for every goal \( g \) of the form 
$\mathit{match}(t_1,t_2)$,
where $t_1$ and $t_2$ are any terms,
$g$ terminates in {\it RegExprMatch} iff $g$ terminates in
{\it RegExprMatch}1 (see also Section~\ref{subsec:correctness_of_ex}).

\subsection{Marking maximal elements}\label{sec:maxmark}

Let us consider the following marking problem. We are given: (i)~a list
$L1$ of the form $[x_0,\ldots,x_r]$, where
for $i\!=\!0,\ldots,r$, $x_i$ is a list of integers, and (ii)~an integer
$n$ $(\geq\!0)$. 
A list $l$ of $s\!+\!1$ elements will also be denoted by
$[l[0],\ldots,l[s]]$.
We assume that for $i\!=\!0,\ldots,r$, the list $x_i$ has at
least $n\!+\!1$ elements (and thus, the element $x_i[n]$ exists)
and we denote by $m$ the
maximum element of the set $\{x_0[n], \ldots, x_r[n]\}$. 
From the list $L1$ we
want to compute a new list $L2$ of the form $[y_0,\ldots,y_r]$
such that, for
$i\!=\!0,\ldots,r$, if $x_i[n]\!=\!m$ then $y_i[n]\!=\!\top$ else
$y_i[n]\!=\!x_i[n]$.

For instance, if $L1 = [[3,8,-2,4],\ [1,3],\ [1,8,1]]$ and
$n\!=\!1$, then $m\!=\!8$, that is, the maximum element in $\{8,3\}$. Thus,
 $L2=[[3,\top,2,4],\ [1,3],\ [1,\top,1]]$.

The following
program {\it MaxMark} computes the desired list $L2$ from the list $L1$ and
the value $N$:

\smallskip
\Feq{1.}\mathit{mmark}(N,L1,L2)\leftarrow
        \mathit{max}\U\mathit{nth}(N,L1,0,M),\
        \mathit{mark}(N,M,L1,L2)
\Nex{2.}\mathit{max}\U\mathit{nth}(N,[~],M,M)\leftarrow
\Nex{3.}\mathit{max}\U\mathit{nth}(N,[X|\mathit{Xs}],A,M)\nc \leftarrow \nc
        \mathit{nth}(N,X,\mathit{XN}),\ \mathit{max}(A,\mathit{XN},B),\
        \bdy \mathit{max}\U\mathit{nth}(N,\mathit{Xs},B,M)
\Nex{4.}\mathit{nth}(0,[H|T],H)\leftarrow
\Nex{5.}\mathit{nth}(s(N),[H|T],E)\leftarrow \mathit{nth}(N,T,E)
\Nex{6.}\mathit{mark}(N,M,[~],[~])\leftarrow
\Nex{7.}\mathit{mark}(N,M,[X|\mathit{Xs}],[Y|\mathit{Ys}])\nc \leftarrow \nc
	\mathit{mark}\U\mathit{nth}(N,M,X,Y),\
	\bdy \mathit{mark}(N,M,\mathit{Xs},\mathit{Ys})
\Nex{8.}\mathit{mark}\U\mathit{nth}(0,M,[H1|T],[H2|T])\leftarrow
	(M\!=\!H1, H2\!=\!\top) \vee (M\!\neq\!H1, H2\!=\!H1)
\Nex{9.}\mathit{mark}\U\mathit{nth}(s(N),M,[H|T1],[H|T2])\leftarrow
	\mathit{mark}\U\mathit{nth}(N,M,T1,T2)
\Nex{10.}\mathit{max}(X,Y,X) \leftarrow X\geq Y
\Nex{11.}\mathit{max}(X,Y,Y) \leftarrow X<Y
\eeq

\medskip

\noindent
When running this program, the input list $L1=[x_0,\ldots,x_r]$ is traversed 
twice: (i)~the first time $L1$
is traversed to compute the maximum $m$ of the set
$\{x_0[n], \ldots, x_r[n]\}$
(see the goal 
$\mathit{max}\U\mathit{nth}(N,L1,0,M)$ in the body of clause 1),
and (ii)~the second time  $L1$ is traversed to construct the list $L2$ by
replacing, for $i\!=\!0,\ldots,r$, the element $x_i[n]$ by $\top$
whenever $x_i[n]\!=\!m$ (see the goal $\mathit{mark}(N,M,L1,L2)$).

\medskip

Now we use the transformation rules of Section~\ref{sec:rules} and from
program {\it MaxMark} we
derive a new program {\it MaxMark}1 
which: (i)~is in continuation passing style, (ii)~preserves termination, and
(iii)~traverses the list $L1$ only once.

By the definition introduction rule 
we introduce the following new predicates with goal arguments:

\smallskip

\Feq{12.}\mathit{newp}1(N,L1,L2,A,M,G,C1,C2)\nc\leftarrow\nc
\nex\hspace*{2.5cm}\mathit{max}\U\mathit{nth}(N,L1,A,M),\
	(G\!=\!(\mathit{mark}(N,M,L1,L2),\ C1)),\ C2
\Nex{13.}\mathit{newp}2(N,X,M,Y,A,B,G1,G2,C)\nc\leftarrow\nc
\nex\hspace*{2.5cm}\mathit{nth}(N,X,\mathit{XN}),\
	(G1\!=\!(\mathit{mark}\U\mathit{nth}(N,M,X,Y), G2)),\
\nex\hspace*{2.5cm}\mathit{max}(A,\mathit{XN},B),\ C
\Nex{14.}\mathit{max}\U\mathit{c}(X,Y,Z,C)\leftarrow \mathit{max}(X,Y,Z),\ C
\eeq

\medskip

\noindent
We unfold clauses 12, 13, and 14, and then we move leftwards term
equalities (see law~3 in Section~\ref{sec:rules}
which allows us to rearrange term equalities).
We get the following clauses:

\smallskip

\Feq{15.}\mathit{newp}1(N,[~],[~],M,M,C1,C1,C2)\nc\leftarrow\nc C2
\Nex{16.}\mathit{newp}1(N,[X|\mathit{Xs}],[Y|\mathit{Ys}],A,M,G,C1,C2)
	\leftarrow
\nex\hspace*{2.5cm}
	\mathit{nth}(N,X,\mathit{XN}),\ \mathit{max}(A,\mathit{XN},B),\
\mathit{max}\U\mathit{nth}(N,\mathit{Xs},B,M),\
\nex\hspace*{2.5cm}(G\!=\!(\mathit{mark}\U\mathit{nth}(N,M,X,Y),\
	\mathit{mark}(N,M,\mathit{Xs},\mathit{Ys}),\ C1)),\
\nex\hspace*{2.5cm}C2
\Nex{17.}\mathit{newp}2(0,[H1|T],M,[H2|T],A,B,G1,G2,C)\nc\leftarrow\nc
\nex\hspace*{2.5cm}
	(G1\!=\!(((M\!=\!H1, H2\!=\!\top) \vee (M\!\neq\!H1, H2\!=\!H1)), G2)),\ 
\nex\hspace*{2.5cm}\mathit{max}(A,H1,B),\ C
\Nex{18.}\mathit{newp}2(s(N),[H|T1],M,[H|T2],A,B,G1,G2,C)\nc\leftarrow\nc
\nopagebreak
\nex\hspace*{2.5cm}\mathit{nth}(N,T1,\mathit{XN}),\
	(G1\!=\!(\mathit{mark}\U\mathit{nth}(N,M,T1,T2), G2)),\ 
\nopagebreak
\nex\hspace*{2.5cm}\mathit{max}(A,\mathit{XN},B),\ C
\Nex{19.}\mathit{max}\U\mathit{c}(X,Y,X,C)\leftarrow X\!\geq\! Y,\ C
\Nex{20.}\mathit{max}\U\mathit{c}(X,Y,Y,C)\leftarrow X\!<\!Y,\ C
\eeq

\smallskip

\noindent
By introducing and rearranging goal equalities, from clause 16 we get:

\smallskip

\Feq{21.}\mathit{newp}1(N,[X|\mathit{Xs}],[Y|\mathit{Ys}],A,M,G,C1,C2)
	\leftarrow \nopagebreak
\nex\hspace*{2.5cm}
	\mathit{nth}(N,X,\mathit{XN}),\
	(G1\!=\!(\mathit{mark}\U\mathit{nth}(N,M,X,Y),\ G2)),
\nex\hspace*{2.5cm}
\mathit{max}(A,\mathit{XN},B),\
\nex\hspace*{2.5cm}
\mathit{max}\U\mathit{nth}(N,\mathit{Xs},B,M),\
	(G2\!=\!(\mathit{mark}(N,M,\mathit{Xs},\mathit{Ys}),\ C1)),
\nex\hspace*{2.5cm}(G\!=\!G1),\ C2
\eeq

\smallskip

\noindent
Finally, by folding steps and goal replacements based on the replacement laws
for the primitive predicates $=$, $\neq$, $\geq$, and $<$, we derive 
the following final program {\it MaxMark}1:

\smallskip

\Feq{22.}\mathit{mmark}(N,L1,L2)\leftarrow
	\mathit{newp}1(N,L1,L2,0,M,G,{\it true},G)
\Nex{15.}\mathit{newp}1(N,[~],[~],M,M,C1,C1,C2)\nc\leftarrow C2
\Nex{23.}\mathit{newp}1(N,[X|\mathit{Xs}],[Y|\mathit{Ys}],A,M,G,C1,C2)
	\leftarrow
\nex\hspace*{2.5cm}
	\mathit{newp}2(N,X,M,Y,A,B,G1,G2),\ 
\nex\hspace*{2.5cm}
	\mathit{newp}1(N,\mathit{Xs},\mathit{Ys},B,M,G2,C1,
	\mathit{eq}\U{c}(G,G1,C2)))
\Nex{24.}\mathit{newp}2(0,[H1|T],M,[H2|T],A,B,G1,G2,C)\nc\leftarrow\nc
\nex\hspace*{2.5cm}
	\mathit{eq}\U{c}(G1,(\mathit{eq}\U{c}(M,H1, 
	\mathit{eq}\U{c}(H2,\top, G2)) \vee 
\nex\hspace*{4.cm}
	\mathit{diff}\U{c}(M,H1, \mathit{eq}\U{c}(H2,H1,G2))),\ 
\nex\hspace*{2.5cm}\mathit{max}\U{c}(A,H1,B,C))
\Nex{25.}\mathit{newp}2(s(N),[H|T1],M,[H|T2],A,B,G1,G2,C)\nc\leftarrow\nc
\nex\hspace*{2.5cm}\mathit{newp}2(N,T1,M,T2,A,B,G1,G2,C)
\Nex{26.}\mathit{max}\U\mathit{c}(X,Y,X,C)\leftarrow {\it geq}\U{c}(X, Y, C)
\Nex{27.}\mathit{max}\U\mathit{c}(X,Y,Y,C)\leftarrow {\it lt}\U{c}(X,Y,C)
\eeq

\smallskip
\noindent
This final program {\it MaxMark}1 is in
continuation passing style and traverses the input list $L1$ only once.
Moreover, by our Theorem~\ref{th:preservation} of 
Section~\ref{sec:correctness},
for every goal \( g \) of the form 
$\mathit{mmark}(t_1,t_2,t_3)$,
where $t_1$, $t_2$, and $t_3$ are any terms, if
$g$ terminates in {\it MaxMark} then $g$ terminates in
{\it MaxMark}1 (see also Section~\ref{subsec:correctness_of_ex}).

\subsection{Correctness of the Program Derivations}
\label{subsec:correctness_of_ex}

Let us briefly comment on the correctness properties
of the program derivations
we have presented in this Section~\ref{sec:examples}.

In all program derivations of Section~\ref{sec:examples}, when using the 
transformation rules, we have complied with 
the restrictions indicated at Point~(1) of
Theorem~\ref{th:correctness} (Weak Correctness). Thus, 
for every program derivation from an initial program $P_{0}$ to
a final program $P_k$, we have that
\( P_{k} \) is a refinement of \( P_{0}\cup \mathit{Def}\! _{k} \),
where ${\it Def}\!_{k}$ is the set of definitions introduced during
the derivation. In particular, for every ordinary goal 
\( g \), if \( g \) terminates in \( P_{0}\), then \( g \) terminates in 
\( P_{k} \)
and the most general answer substitutions for $g$ computed by \( P_{0}\)
are the same as those computed by \( P_{k} \).

In the examples of Sections~\ref{sec:flipcheck}, 
\ref{sec:treesum}, and \ref{sec:matching} we have also complied with
the restrictions of Point~(2) of
Theorem~\ref{th:correctness} (Strong Correctness),
because all applications of the goal replacement rule
are based on strong replacement laws. Thus, in these examples
we have that \( P_{k} \) is equivalent to 
\( P_{0}\cup \mathit{Def}\! _{k} \).
In particular, for every ordinary goal 
\( g \), if  \( g \) terminates in \( P_{k} \) then
\( g \) terminates in \( P_{0}\cup \mathit{Def}\! _{k} \).

However, in the derivation of 
Section~\ref{sec:maxmark} we have not complied with
the restrictions of Point~(2) of Theorem~\ref{th:correctness}.
In particular, after unfolding clauses 12, 13, and 14,
we have made leftward moves of term equalities by using
law~3 of Section~\ref{sec:rules}, and law~3 is not a strong replacement
law.
Thus, there may be an ordinary goal which does not terminate
in the initial program {\it MaxMark} and terminates in the
final program {\it MaxMark}1. Indeed, the goal
${\it mmark}(0,[H|T],[~])$ does not terminate in {\it MaxMark}
and terminates in {\it MaxMark}1.

Finally, in all program derivations of this Section~\ref{sec:examples},
we have complied with the restrictions of Theorem~\ref{th:safety} 
(Preservation of Safety), because all replacement laws we have applied 
preserve safety.
Thus, since every ordinary goal is safe in the ordinary
initial program $P_{0}$, we have that every ordinary goal is safe 
in the final program~\(P_{k} \).

\subsection{Experimental Results\label{sec:speedups}}

In Table~\ref{table:speed-ups} below we have reported the 
speed-ups achieved in the examples
presented in this paper. The {\it speed-up} (see Column~D) is defined as
the ratio between the run-time of the initial program (see Column~A)
and the run-time of the derived, final program (see Column~B).
In Columns~A and B we have also indicated the asymptotic worst-case time
complexity of the initial and final programs, respectively. For each program
the complexity is measured in terms of the size of the proofs 
relative to that program (or,
equivalently, the number of LD-resolution steps performed 
using that program). 
The input goal is indicated in Column~C. We performed our
measurements by using BinProlog on a SUN workstation. This use is justified
by the fact that 
every ordinary goal $g$ is safe both in the initial program \( P_{0} \) and
in the final program $P_k$. Thus, we can use any Prolog system which
implements LD-resolution (and, in particular, the BinProlog system)
for computing the relations \( P_{0}\, \vdash \, g\mapsto A \) and
\( P_{k}\, \vdash \, g\mapsto A \) defined by our operational semantics.
\begin{table}[h]
  \caption{{\it Speed-ups of the Final Programs with respect to the Initial
           Programs}}
  \label{table:speed-ups}
  \begin{minipage}{\textwidth}
    \begin{tabular}{lllc}
      \hline\hline
      A. Initial  Program\,:& B. Final Program\,:&C. Input goal
      &D. Speed-up\,:\footnote{run-time$({\rm A})$ denotes the
       run-time of the program in Column~A for the input goal in Column~C.
       run-time$({\rm B})$ denotes the
       run-time of the program in Column~B for the input goal in Column~C.}\\
     ~~~{\scriptsize Asymptotic Complexity}&~~~{\scriptsize Asymptotic
         Complexity} &
    &~~~$\frac{\textup {\footnotesize{run-time(A)}}}
             {\textup {\footnotesize{run-time(B)}}}$\\
   \hline
1. {\it Deepest} $:O(n^2)$\footnote{$n$ is the number of nodes of the tree
$t_1$.}
   &{\it Deepest}2 $:O(n)$
   &$\mathit{deepest}(t_1,\!N)$
   &  5.2     \\
2. {\it DeepestOr} $:O(n^2)$\footnote{$n$ is the number of nodes of the tree
$t_2$.}
   & {\it Deepest}2 $:O(n)$
   & $\mathit{deepest}(t_2,\!N)$
   &  2.7     \\
3. {\it FlipCheck} $:O(n)$\footnote{$n$ is the number of nodes of the tree
$t_3$. For the goal $\mathit{flipcheck}(t_3,\!T)$, the program {\it FlipCheck}
visits the tree $t_3$ twice, while the program {\it FlipCheck}2 visits $t_3$ only once.}
   & {\it FlipCheck}2 $:O(n)$
   & $\mathit{flipcheck}(t_3,\!T)$
   &  1.0     \\
4. {\it TreeSum} $:O(n^2)$\footnote{$n$ is the sum of the leaves of
   the tree $t_4$.}
   & {\it TreeSum}2$:O(n)$&$\mathit{treesum}(t_4,\!N)$
   &  9.2   \\
5. {\it RegExprMatch} $:O(n^2)$\footnote{$n$
    is the length of the string $s$.}
   & {\it RegExprMatch}1 $:O(n)$ & ${\mathit match}(s,\!N)$
   & 1.8 \\
6. {\it MaxMark} $:O(n)$\footnote{$n$
    is the sum of the lengths of the lists in $l_1$.}
   & {\it MaxMark}1 $:O(n)$ & ${\mathit mmark}(n_1,\!l_1,\!L_2)$
   & 1.8 \\
\hline\hline
    \end{tabular}
    \vspace{-2\baselineskip}
  \end{minipage}
\end{table}

In Column~C of Table~\ref{table:speed-ups} we have that:\\ 
(1) $t_1$ is a random binary tree with 100,000 nodes;\\
(2) $t_2$ is a random binary tree with 100,000 nodes;\\
(3) \( t_3 \)
is a random binary tree with 20,000 nodes and each node is labeled by
a numeral of the form $s^{k}(0)$, where $0\!\leq\! k\!\leq\!500$;\\
(4) \( t_4 \) is a
random binary tree with 20,000 nodes whose leaves are labeled by
numerals of the form $s^{k}(0)$, where \( 0\!\leq\! k\!\leq\!500 \); \\
(5) \( s \) is a random
sequence of integers of the form: $\{0,2\}^{50000}1$; and \\
(6) $n_1$ is 700, $l_1$ is a random list of 1000 lists, and
each of these lists consists of
800 integers.

When measuring the speed-ups for the programs {\it Deepest} and {\it
DeepestOr}\/  in Rows~1 and 2 we have computed the set of all answer
substitutions, while for the programs {\it FlipCheck}, {\it TreeSum}, 
{\it RegExprMatch}, and {\it MaxMark} in Rows~3--6 we have computed one 
answer substitution only.

As already mentioned at the end of Section~\ref{sec:motivating_ex},
the value of the speed-up relative to the initial program {\it Deepest}
(see Row~1) is higher than the 
value of the speed-up relative to the initial program {\it
DeepestOr}\/ (see Row~2), and this is not 
due to the use of goals as arguments, but
to the introduction of a disjunction, thereby clauses~2 and 3 have
been replaced by clause~16.

The absence of speed-up for the final program {\it FlipCheck}2 
(see Row~3) with
respect to the initial program {\it FlipCheck}, is caused by the
fact that the efficiency improvements due 
to the elimination of the double traversal
of the input tree $t_4$ are cancelled out
by the slowdown due to the introduction
of multiple continuation arguments.
However, the experimental results for the initial program
{\it MaxMark} and the final program {\it MaxMark}1  (see Row~6) 
show that the elimination of 
double traversals of data structures may yield a significant speed-up,
especially when the access to the data structure is very costly.
Recall that the program {\it MaxMark} traverses twice the list 
$l_1$, and for each list $l$ in the list $l_1$, the program has to access
$n_1$ elements of $l$. We have verified that
the speed-up obtained by eliminating the
double traversal of $l_1$ increases with the value of $n_1$.

\section{Final Remarks and Related Work\label{sec:final}}

We have shown that a simple extension of logic programming, where
variables may range over goals and goals may appear as arguments of
predicate symbols, can be very useful for transforming programs and
improving their efficiency. 

We have presented a set of transformation rules for our extended logic
language and we have shown their correctness with respect to~the
operational semantics given in Section~\ref{sec:semantics}. 
In particular, in Section~\ref{sec:correctness}
we have shown that, under suitable conditions, our transformation
rules preserve termination (see Theorem~\ref{th:preservation}),
most general answer substitutions (see Theorem~\ref{th:correctness}),
and safety (see Theorem~\ref{th:safety}). As in~\cite{BoC94}, for
our logic programs we consider an operational semantics based on universal
termination (that is, the operational semantics of a goal is defined
iff all LD-derivations starting from that goal are finite). 
Theorem~\ref{th:correctness}
extends the results presented in~\cite{BoC94} for definite logic
programs in that: (i)~our language is an extension of definite logic
programs, and (ii)~our folding rule is more powerful. Indeed, even
restricting ourselves to programs that do not contain goal variables
and goal arguments, we allow folding steps which use clauses whose
bodies contain disjunctions, and this is not possible in~\cite{BoC94},
where for applying the folding rule one is required to use exactly
one clause whose body is a conjunction of atoms. However,
one should notice that
the transformations presented in~\cite{BoC94} preserve {\it all}
computed answer substitutions, while ours preserve the {\it most}
{\it general} answer substitutions only. 

Our logic language has some higher order capabilities because 
goals may occur as arguments, but 
these capabilities are limited by the fact that the quantification of
function or predicate variables is not allowed. 
However, the objective of this paper is not the design of a 
new higher order logic language,
such as the ones presented in~\cite{Ch&93,HiG98,NaM98}.
Rather, our aim was to demonstrate
the usefulness of some higher order constructs for deriving
efficient logic programs by transformation. 
Indeed, we have shown
that variables which range over goals are useful in the context of
program transformation.
Moreover, the use of these variables may avoid the need for goal 
rearrangements
which could generate programs that do not preserve termination. 

The approach we have proposed in this paper for avoiding incorrect
goal rearrangements, is complementary to the approach described 
in~\cite{Bo&95},
where the authors give sufficient conditions for goal rearrangements
to preserve {\it left termination.} (Recall that a program \( P \)
is said to be left terminating iff all {\it ground} goals universally
terminate in \( P \).) Thus, when these sufficient conditions are
not met or their validity cannot be proved, one may apply our technique
which avoids incorrect goal rearrangements by the introduction and
the rearrangement of goal equalities. Indeed, we have proved
that the application of our technique preserves universal termination, and
thus, it preserves left termination as well.

The theory we have presented may also be used to give sound semantic
foundations to the development of logic programs which use {\it higher
order generalizations} and {\it continuations}. In~\cite{PeP97b,TaB90}
and~\cite{PeS87,Wan80} the reader may find some examples of use of these
techniques in the case of logic and functional programs, respectively.

We leave for future work the development of suitable strategies for
directing the use of the transformation rules we have proposed in
this paper.

\section*{Acknowledgements}

We would like to thank Michael Leuschel for pointing out an error
in a preliminary version of this paper and for his helpful comments.
We also thank the anonymous referees of the LoPSTr '99 Workshop,
where a preliminary version of this paper was presented~\cite{PeP99b}, and the
referees of the Theory and Practice of Logic Programming Journal for their
suggestions. 

This work has been partially supported by MURST Progetto
Cofinanziato `Tecniche Formali per la Specifica, l'Analisi, la Verifica, la
Sintesi e la Trasformazione di Sistemi Software' (Italy), and Progetto
Coordinato CNR `Verifica, Analisi e Trasformazione dei Programmi Logici'
(Italy).

\section*{Appendix}

\noindent This Appendix contains:\\
(i)~Proposition \ref{proposition:context}
and its proof,\\
(ii)~the proofs of Lemmata~\ref{lemma:improvement}, 
\ref{lemma:improving_rules},
\ref{lemma:soundness}, and \ref{lemma:sound_rules}  (based on
Propositions~\ref{proposition:repl} and \ref{proposition:context}),\\
(iii)~Lemmata~\ref{lemma:safety} and \ref{lemma:safe_rules} and
their proofs (based on
Proposition~\ref{proposition:context}), and\\
(iv)~the proofs of the main results, that is, (iv.1)~the proof of
Theorem~\ref{th:preservation} (based on Proposition~\ref{prop:ordered},
Lemmata~\ref{lemma:improvement}, \ref{lemma:improving_rules}, 
\ref{lemma:soundness},
and \ref{lemma:sound_rules}), (iv.2)~the proof of 
Theorem~\ref{th:correctness}
(based on Proposition~\ref{proposition:cas} and 
Theorem~\ref{th:preservation}),
and (iv.3)~the proof of Theorem~\ref{th:safety} (based on 
Lemmata~\ref{lemma:safety}
and \ref{lemma:safe_rules}). 

For the proofs of Proposition~\ref{proposition:context}
and Lemma~\ref{lemma:improvement} given below, we need the following
definition.

\begin{definition}[Size and \( \mu  \)-measure of a Deduction Tree]
\label{def:tau_mu_measure}Let \( \tau  \) be a finite deduction
tree. The {\it size} of \( \tau  \) is the number of its nodes,
and the {\it \(\mu\)-measure} of \( \tau  \), denoted \( \mu (\tau ) \),
is the pair \( \langle m,s\rangle  \), where \( m \) is the depth
of \( \tau  \) and \( s \) is the size of \( \tau  \).
\end{definition}

The values of the \( \mu  \)-measure can be lexicographically ordered,
and we stipulate that: 
\( \langle m_{1},s_{1}\rangle <\langle m_{2},s_{2}\rangle  \)
iff either \( m_{1}\! <\! m_{2} \) or (\( m_{1}\! =\! m_{2} \) and
\( s_{1}\! <\! s_{2} \)).

\begin{proposition}
\noindent\label{proposition:context}Let \( P \)
be a program, \( g_{1} \), \( g_{2} \) be goals and let \( V \)
be a set of variables. 

\smallskip

\noindent (i)~\( P\vdash \, \forall V\, 
(g_{1}\ar g_{2}) \)
holds iff for every idempotent substitution \( \vartheta  \) such
that \( \mathit{vars}(\vartheta )\cap {\it vars}(g_1,g_2) \subseteq V \),
for every goal \( g \) such that 
\( \mathit{vars}(g)\cap {\it vars}(g_1,g_2) \subseteq V \),
and for every \( b\in \{\mathit{true},\mathit{false}\} \), we have
that:

if \( P\vdash \, (g_{1}\vartheta \wedge g)\downarrow b \) then
\( P\vdash \, (g_{2}\vartheta \wedge g)\downarrow b \).

\smallskip

\noindent (ii)~\( P\vdash \, \forall V\, 
(g_{1}\arr g_{2}) \)
holds iff for every idempotent substitution \( \vartheta  \) such
that \( \mathit{vars}(\vartheta )\cap {\it vars}(g_1,g_2) \subseteq V \),
for every goal \( g \) such that 
\( \mathit{vars}(g)\cap {\it vars}(g_1,g_2) \subseteq V \),
and for every \( b\in \{\mathit{true},\mathit{false}\} \), we have
that:

if \( P\vdash \, (g_{1}\vartheta \wedge g)\downarrow _{m}b \)
then \( P\vdash \, (g_{2}\vartheta \wedge g)\downarrow _{n}b \) and
\( m\geq n \).

\smallskip

\noindent (iii)~The following two properties are equivalent:

\smallskip

\noindent (iii.1)~for every goal context \( h[\_] \) such
that \( \mathit{vars}(h[\_])\cap {\it vars}(g_1,g_2) \subseteq V  \),

if \( h[g_{1}] \) is safe in \( P \) then \( h[g_{2}] \)
is safe in \( P \), and

\smallskip

\noindent (iii.2)~for every idempotent substitution \( \vartheta  \)
such that \( \mathit{vars}(\vartheta )\cap {\it vars}(g_1,g_2) \subseteq V  \)
and for every goal \( g \)  such that \( \mathit{vars}(g)\cap {\it vars}(g_1,g_2) 
\subseteq V\),

if \( g_{1}\vartheta \wedge g \) is safe in \( P \) then \( g_{2}\vartheta
\wedge g \) is safe in \( P \).
\end{proposition}

\begin{proof}[Proof of Proposition~\ref{proposition:context}]
(i){\it ~only-if} part. Let us consider
an idempotent substitution \( \vartheta  \) such that \( 
\mathit{vars}(\vartheta )\cap {\it vars}(g_1,g_2) \subseteq V \).
Let \( \vartheta  \) be \( \{U_{1}/u_{1},\ldots ,U_{k}/u_{k}\} \).
Since \( \vartheta  \) is idempotent we have that for \( i=1,\ldots ,k \),
\( U_{i}\not \in u_{i} \). Assume that for every goal \( g \) such
that \( \mathit{vars}(g)\cap {\it vars}(g_1,g_2) \subseteq V \), and
for every \( b\in \{\mathit{true},\mathit{false}\} \), there exists
\( A_{1}\in \mathcal{P}(\mathit{Subst}) \) such that \( P\vdash \, 
(g_{1}\vartheta \wedge g)\mapsto A_{1} \).
We have to show that there exists \( A_{2}\in \mathcal{P}(\mathit{Subst}) \)
such that \( P\vdash \, (g_{2}\vartheta \wedge g)\mapsto A_{2} \)
and \( A_{1}\! =\! \emptyset  \) iff \( A_{2}\! =\! \emptyset  \).

\noindent By suitably renaming the variables of the goal \( g_{1} \),
without loss of generality we may assume that, for \( i=1,\ldots ,k \),
\( U_{i}\not \in \mathit{vars}(g) \). Since \( \vartheta  \) is
idempotent, by using rules~({\it teq}2) and ({\it geq}) we may
construct a proof of $ P\vdash \, U_{1}\! =\! u_{1}\wedge \ldots 
\wedge U_{k}\! =\! u_{k}\wedge$ $ g_{1}\wedge g\mapsto B_{1}$,
where \( B_{1}\! =\! (\vartheta \circ A_{1}) \). By the hypothesis
that \( P\vdash \, \forall V\, (g_{1}\ar g_{2}) \)
holds and the hypotheses that \( \mathit{vars}(\vartheta )\cap 
{\it vars}(g_1,g_2) \subseteq V \)
and \( \mathit{vars}(g)\cap {\it vars}(g_1,g_2) \subseteq V\), we have
that there exists \( B_{2}\in \mathcal{P}(\mathit{Subst}) \) such
that \( P\vdash \, U_{1}\! =\! u_{1}\wedge \ldots \wedge U_{k}\! =\! u_{k}
\wedge\) \(g_{2}\wedge g\mapsto B_{2} \)
has a proof and \( B_{1}\! =\! \emptyset  \) iff \( B_{2}\! =\! \emptyset  \).
The only way of constructing this proof is by using \( k \) times
the rules~({\it teq}2) or ({\it geq}) and constructing a proof
of \( P\vdash \, g_{2}\vartheta \wedge g\mapsto A_{2} \), where 
\( B_{2}\! =\! (\vartheta \! \circ \! A_{2}) \).
Thus, \( A_{1}\! =\! \emptyset  \) iff \( B_{1}\! =\! \emptyset  \)
iff \( B_{2}\! =\! \emptyset  \) iff \( A_{2}\! =\! \emptyset  \).
\smallskip{}   

\noindent (i){\it ~if} part. We show a slightly more general fact
than the {\it if} part of~(i). We assume that for every idempotent
substitution \( \vartheta  \) such that \( \mathit{vars}(\vartheta )\cap 
\mathit{vars}(g_1,g_2)\subseteq V\),
and for every goal \( g \) such that \( \mathit{vars}(g)\cap 
\mathit{vars}(g_1,g_2)\subseteq V\),
if there exists \( A_{1}\in \mathcal{P}(\mathit{Subst}) \) such that
\( P\vdash \, (g_{1}\vartheta \wedge g)\mapsto A_{1} \), then there
exists \( A_{2}\in \mathcal{P}(\mathit{Subst}) \) such that \( P\vdash \, 
(g_{2}\vartheta \wedge g)\mapsto A_{2} \)
and \( A_{1}\! =\! \emptyset  \) iff \( A_{2}\! =\! \emptyset  \).
Then we show that, for every goal context \( h[\_] \) and substitution
\( \vartheta  \) such that \( \mathit{vars}(h[\_]\vartheta )\cap 
\mathit{vars}(g_1,g_2)\subseteq V\), 

if there exists \( B_{1}\in \mathcal{P}(\mathit{Subst}) \) such that
\( P\, \vdash \, h[g_{1}]\vartheta \mapsto B_{1} \)

then there exists \( B_{2}\in \mathcal{P}(\mathit{Subst}) \) such
that \( P\, \vdash \, h[g_{2}]\vartheta \mapsto B_{2} \)

\noindent and \( B_{1}\! =\! \emptyset  \) iff \( B_{2}\! =\! \emptyset  \).
\smallskip{}

\noindent We prove our thesis by induction on the measure \( \mu (\pi ) \)
(see Definition~\ref{def:tau_mu_measure}) of the proof \( \pi  \)
of \( P\, \vdash \, h[g_{1}]\vartheta \mapsto B_{1} \)(recall that
a proof is a particular finite deduction tree). We reason by cases
on the structure of the goal context \( h[\_] \). We consider the
following four cases only. The others are similar and we omit them.
\smallskip{}

\noindent - Case 1: \( h[\_] \) is \( \_\wedge g_{3} \).

\noindent Assume that \( \mathit{P}\, \vdash \, g_{1}\vartheta \wedge 
g_{3}\vartheta \mapsto B_{1} \).
Then, by hypothesis, we get: \( \mathit{P}\, \vdash \, g_{2}\vartheta 
\wedge g_{3}\vartheta \mapsto B_{2} \)
for some \( B_{2}\in \mathcal{P}(\mathit{Subst}) \) such that 
\( B_{1}\! =\! \emptyset  \)
iff \( B_{2}\! =\! \emptyset  \). 
\smallskip{}

\noindent - Case 2: \( h[\_] \) is \( t_{1}\! =\! t_{2}\wedge g_{3}[\_] \).

\noindent Assume that there exists a proof \( \pi _{1} \) of \( \mathit{P}\, 
\vdash \, t_{1}\vartheta \! =\! t_{2}\vartheta \wedge g_{3}[g_{1}]\vartheta 
\mapsto B_{1} \).

\noindent If \( t_{1}\vartheta  \) and \( t_{2}\vartheta  \) are
not unifiable then, by rule ({\it teq}1), \( B_{1} \) is \( \emptyset  \)
and there exists a proof of \( \mathit{P}\, \vdash \, t_{1}\vartheta \! =\! 
t_{2}\vartheta \wedge g_{3}[g_{2}]\vartheta \mapsto \emptyset  \). 

\noindent If \( t_{1}\vartheta  \) and \( t_{2}\vartheta  \) are
unifiable then, by rule ({\it teq}2), \( B_{1} \) is of the form
\( (\mathit{mgu}(t_{1}\vartheta ,t_{2}\vartheta )\circ C_{1}) \)
for some \( C_{1}\in \mathcal{P}(\mathit{Subst}) \) and there exists
a proof \( \pi _{2} \) of
$\mathit{P}\,\vdash \,g_{3}[g_{1}]\vartheta \,
\mathit{mgu}(t_{1}\vartheta,t_{2}\vartheta)\mapsto C_{1}$.
Since \( \mu
(\pi _{2})<\mu (\pi _{1}) \), by induction hypothesis \( \mathit{P}\, \vdash
\, g_{3}[g_{2}]\vartheta \, \mathit{mgu}(t_{1}\vartheta ,t_{2}\vartheta
)\mapsto C_{2} \) has a proof for some \( C_{2}\in \mathcal{P}(\mathit{Subst})
\) and \( C_{1}\! =\! \emptyset  \) iff \( C_{2}\! =\! \emptyset  \). Thus, by
rule~({\it teq}2), there exists \( B_{2}\in \mathcal{P}(\mathit{Subst}) \)
such that \( \mathit{P}\, \vdash \, t_{1}\vartheta \! =\! t_{2}\vartheta
\wedge g_{3}[g_{2}]\vartheta \mapsto B_{2} \) where \(
B_{2}=\mathit{mgu}(t_{1}\vartheta ,t_{2}\vartheta )\circ C_{2} \) and \(
B_{1}\! =\! \emptyset  \) iff \( C_{1}\! =\! \emptyset  \) iff \( C_{2}\! =\!
\emptyset  \) iff \( B_{2}\! =\! \emptyset  \). \smallskip{}

\noindent - Case 3: \( h[\_] \) is \( (G\! =\! g_{3}[\_])\wedge g_{4} \).

\noindent Assume that \( \mathit{P}\, \vdash \, ((G\! =\! g_{3}[g_{1}])\wedge 
g_{4})\vartheta \mapsto B_{1} \)
has a proof of depth~\( m \) and size~\( s \). Then, \( G\vartheta  \)
is a goal variable not occurring in \( g_{3}[g_{1}]\vartheta  \),
the node \( \mathit{P}\, \vdash \, (G\vartheta \! =\! g_{3}[g_{1}]\vartheta
)\wedge g_{4}\vartheta \mapsto B_{1} \) has been obtained by applying rule \(
(\mathit{geq}) \), \( B_{1} \) is \( \{G\vartheta /g_{3}[g_{1}]\vartheta \}\!
\circ \! C_{1} \) for some \( C_{1}\in \mathcal{P}(\mathit{Subst}) \), and \(
\mathit{P}\, \vdash \, g_{4}\vartheta \{G\vartheta /g_{3}[g_{1}]\vartheta
\}\mapsto C_{1} \) has a proof of depth \( m \) and size \( s\! -\! 1 \).
Now, suppose that $G\vartheta$ occurs in $g_{4}\vartheta$ $n$ times.
Thus, also $g_{1}$ will occur $n$ times in
$g_{4}\vartheta \{G\vartheta /g_{3}[g_{1}]\vartheta \}$.
Since \( \langle m,s\! -\!
1\rangle < \langle m,s\rangle \), by applying the induction hypothesis $n$
times, we have that there exists \( C_{2}\in \mathcal{P}(\mathit{Subst}) \)
such that \( \mathit{P}\, \vdash \, g_{4}\vartheta \{G\vartheta
/g_{3}[g_{2}]\vartheta \}\mapsto C_{2} \) has a proof and \( C_{1}\! =\!
\emptyset  \) iff \( C_{2}\! =\! \emptyset \). By using rule \( (\mathit{geq})
\), we can construct a proof of \( \mathit{P}\, \vdash \, G\vartheta \! =\!
g_{3}[g_{2}]\vartheta \wedge g_{4}\vartheta \mapsto B_{2} \), where \( B_{2}
\) is \( \{G\vartheta /g_{3}[g_{2}]\vartheta \}\! \circ \! C_{2} \). Thus, \(
B_{1}\! =\! \emptyset \) iff \( C_{1}\! =\! \emptyset  \) iff \( C_{2}\! =\!
\emptyset  \) iff \( B_{2}\! =\! \emptyset \). \smallskip{}

\noindent - Case 4: \( h[\_] \) is \( p(u_{1},\ldots ,u_{i}[\_],\ldots ,u_{k})
\wedge g_{3} \).

\noindent Assume that \( \mathit{P}\, \vdash \, p(u_{1}\vartheta ,
\ldots ,u_{i}[g_{1}]\vartheta ,\ldots ,u_{k}\vartheta )\wedge g_{3}\vartheta 
\mapsto B_{1} \)
has a proof of depth~\( m \) and size~\( s \). Then, in the last
step of this proof rule \( (\mathit{at}) \) has been used, \( B_{1} \)
is of the form \( C_{1}\! \upharpoonright \! \mathit{vars}(p(u_{1}\vartheta ,
\ldots ,u_{i}[g_{1}]\vartheta ,\ldots ,u_{k}\vartheta )\wedge g_{3}\vartheta ) 
\)
for some \( C_{1}\in \mathcal{P}(\mathit{Subst}) \), and \( \mathit{P}\, 
\vdash \, \mathit{body}\{U_{1}/u_{1}\vartheta ,\ldots ,
U_{i}/u_{i}[g_{1}]\vartheta ,\ldots ,U_{k}/u_{k}\vartheta \}\wedge 
g_{3}\vartheta \, \mapsto \, C_{1} \)
has a proof of depth \( m\! -\! 1 \) and size \( s\! -\! 1 \), where
\( p(U_{1},\ldots ,U_{i},\ldots ,U_{k})\leftarrow \mathit{body} \)
is a renamed apart clause of \( \mathit{P} \). 
Since \( \langle m\! -\! 1,s\! -\! 1\rangle <\langle m,s\rangle  \),
by induction hypothesis we have that there exists \( C_{2}\in 
\mathcal{P}(\mathit{Subst}) \)
such that \( \mathit{P}\, \vdash \, 
\mathit{body}\{U_{1}/u_{1}\vartheta ,\ldots ,
U_{i}/u_{i}[g_{2}]\vartheta ,\ldots ,U_{k}/u_{k}\vartheta \}\wedge 
g_{3}\vartheta \, \mapsto \, C_{2} \)
has a proof and \( C_{1}\! =\! \emptyset  \) iff \( C_{2}\! =\! \emptyset  \).
Thus, by using rule \( (\mathit{at}) \), we can construct a proof of
\( \mathit{P}\, \vdash \, p(u_{1}\vartheta ,\ldots ,u_{i}[g_{2}]\vartheta ,
\ldots ,u_{k}\vartheta )\wedge g_{3}\vartheta \mapsto B_{2} \),
where \( B_{2} \) is \( C_{2}\! \upharpoonright \! 
\mathit{vars}(p(u_{1}\vartheta ,\ldots ,u_{i}[g_{2}]\vartheta ,
\ldots ,u_{k}\vartheta )\wedge g_{3}\vartheta ) \)
and \( B_{1}\! =\! \emptyset  \) iff \( C_{1}\! =\! \emptyset  \)
iff \( C_{2}\! =\! \emptyset  \) iff \( B_{2}\! =\! \emptyset  \).
\smallskip{}

\noindent (ii)~The proof is similar to the one of (i) and we omit it.
\smallskip{}

\noindent (iii)~Suppose that (iii.1) holds and suppose also that
\( \vartheta  \) is an idempotent substitution such that 
\( \mathit{vars}(\vartheta )\cap \mathit{vars}(g_1,g_2)\subseteq V  \),
\( g \) is a goal such that \( \mathit{vars}(g)\cap 
\mathit{vars}(g_1,g_2)\subseteq V \),
and \( g_{1}\vartheta \wedge g \) is safe in \( P \). We have to
prove that \( g_{2}\vartheta \wedge g \) is safe in \( P \).

\noindent Suppose that \( g_{2}\vartheta \wedge g \) is not safe
in \( P \). Then there exist \( A\in \mathcal{P}(\mathit{Subst}) \)
and a deduction tree \( \tau _{1} \) for \( P\vdash g_{2}\vartheta 
\wedge g\mapsto A \)
such that a leaf of \( \tau _{1} \) is of the form \( P\, \vdash \, 
g_{3}\mapsto B \)
and \( g_{3} \) is stuck. Let \( \vartheta  \) be the substitution
\( \{U_{1}/u_{1},\ldots ,U_{k}/u_{k}\} \) such that, 
for \( i=1,\ldots ,k \),
\( U_{i}\not \in u_{i} \). Without loss of generality, we may assume
that, for \( i=1,\ldots ,k \), \( U_{i}\not \in \mathit{vars}(g) \).
By using rules ({\it teq}2) and ({\it geq}), we can construct a
deduction tree \( \tau _{2} \) for \( P\vdash U_{1}\! =\! u_{1}\wedge 
\ldots \wedge U_{k}\! =\! u_{k}\wedge g_{2}\wedge g\mapsto A \)
such that \( \tau _{2} \) has \( P\, \vdash \, g_{3}\mapsto B \)
at a leaf. Thus, \( U_{1}\! =\! u_{1}\wedge \ldots \wedge U_{k}\! =\! u_{k}
\wedge g_{2}\wedge g \)
is not safe in \( P \). Since \( \mathit{vars}(\vartheta )\cap 
\mathit{vars}(g_1,g_2)\subseteq V \)
and \( \mathit{vars}(g)\cap \mathit{vars}(g_1,g_2)\subseteq V \), we have
that \( \mathit{vars}(U_{1}\! =\! u_{1}\wedge \ldots 
\wedge U_{k}\! =\! u_{k}\wedge g)\cap \mathit{vars}(g_1,g_2)\subseteq V \)
and, thus, by (iii.1) \( U_{1}\! =\! u_{1}\wedge \ldots \wedge 
U_{k}\! =\! u_{k}\wedge g_{1}\wedge g \)
is not safe in \( P \). None of the goals \( U_{1}\! =\! u_{1},\ldots ,
U_{k}\! =\! u_{k} \)
is stuck and, thus, a descendant node of \( g_{1}\vartheta \wedge g \)
is stuck, that is, \( g_{1}\vartheta \wedge g \) is not safe in \( P \). 
\smallskip{}

\noindent The proof that (iii.2) implies (iii.1) can be done by induction
on deduction trees ordered by the \( \mu  \)-measure. We omit this
proof. \hfill\end{proof}

\begin{proof}
[Proof of Lemma~\ref{lemma:improvement}]Recall
that, by definition, for every \( b\in \{\mathit{true},\mathit{false}\} \),
\( P\, \vdash \, g\downarrow _{m}b \) means that there exists \( A\in 
\mathcal{P}(\mathit{Subst}) \)
such that \( \mathit{P}\, \vdash \, g\mapsto A \) has a proof of
depth \( m \) and \( b\! =\! \mathit{true} \) iff \( A\! \neq \! 
\emptyset  \).
We prove the thesis by induction on the \( \mu  \)-measure (see 
Definition~\ref{def:tau_mu_measure})
of the proof of \( \mathit{P}\, \vdash \, g\mapsto A \) which, by
hypothesis, has depth \( m \) and size~\( s \).

\noindent
Our induction hypothesis is that, for all \( \langle m1,s1\rangle <\langle 
m,s\rangle  \),
for all goals \( g \), and for all \( A_{1}\in \mathcal{P}(\mathit{Subst}) \),
if \( P\, \vdash \, g\mapsto A_{1} \) has a proof of depth \( m1 \)
and size \( s1 \), then there exists 
\( B_{1}\in \mathcal{P}(\mathit{Subst}) \)
such that \( \mathit{NewP}\, \vdash \, g\mapsto B_{1} \) has a proof
of depth \( n1 \), with \( m1\geq n1 \), and \( A_{1}\! =\! \emptyset  \)
iff \( B_{1}\! =\! \emptyset  \). We have to show that there exists
\( B\in \mathcal{P}(\mathit{Subst}) \) such that \( \mathit{NewP}\, \vdash 
\, g\mapsto B \)
has a proof of depth \( n \), with \( m\geq n \), and \( A\! =\! \emptyset  \)
iff \( B\! =\! \emptyset  \). We proceed by cases on the structure
of \( g \).  We first notice that, since \( \wedge  \) is associative
with neutral element {\it true}, the grammar for generating goals
given in Section 2 can be replaced by the following one:

\smallskip{}
$g::=G\wedge g_{1}\ |\ \mathit{true}\, |\ \mathit{false}\wedge
g_{1}\ |\ (t_{1}\! =\! t_{2})\wedge g_{1}\ |\ (g_{1}\! =\! g_{2})\wedge
g_{3}\ |$

~~~~~~~~$p(u_{1},\ldots ,u_{m})\wedge g_{1}\ |\ (g_{1}\vee g_{2})\wedge
g_{3}$

\smallskip{}
\noindent
We consider the following two cases only. The others are similar
and we omit them. 

\noindent - Case 1: \( g \) is \( (g_{1}\! =\! g_{2})\wedge g_{3} \).
Assume that \( \mathit{P}\, \vdash \, (g_{1}\! 
=\! g_{2})\wedge g_{3}\mapsto A \)
has a proof of depth \( m \) and size \( s \). Then, \( g_{1} \)
is a goal variable, say \( G \), \( G\not \in \mathit{vars}(g_{2}) \),
\( \mathit{P}\, \vdash \, (G\! =\! g_{2})\wedge g_{3}\mapsto A \)
has been derived by applying rule \( (\mathit{geq}) \), and there
exists \( A_{1}\in \mathcal{P}(\mathit{Subst}) \) such that \( A\! =\! 
(\{G/g_{2}\}\circ A_{1}) \)
and \( \mathit{P}\, \vdash \, g_{3}\{G/g_{2}\}\mapsto A_{1} \) has
a proof of depth \( m \) and size \( s-1 \). 
Since \( \langle m,s\! -\! 1\rangle <\langle m,s\rangle  \),
by induction hypothesis there exists \( B_{1}\in 
\mathcal{P}(\mathit{Subst}) \)
such that \( \mathit{NewP}\, \vdash \, g_{3}\{G/g_{2}\}\mapsto B_{1} \)
has a proof of depth \( n \) with \( m\geq n \) and \( A_{1}\! =\! 
\emptyset  \)
iff \( B_{1}\! =\! \emptyset  \). By rule \( (\mathit{geq}) \),
we have that \( \mathit{NewP}\, \vdash \, (G\! =\! g_{2})\wedge 
g_{3}\mapsto B \),
where \( B=(\{G/g_{2}\}\circ B_{1}) \), has a proof of depth \( n \)
with \( m\geq n \). By the definition of the \( \circ  \) operator,
we have that \( A\! =\! \emptyset  \) iff \( A_{1}\! =\! \emptyset  \)
iff \( B_{1}\! =\! \emptyset  \) iff \( B\! =\! \emptyset  \).
\smallskip{}

\noindent - Case 2: \( g \) is \( p(u_{1},\ldots ,u_{m})\wedge g_{1} \).
Assume that \( \mathit{P}\, \vdash \, p(u_{1},\ldots ,u_{m})\wedge 
g_{1}\mapsto A \)
has a proof of depth \( m \) and size \( s \). Then, \( \mathit{P}\, 
\vdash \, p(u_{1},\ldots ,u_{m})\wedge g_{1}\mapsto A \)
has been derived by using rule \( (\mathit{at}) \), and there exists
\( A_{1}\in \mathcal{P}(\mathit{Subst}) \) such that \( A\! =\! (A_{1}\! 
\upharpoonright \! \mathit{vars}(p(u_{1},\ldots ,u_{k})\wedge g_{1})) \)
and \( \mathit{P}\, \vdash \, \mathit{bd}_{r}\{V_{1}/u_{1},\ldots ,
V_{m}/u_{m}\}\wedge g_{1}\, \mapsto A_{1} \)
has a proof of depth \( m\! -\! 1 \) and size \( s\! -\! 1 \), where
\( p(V_{1},\ldots ,V_{m})\leftarrow \mathit{bd}_{r} \) is a renamed
apart clause of \( \mathit{P} \). Now, by the hypothesis that 
\( P\, \vdash \, \forall V_{1},\ldots ,V_{m}\, (bd_{r}\arr \mathit{newbd}_{r}) \),
by the fact that \( \mathit{vars}(\{\mathit{V}_{1}/\mathit{u}_{1},\ldots ,
\mathit{V}_{m}/\mathit{u}_{m}\})\cap 
\mathit{vars}(\mathit{bd}_{r},\mathit{newbd}_{r}) 
\subseteq \{V_{1},\ldots ,V_{m}\}  \)
and \( \mathit{vars}(g_{1})\cap 
\mathit{vars}(\mathit{bd}_{r},\mathit{newbd}_{r}) 
\subseteq \{V_{1},\ldots ,V_{m}\}\),
and by Proposition~\ref{proposition:context}~(ii), we have that
there exists \( A_{2}\in \mathcal{P}(\mathit{Subst}) \) such that
\( \mathit{P}\, \vdash \, \mathit{newbd}_{r}\{V_{1}/u_{1},\ldots , \)
\( V_{m}/u_{m}\}\wedge g_{1}\, \mapsto \, A_{2} \) has a proof of
depth \( n1 \) and size \( s1 \), with \( m\! -\! 1\geq n1 \) and
\( A_{1}\! =\! \emptyset  \) iff \( A_{2}\! =\! \emptyset  \). Since
\( \langle n1,s1\rangle <\langle m,s\rangle  \), by induction hypothesis
there exists \( B_{1}\in \mathcal{P}(\mathit{Subst}) \) such that
\( \mathit{NewP}\, \vdash \, \mathit{newbd}_{r}\{V_{1}/u_{1},\ldots , \)
\( V_{m}/u_{m}\}\wedge g_{1}\mapsto \, B_{1} \) has a proof of depth
\( n2 \) with \( n1\geq n2 \) and \( A_{2}\! =\! \emptyset  \)
iff \( B_{1}\! =\! \emptyset  \). Since \( \mathit{hd}_{r} \) is
\( p(V_{1},\ldots ,V_{m}) \), by using rule \( (\mathit{at}) \)
we can construct a proof for \( \mathit{NewP}\, \vdash \, p(u_{1},\ldots ,
u_{m})\wedge g_{1}\mapsto B \)
of depth \( n=n2\! +\! 1 \) where \( B\! =\! (B_{1}\! \upharpoonright \! 
\mathit{vars}(p(u_{1},\ldots ,u_{k})\wedge g_{1})) \).
Thus, \( m\geq n \) and, by the definition of the~\( \upharpoonright  \)
operator, \( A\! =\! \emptyset  \) iff \( A_{1}\! =\! \emptyset  \)
iff \( A_{2}\! =\! \emptyset  \) iff \( B_{1}\! =\! \emptyset  \)
iff \( B\! =\! \emptyset  \). \hfill\end{proof}

\begin{proof}[Proof of Lemma~\ref{lemma:improving_rules}]
(i) Let us consider the transformation sequence \(
P_{i},\ldots ,P_{j} \).
Let us also consider any index $h$ in $\{i,\ldots,j\!-\!1\}$ 
and any two clauses
$c_1$: $\mathit{hd}\leftarrow \mathit{bd}$ in program $P_h$ and
$c_2$:~$\mathit{hd}\leftarrow \mathit{newbd}$ in program $P_{h\!+\!1}$.
Since  \( P_{i},\ldots ,P_{j} \) is constructed by using the unfolding rule
only, we have that:

\smallskip{}
\( \mathit{bd} \) = \( b[p(u_{1},\ldots ,u_{m})] \) ~~and~~
\( \mathit{newbd} \)
= \( b[g\{V_{1}/u_{1},\ldots ,V_{m}/u_{m}\}] \)
\smallskip{}

\noindent 
for some clause 
\( p(V_{1},\ldots ,V_{m})\leftarrow g \)
in \( P_{i}\), some goal context \( b[\_] \), and 
some $m$-tuple of arguments
$(u_1,\ldots,u_m)$. To prove this lemma we have to show
that:

\smallskip{}
\( P_{i}\, \vdash \, \forall V\,
(b[p(u_{1},\ldots ,u_{m})]\arr
b[g\{V_{1}/u_{1},\ldots ,V_{m}/u_{m}\}]) \)\hfill{}(\( \alpha  \))
\smallskip{}

\noindent where \( V=\mathit{vars}(\mathit{hd}) \).
Now, for every clause \( p(V_{1},\ldots ,V_{m})\leftarrow g \) in
\( P_{i}\) we have that: 

\smallskip{}
\( P_{i}\, \vdash \, \forall V_1,\ldots,V_m \, 
(p(V_{1},\ldots ,V_{m})\arr g) \)\hfill{}(\( \beta  \))

\smallskip{}

\noindent From~(\( \beta  \)), by Point~(iv$'$) of 
Proposition~\ref{proposition:repl}
we get:

\smallskip

\( P_{i}\, \vdash \, \forall W\,
(p(u_{1},\ldots ,u_{m})\arr g\{V_{1}/u_{1},\ldots ,
V_{m}/u_{m}\}) \)\hfill{}(\( \gamma  \))

\smallskip

\noindent where $W=\mathit{vars}(u_{1},\ldots ,u_{m})$.
From~(\( \gamma  \)), by Point~(i$'$) of 
Proposition~\ref{proposition:repl}
we get:

\smallskip

\( P_{i}\, \vdash \, 
\forall Z\, (b[p(u_{1},\ldots ,u_{m})]\arr 
b[g\{V_{1}/u_{1},\ldots,V_{m}/u_{m}\}] )\)\hfill{}(\( \delta  \))

\smallskip

\noindent
where $Z= \mathit{vars}(b[p(u_{1},\ldots ,u_{m})])$.
From~(\( \delta  \)), by Points~(ii$'$) and (iii$'$) of 
Proposition~\ref{proposition:repl}
we get (\( \alpha  \)), as desired.

\smallskip{}

\noindent (ii)~In order to prove Point~(ii) of the thesis, we first
show the following property.

\smallskip{}
\noindent {\it Property}~(A): For every clause \( d \):
\( \mathit{newp}(V_{1},\ldots ,V_{m})\leftarrow g \)
in \( \mathit{Def}\! _{k} \) which is used for folding during the
construction of the sequence
\( P_{j},\ldots ,P_{k} \), we have that the replacement law
\( P_{j}\, \vdash \,  \forall V_{1},\ldots ,V_{m}
\,(\mathit{newp}(V_{1},\ldots ,V_{m})\llarr g) \) holds. 

\smallskip{}
\noindent Property~(A) is a consequence of the fact that during the
sequence \( P_{i},\ldots ,P_{j} \) we have performed the parallel
leftmost unfolding of every clause which is used for folding during
\( P_{j},\ldots ,P_{k} \). 
\smallskip{}

Now we prove Point~(ii) of the thesis by cases with respect to~the
transformation rule which is used to derive program \( P_{h+1} \)
from program \( P_{h} \), for \( h=j,\ldots ,k\! -\! 1 \).

\smallskip{}
\noindent - Case 1: \( P_{h+1} \) is derived from \( P_{h} \) by
the unfolding rule using a clause which is among those also used for
folding (in a previous transformation step). The thesis follows from
Property~(A) and Points~(i$'$), (ii$'$), (iii$'$), and (iv$'$) 
of Proposition~\ref{proposition:repl}. 

\smallskip{}
\noindent - Case 2: \( P_{h+1} \) is derived from \( P_{h} \) by
the unfolding rule using a clause \( c \) which is {\it not} among
those used for folding. Thus, \( c \) belongs to \( P_{0} \) because
the only way of introducing in the body of a clause an occurrence
of a non-primitive predicate which is not defined in \( P_{0} \),
is by an application of the folding rule. Hence, \( c \) belongs
to \( P_{j} \) as well. Now, for every clause \( c \) of the form:
\( p(V_{1},\ldots ,V_{m})\leftarrow g \) in \( P_{j} \) we have
that: 

\smallskip{}

\( P_{j}\, \vdash \, \forall V_{1},\ldots ,V_{m} \, 
(p(V_{1},\ldots ,V_{m})\arr g) \)

\smallskip{}

\noindent The thesis follows from Property~(A)
and Points~(i$'$), (ii$'$), (iii$'$), and (iv$'$) of 
Proposition~\ref{proposition:repl}. 

\smallskip{}
\noindent - Case 3: \( P_{h+1} \) is derived from \( P_{h} \) by
the folding rule. The thesis follows from Property~(A)
and Points~(i$'$), (ii$'$), (iii$'$), and (iv$'$) of 
Proposition~\ref{proposition:repl}. 

\smallskip{}
\noindent - Case 4: \( P_{h+1} \) is derived from \( P_{h} \) by
the goal replacement rule based on a replacement law of the form 
\( P_{0}\, \vdash \, \forall V\, (g_{1}\arr g_{2}) \).
The thesis follows from Points~(i$'$), (ii$'$), and (iii$'$) 
of Proposition~\ref{proposition:repl}
and the fact that also \( P_{j}\, \vdash \, \forall\, V (g_{1}\arr g_{2}) \)
holds, because the non-primitive predicates of \( \{g_{1},g_{2}\} \)
are defined in \( P_{0} \), and for each predicate \( p \) defined
in \( P_{0} \), the definition of \( p \) in \( P_{0} \) is equal
to the definition of \( p \) in \( P_{j} \). \hfill\end{proof}

\begin{proof}[Proof of Lemma~\ref{lemma:soundness}]
We assume that there exists \( A\in \mathcal{P}(\mathit{Subst}) \) such that
\( \mathit{NewP}\, \vdash \, g\mapsto A \) has a proof of size \( n \).
We have to show that there exists \( B\in \mathcal{P}(\mathit{Subst}) \)
such that \( \mathit{P}\, \vdash \, g\mapsto B \) holds, and \( A\! =\! 
\emptyset  \)
iff \( B\! =\! \emptyset  \). We proceed by induction on \( n \).
We assume that, for all \( m<n \), for all goals \( h \), and for
all \( A_{1}\in \mathcal{P}(\mathit{Subst}) \), if \( \mathit{NewP}\, 
\vdash \, h\mapsto A_{1} \)
has a proof of size \( m \), then~ \( P\, \vdash \, h\mapsto B_{1} \)
has a proof for some \( B_{1}\in \mathcal{P}(\mathit{Subst}) \) such
that \( A_{1}\! =\! \emptyset  \) iff \( B_{1}\! =\! \emptyset  \).
Now we proceed by cases on the structure of \( g \). We consider
the following two cases. The other cases are similar and we omit them. 
\smallskip{}

\noindent - Case 1: \( g \) is \( (g_{1}\! =\! g_{2})\wedge g_{3} \).
Assume that \( \mathit{NewP}\, \vdash \, (g_{1}\! =\! g_{2})\wedge g_{3}
\mapsto A \)
has a proof of size \( n \). Then, \( g_{1} \) is a goal variable,
say \( G \), \( G\not \in \mathit{vars}(g_{2}) \), and \( \mathit{NewP}\, 
\vdash \, (G\! =\! g_{2})\wedge g_{3}\mapsto A \)
has been derived by applying rule \( (\mathit{geq}) \). Thus, there
exists \( A_{1}\in \mathcal{P}(\mathit{Subst}) \) such that \( A \)
is \( (\{G/g_{2}\}\circ A_{1}) \) and \( \mathit{NewP}\, \vdash \, 
g_{3}\{G/g_{2}\}\mapsto A_{1} \)
has a proof of size \( n\! -\! 1 \). By induction hypothesis there
exists \( B_{1}\in \mathcal{P}(\mathit{Subst}) \) such that \( P\, 
\vdash \, g_{3}\{G/g_{2}\}\mapsto B_{1} \)
has a proof and \( A_{1}\! =\! \emptyset  \) iff \( B_{1}\! =\! \emptyset  \).
By using rule \( (\mathit{geq}) \), we can construct a proof of 
\( \mathit{P}\, \vdash \, (G\! =\! g_{2})\wedge g_{3}\mapsto B \)
where \( B \) is \( \{G/g_{2}\}\circ B_{1} \). By the definition of the
\( \circ  \) operator, we have that \( A\! =\! \emptyset  \) iff
\( A_{1}\! =\! \emptyset  \) iff \( B_{1}\! =\! \emptyset  \) iff
\( B\! =\! \emptyset  \).
\smallskip{}

\noindent - Case 2: \( g \) is \( p(u_{1},\ldots ,u_{m})\wedge g_{1} \).
Assume that \( \mathit{NewP}\, \vdash \, p(u_{1},\ldots ,u_{m})\wedge 
g_{1}\mapsto A \)
has a proof of size \( n \). Then, \( \mathit{NewP}\, \vdash \, 
p(u_{1},\ldots ,u_{m})\wedge g_{1}\mapsto A \)
has been derived by applying rule \( (\mathit{at}) \), and there
exists a proof of size \( n-1 \) of \( \mathit{NewP}\, \vdash \, 
newbd_{r}\{V_{1}/u_{1},\ldots , \)
\( V_{m}/u_{m}\}\wedge g_{1}\, \mapsto \, A_{1} \) where 
\( p(V_{1},\ldots ,V_{m})\leftarrow newbd_{r} \)
is a renamed apart clause of \( \mathit{NewP} \) and \( A \) is
\( (A_{1}\! \upharpoonright \! \mathit{vars}(p(u_{1},\ldots ,u_{k})\wedge 
g_{1})) \).
By induction hypothesis there exists a proof of \( \mathit{P}\, \vdash \, 
newbd_{r}\{V_{1}/u_{1},\ldots , \)
\( V_{m}/u_{m}\}\wedge g_{1} \) \( \mapsto \, B_{1} \) such that
\( A_{1}\! =\! \emptyset  \) iff \( B_{1}\! =\! \emptyset  \). 
Now,
by the hypothesis that \( P\, \vdash \, \forall V_{1},\ldots ,V_{m}\,
(newbd_{r}\ar bd_{r}) \),
by the fact that \( \mathit{vars}(\{V_{1}/u_{1},\ldots ,V_{m}/u_{m}\})
\cap \mathit{vars}(bd_{r},newbd_{r}) \subseteq 
\{V_{1},\ldots ,V_{m}\} \)
and \( \mathit{vars}(g_1)
\cap \mathit{vars}(bd_{r},newbd_{r}) \subseteq 
\{V_{1},\ldots ,V_{m}\} \),
and by Proposition~\ref{proposition:context}~(i), we have that
\( \mathit{P}\, \vdash \, bd_{r}\{V_{1}/u_{1},\ldots ,V_{m}/u_{m}\}
\wedge g_{1}\, \mapsto \, B_{2} \)
has a proof for some \( B_{2}\in \mathcal{P}(\mathit{Subst}) \) such
that \( B_{1}\! =\! \emptyset  \) iff \( B_{2}\! =\! \emptyset  \).
Since \( \mathit{hd}_{r} \) is \( p(V_{1},\ldots ,V_{m}) \), by
using rule \( (\mathit{at}) \) we can construct a proof for \( \mathit{P}\,
\vdash \, p(u_{1},\ldots ,u_{m})\wedge g_{1}\mapsto B \)
where \( B \) is \( (B_{2}\! \upharpoonright \! \mathit{vars}(p(u_{1},
\ldots ,u_{k})\wedge g_{1})) \).
By the definition of the \( \upharpoonright  \) operator, we have
that \( A\! =\! \emptyset  \) iff \( A_{1}\! =\! \emptyset  \) iff
\( B_{1}\! =\! \emptyset  \) iff \( B_{2}\! =\! \emptyset  \) iff
\( B\! =\! \emptyset  \). \hfill\end{proof}

\sloppy 
\begin{proof}[Proof of Lemma~\ref{lemma:sound_rules}]
\nopagebreak 
If \( P_{h+1} \)
is derived from \( P_{h} \) by the unfolding rule using a clause
of the form \mbox{$p(V_{1},\ldots ,V_{m})\leftarrow g$} in \( P_{0}\cup
\mathit{Def}\! _{k} \), then the thesis follows from Points~(i), (ii),
(iii), and (iv) of Proposition~\ref{proposition:repl}, 
and the fact that the replacement law  \( P_{0}\cup
\mathit{Def}\! _{k}\, \vdash \, \forall V_{1},\ldots ,V_{m}\, 
(g\ar p(V_{1},\ldots ,V_{m})) \) holds. Similarly, if \(
P_{h+1} \) is derived from \( P_{h} \) by the folding rule using a clause of
the form  \( \mathit{newp}(V_{1},\ldots ,V_{m})\leftarrow g \) in \(
\mathit{Def}\! _{k} \), then the thesis follows from Points~(i), (ii), 
(iii), and (iv) of
Proposition~\ref{proposition:repl}, and the fact that the replacement law 
\( P_{0}\cup
\mathit{Def}\! _{k}\, \vdash \, \forall V_{1},\ldots ,V_{m}\, 
(\mathit{newp}(V_{1},\ldots ,V_{m})\ar  g) \) holds. Finally, if
\( P_{h+1} \) is derived from \( P_{h} \) by the goal replacement rule, then
the thesis follows from the fact that it is based on a strong replacement law
and from Points~(i), (ii), and (iii) 
of Proposition~\ref{proposition:repl}.\hfill\end{proof}
\fussy

\noindent The following Lemma~\ref{lemma:safety} and 
Lemma~\ref{lemma:safe_rules}
are necessary for proving that a transformation 
sequence preserves safety
(see Theorem~\ref{th:safety}).

\begin{lemma}
\label{lemma:safety}Let \( P \) and \(
\mathit{NewP} \) be programs of the form:

\smallskip{}
\textup{\begin{tabular}{ccccc}
\( P: \) &
\( hd_{1}\leftarrow bd_{1} \) &
~~~~~~~~~~~~~~&
\( \mathit{NewP}: \) &
\( hd_{1}\leftarrow \mathit{newbd}_{1} \)\\
&
~\( \vdots  \)&
&
&
\( \vdots  \)~~~~~\\
&
\( hd_{s}\leftarrow bd_{s} \)&
&
&
\( hd_{s}\leftarrow \mathit{newbd}_{s} \)\\
\end{tabular}}
\smallskip{}

\noindent 
Suppose that for \( r=1,\ldots ,s \) and for every goal context \( b[\_] \)
such that \( \mathit{vars}(b[\_])\cap 
\mathit{vars}({\it bd_{r}},{\it newbd_{r}})\subseteq\mathit{vars}(hd_r)  \),
we have that if \( b[\mathit{bd}_{r}] \) is safe in \( P \) then
\( b[\mathit{newbd}_{r}] \) is safe in \( \mathit{P}
\). Then, for every goal \( g \), if \( g \) is safe in \( P \) then
\( g \) is safe in \( \mathit{NewP} \).
\end{lemma}
\begin{proof}[Proof of Lemma~\ref{lemma:safety}]
We assume that \( g \) is not safe in \( \mathit{NewP} \)
and we prove that \( g \) is not safe in \( P \). Since \( g \)
is not safe in {\it NewP}, there exist \( A\in \mathcal{P}(\mathit{Subst}) \)
and a deduction tree \( \tau  \) for \( \mathit{NewP}\vdash g\mapsto A \)
such that a leaf of \( \tau  \) is of the form \( \mathit{NewP}\, \vdash \,
g_{\scriptstyle{stuck}}\mapsto B \)
and the goal \( g_{\scriptstyle{stuck}} \) is stuck. 
We proceed by induction on
the size of \( \tau  \). We consider the following two cases only.
The others are similar and we omit them. 

\noindent - Case 1: \( g \) is \( (g_{1}\! =\! g_{2})\wedge g_{3} \).
Assume that the deduction tree \( \tau  \) for
\( \mathit{NewP}\, \vdash \, $\mbox{$(g_{1}\!=\!g_{2})$}$ \wedge g_{3}\mapsto
A \) has size \( s \). If \( g_{1} \) is not a goal variable or it is
a goal variable occurring in \( g_{2} \), then \( (g_{1}\! =\! g_{2})\wedge 
g_{3} \)
is not safe in \( P \). Otherwise, \( g_{1} \) is a goal variable,
say \( G \), and \( G\not \in \mathit{vars}(g_{2}) \). 
Thus, \( \mathit{NewP}\, \vdash \, (G\! =\! g_{2})\wedge g_{3}\mapsto A \)
has been derived by applying rule~\( (\mathit{geq}) \), and there
exists \( A_{1}\in \mathcal{P}(\mathit{Subst}) \) such that: (a)
the subtree \( \tau _{1} \) of \( \tau  \) rooted at \( \mathit{NewP}\, 
\vdash \, g_{3}\{G/g_{2}\}\mapsto A_{1} \)
has size \( s\! -\! 1 \), and (b) \( \mathit{NewP}\, \vdash \,
g_{\scriptstyle{stuck}}\mapsto B \)
is a leaf of \( \tau _{1} \). By induction hypothesis \( g_{3}\{G/g_{2}\} \)
is not safe in {\it P} and, by rule~\( (\mathit{geq}) \), also
\( (G\! =\! g_{2})\wedge g_{3} \) is not safe in \( P \). 
\smallskip{}

\sloppy
\noindent - Case 2: \( g \) is \( p(u_{1},\ldots ,u_{m})\wedge g_{1} \).
Assume that the deduction tree \( \tau  \) for \( \mathit{NewP}\, \vdash \, 
p(u_{1},\ldots ,u_{m})\wedge g_{1}\mapsto A \)
has size \( s \). Thus, \( \mathit{NewP}\, \vdash \, p(u_{1},\ldots ,
u_{m})\wedge g_{1}\mapsto A \)
has been derived by using rule~\( (\mathit{at}) \), and there exist
\( A'\in \mathcal{P}(\mathit{Subst}) \) and a renamed apart clause
\( p(V_{1},\ldots ,V_{m})\leftarrow \mathit{newbd}_{r} \) of \( 
\mathit{NewP} \)
such that: (a) the subtree \( \tau _{1} \) of \( \tau  \) rooted
at \( \mathit{NewP}\, \vdash \, \mathit{newbd}_{r}\{V_{1}/u_{1},\ldots ,
V_{m}/u_{m}\}\wedge g_{1}\, \mapsto A' \)
has size \( s\! -\! 1 \) and (b) \( \mathit{NewP}\, \vdash \,
g_{\scriptstyle{stuck}}\mapsto B \) is a leaf of \( \tau _{1} \). 
By induction
hypothesis \( \mathit{newbd}_{r}\{V_{1}/u_{1},\ldots ,V_{m}/u_{m}\}\wedge
g_{1} \) is not safe in \( P \). 
Now, by hypothesis, by the fact that 
\(\mathit{vars}(\{\mathit{V}_{1}/\mathit{u}_{1},\ldots,
\mathit{V}_{m}/\mathit{u}_{m}\})\cap 
\mathit{vars}(bd_{r},{\it newbd_r})\subseteq \{V_{1},\ldots ,V_{m}\}\)
and \(\mathit{vars}(g_1)\cap 
\mathit{vars}(bd_{r},{\it newbd_r})\subseteq \{V_{1},\ldots ,V_{m}\}\), and by
Proposition~\ref{proposition:context}~(iii), we have that \(
\mathit{bd}_{r}\{V_{1}/u_{1},\ldots ,V_{m}/u_{m}\}\wedge g_{1} \) is not safe
in \( P \). Since \( p(V_{1},\ldots ,V_{m})\leftarrow \mathit{bd}_{r} \) is a
renamed apart clause of \( \mathit{P} \), by rule~\( (\mathit{at}) \), also \(
p(u_{1},\ldots ,u_{m})\wedge g_{1} \) is not safe in \( P \).~\hfill\end{proof}

\fussy

\begin{lemma}
\noindent\label{lemma:safe_rules}Let
\( P_{0},\ldots ,P_{k} \) be a transformation sequence and let 
\( \mathit{Def}\! _{k} \)
be the set of definitions introduced during that sequence. 
For \( h=0,\ldots ,k\! -\! 1 \), for any pair of clauses
$c_1$: \( \mathit{hd}\leftarrow \mathit{bd} \)
in program $P_h$ and 
$c_2$: \( \mathit{hd}\leftarrow \mathit{newbd} \) in program
$P_{h+1}$, such that $c_2$ is derived from $c_1$
by an application of the unfolding rule, or folding
rule, or goal replacement rule which preserves safety, and for every
goal context \( b[\_] \) such that 
\( \mathit{vars}(b[\_])\cap \mathit{vars}(\mathit{bd},\mathit{newbd})
\subseteq \mathit{vars}(\mathit{hd}) \), we have that:

\smallskip

if \( b[\mathit{bd}] \) is safe in \( P_{0}\cup \mathit{Def}\! _{k} \)
then \( b[\mathit{newbd}] \) is safe in
\( P_{0}\cup \mathit{Def}\! _{k} \).
\end{lemma}

\begin{proof}[Proof of Lemma~\ref{lemma:safe_rules}]
First we notice that,
for every clause \( hd_{0}\leftarrow bd_{0} \) in \( P_{0}\cup \mathit{Def}\!
_{k} \) and for every goal context \( b[\_] \) such that
\( \mathit{vars}(b[\_])\cap \mathit{vars}(bd_{0})
\subseteq \mathit{vars}(hd_{0}) \), we have the following: 

\smallskip{}
\noindent {\it Property} (S):~~\( b[hd_{0}] \) is safe in \( P_{0}\cup
\mathit{Def}\! _{k} \)
iff \( b[bd_{0}] \) is safe in \( P_{0}\cup \mathit{Def}\! _{k} \).
\smallskip{}

\noindent Now, take any \( h=0,\ldots ,k\! -\! 1 \). We reason by
cases on the transformation rule applied for deriving the clause \( hd
\leftarrow newbd \) in $P_{h+1}$ 
from the clause \( hd\leftarrow bd\) in $P_h$. 

If \( hd\leftarrow newbd \) is derived from 
\( hd\leftarrow bd\)
by the unfolding rule using a clause \( hd_{0}\leftarrow bd_{0} \)
in \( P_{0}\cup \mathit{Def}\! _{k} \), then for some goal context
\( g[\_] \), \( bd \) is of the form \( g[hd_{0}\vartheta ] \)
and \( newbd \) is of the form \( g[bd_{0}\vartheta ] \). Then
the thesis follows from the {\it only}-{\it if} part of Property~(S).

Similarly, if \( hd\leftarrow newbd \) is derived 
from \( hd\leftarrow bd \)
by the folding rule using a clause \( hd_{0}\leftarrow bd_{0} \)
in \( P_{0}\cup \mathit{Def}\! _{k} \), then for some goal context
\( g[\_] \), \( bd \) is of the form \( g[bd_{0}\vartheta ] \)
and \( newbd \) is of the form \( g[hd_{0}\vartheta ] \). Then
the thesis follows from the {\it if} part of Property~(S).

Finally, if \( hd\leftarrow newbd \) is derived from 
\( hd\leftarrow bd\)
by the goal replacement rule, then the thesis follows from the hypothesis
that every application of the goal replacement rule preserves safety.
\hfill\end{proof}

\begin{proof}[Proof of Theorem~\ref{th:preservation} \rm{(Preservation
of Successes and Failures)}.]By
Proposition~\ref{prop:ordered}, without loss of generality we may assume that
the admissible sequence \( P_{0},\ldots ,P_{k} \) is ordered. Let \(
P_{j} \) be the program obtained at the end of the second subsequence of
\( P_{0},\ldots ,P_{k} \), that is, after unfolding every clause in
\( \mathit{Def}\! _{k} \) which is used for folding. 
Point~(1) of this theorem is a
consequence of the following two facts:

\noindent (F1)~by Lemma~\ref{lemma:improvement} and Point~(i)
of Lemma~\ref{lemma:improving_rules}, we have that, for every goal
\( g \) and for every \( b\in \{\mathit{true},\mathit{false}\} \),
if~ \( P_{0}\cup \mathit{Def}\! _{k}\, \vdash \, g\downarrow _{m}b \)
~then~ \( \mathit{P}_{j}\, \vdash \, g\downarrow _{n1}b \) with
\( m\geq n1 \), and 

\noindent (F2)~by Lemma~\ref{lemma:improvement} and Point~(ii)
of Lemma~\ref{lemma:improving_rules}, we have that: for every goal
\( g \) and for every \( b\in \{\mathit{true},\mathit{false}\} \),
if~ \( P_{j}\, \vdash \, g\downarrow _{n1}b \) ~then~ 
\( \mathit{P}_{k}\, \vdash \, g\downarrow _{n}b \)
with \( n1\geq n \).

\noindent Point~(2) of this theorem is a straightforward consequence of 
Lemmata~\ref{lemma:soundness}
and \ref{lemma:sound_rules}.\hfill\end{proof}

\begin{proof}[Proof of Theorem~\ref{th:correctness} \rm{(Correctness
Theorem)}.](1) First we prove that \( P_{0}\cup 
\mathit{Def}\! _{k}\sqsubseteq P_{k} \).
Let \( g \) be an ordinary goal and let \( A \) be a set of substitutions
such that \( P_{0}\cup \mathit{Def}\! _{k}\vdash g\mapsto A \). We
have to prove that there exists \( B\in \mathcal{P}(\mathit{Subst}) \)
such that \( P_{k}\, \vdash \, g\mapsto B \) and \( A \) and \( B \)
are equally general with respect to~\( g \).

\noindent Since~\( P_{0}\cup \mathit{Def}\! _{k}\vdash g\mapsto A \),
by definition there exists \( b\in \{\mathit{true},\mathit{false}\} \)
such that \( P_{0}\cup \mathit{Def}\! _{k}\vdash g\downarrow b \).
By Point~(1) of Theorem~\ref{th:preservation}, 
we have that \( P_{k}\vdash g\downarrow b \)
and, thus, there exists \( B\in \mathcal{P}(\mathit{Subst}) \) such
that \( P_{k}\vdash g\mapsto B \).

\noindent In order to prove that \( A \) and \( B \) are equally
general with respect to~\( g \), we have to show that: (a) for every
substitution \( \alpha \in A \) there exists a substitution \( \beta \in B \)
such that \( g\alpha  \) is an instance of \( g\beta  \), and
(b) for every \( \beta \in B \) there exists \( \alpha \in A \)
such that \( g\beta  \) is an instance of \( g\alpha  \).

\noindent (a)~Let \( \alpha  \) be a substitution in \( A \). From
\( P_{0}\cup \mathit{Def}\! _{k}\vdash g\mapsto A \), by 
Proposition~\ref{proposition:cas}~(ii.1),
we have that \( P_{0}\cup \mathit{Def}_{\! k}\vdash g\alpha 
\downarrow \mathit{true} \).
Thus, by Point~(1) of Theorem~\ref{th:preservation}, we have that
\( P_{k}\vdash g\alpha \downarrow \mathit{true} \). Since \( P_{k}\vdash 
g\mapsto B \)
holds, by Proposition~\ref{proposition:cas}~(ii.1), there exists
a substitution \( \beta \in B \) such that \( g\alpha  \) is an
instance of \( g\beta  \). 

\noindent (b)~Let \( \beta  \) be a substitution in \( B \). From
\( P_{k}\vdash g\mapsto B \), by Proposition~\ref{proposition:cas}~(ii.1),
we have that \( P_{k}\vdash g\beta \downarrow \mathit{true} \). From
\( P_{0}\cup \mathit{Def}\! _{k}\vdash g\mapsto A \), by 
Proposition~\ref{proposition:cas}~(i),
we have that {\it either} \( P_{0}\cup \mathit{Def}\! _{k}\vdash g\beta
\downarrow \mathit{true} \)
{\it or} \( P_{0}\cup \mathit{Def}\! _{k}\vdash g\beta \downarrow
\mathit{false} \).
Now \( P_{0}\cup \mathit{Def}\! _{k}\vdash g\beta \downarrow \mathit{false} \)
is impossible because by Point~(1) of Theorem~\ref{th:preservation},
we would have \( P_{k}\vdash g\beta \downarrow \mathit{false} \).
Thus, \( P_{0}\cup \mathit{Def}\! _{k}\vdash g\beta \downarrow 
\mathit{true} \).
Since~\( P_{0}\cup \mathit{Def}_{\! k}\vdash g\mapsto A \), by 
Proposition~\ref{proposition:cas}~(ii.1),
there exists \( \alpha \in A \) such that \( g\beta  \) is an instance
of \( g\alpha  \). 
\smallskip{}

\noindent (2)~We have to prove that if all applications of the goal
replacement rule in the sequence \( P_{0},\ldots ,P_{k} \) are based
on strong replacement laws, then 
\( P_{0}\cup \mathit{Def}_{\! k}\equiv P_{k} \).
Since \( P_{0}\cup \mathit{Def}_{\! k}\sqsubseteq P_{k} \) has been
shown at Point~(1) of this proof, it remains to show that: \( P_{k}\sqsubseteq 
P_{0}\cup \mathit{Def}_{\! k} \).
The proof is similar to that of Point~(1) and it is based on Point~(2)
of Theorem~\ref{th:preservation} and
Proposition~\ref{proposition:cas}~(ii.1).
\hfill\end{proof}

\begin{proof}[Proof of Theorem~\ref{th:safety} \rm{(Preservation
of Safety)}.]
Let \( hd\leftarrow bd \) be a clause in \( P_{0}\cup \mathit{Def}\! _{k}
\) and let \( hd\leftarrow newbd \) be the clause in \( P_{k} \) with
the same head. By Lemma~\ref{lemma:safe_rules} we have that, for every goal context
\( b[\_] \) such that \( \mathit{vars}(b[\_])\cap 
\mathit{vars}({\it bd},{\it newbd})\subseteq \mathit{vars}(hd) \), 
if \( b[\mathit{bd}] \) is safe in \( P_{0}\cup \mathit{Def}_{k} \)
then \( b[\mathit{newbd}] \) is safe in \( P_{0}\cup \mathit{Def}\! _{k} \).
Then, by Lemma~\ref{lemma:safety}, for every goal \( g \), if \( g \)
is safe in \( P_{0}\cup \mathit{Def}\! _{k} \) then \( g \) is safe
in \( P_{k} \).\hfill\end{proof}


\end{document}